\RequirePackage[colorlinks,allcolors=blue]{hyperref} % load 'hyperref' first

\documentclass[]{agujournal2019}
\usepackage{siunitx}
\usepackage[export]{adjustbox}
\usepackage{url} %this package should fix any errors with URLs in refs.
\usepackage{lineno}
\usepackage[inline]{trackchanges} %for better track changes. finalnew option will compile document with changes incorporated.
\usepackage{soul}
\draftfalse

\journalname{JGR: Planets}

\begin{document}

%%%%%%%%%%%%%%%%%%%%%%%%%%%%%%%%%%%%%%%%%%%%%%%
%  TITLE
%
% (A title should be specific, informative, and brief. Use
% abbreviations only if they are defined in the abstract. Titles that
% start with general keywords then specific terms are optimized in
% searches)
%
%%%%%%%%%%%%%%%%%%%%%%%%%%%%%%%%%%%%%%%%%%%%%%%

% Example: \title{This is a test title}

\title{Saturn's Atmosphere in Northern Summer Revealed by JWST/MIRI}

%%%%%%%%%%%%%%%%%%%%%%%%%%%%%%%%%%%%%%%%%%%%%%%
%
%  AUTHORS AND AFFILIATIONS
%
%%%%%%%%%%%%%%%%%%%%%%%%%%%%%%%%%%%%%%%%%%%%%%%

% Authors are individuals who have significantly contributed to the
% research and preparation of the article. Group authors are allowed, if
% each author in the group is separately identified in an appendix.)

% List authors by first name or initial followed by last name and
% separated by commas. Use \affil{} to number affiliations, and
% \thanks{} for author notes.
% Additional author notes should be indicated with \thanks{} (for
% example, for current addresses).

% Example: \authors{A. B. Author\affil{1}\thanks{Current address, Antartica}, B. C. Author\affil{2,3}, and D. E.
% Author\affil{3,4}\thanks{Also funded by Monsanto.}}

\authors{Leigh N. Fletcher\affil{1}, 
    Oliver R.T. King\affil{1},
    Jake Harkett\affil{1},
    Heidi B. Hammel\affil{2},
    Michael T. Roman\affil{1},
    Henrik Melin\affil{1},
    Matthew M. Hedman\affil{3},
    Julianne I. Moses\affil{4},
    Sandrine Guerlet\affil{5,6},
    Stefanie N. Milam\affil{7},
    Matthew S. Tiscareno\affil{8}
    }

% \affiliation{1}{First Affiliation}
% \affiliation{2}{Second Affiliation}
% \affiliation{3}{Third Affiliation}
% \affiliation{4}{Fourth Affiliation}

\affiliation{1}{School of Physics and Astronomy, University of Leicester, University Road, Leicester, LE1 7RH, UK}
%(repeat as many times as is necessary)
\affiliation{2}{Association of Universities for Research in Astronomy, Suite 1475, 1331 Pennsylvania Avenue NW, Washington DC 20004, USA}
\affiliation{3}{Department of Physics, University of Idaho, Moscow, Idaho 83844, USA}
\affiliation{4}{Space Science Institute, Boulder, CO, USA}
\affiliation{5}{Laboratoire de M\'{e}t\'{e}orologie Dynamique/Institut Pierre-Simon Laplace (LMD/IPSL), Sorbonne Universit\'{e}, CNRS, \'{E}cole Polytechnique, Institut Polytechnique de Paris, \'{E}cole Normale Sup\'{e}rieure (ENS), PSL Research University, Paris, France}
\affiliation{6}{LESIA, Observatoire de Paris, Universit\'{e} PSL, CNRS, Sorbonne Universit\'{e}, Univ. Paris-Diderot, Sorbonne Paris-Cit\'{e}, Meudon, France}
\affiliation{7}{Astrochemistry Laboratory Code 691, NASA Goddard Space Flight Center, 8800 Greenbelt Road, Greenbelt MD 20771, USA}
\affiliation{8}{SETI Institute, Mountain View, CA 94043, USA}

% Corresponding author mailing address and e-mail address:

% (include name and email addresses of the corresponding author.  More
% than one corresponding author is allowed in this LaTeX file and for
% publication; but only one corresponding author is allowed in our
% editorial system.)

% Example: \correspondingauthor{First and Last Name}{email@address.edu}

\correspondingauthor{Leigh N. Fletcher}{leigh.fletcher@le.ac.uk}

%%%%%%%%%%%%%%%%%%%%%%%%%%%%%%%%%%%%%%%%%%%%%%%
% KEY POINTS
%%%%%%%%%%%%%%%%%%%%%%%%%%%%%%%%%%%%%%%%%%%%%%%
%  List up to three key points (at least one is required)
%  Key Points summarize the main points and conclusions of the article
%  Each must be 140 characters or fewer with no special characters or punctuation and must be complete sentences

% Example:
% \begin{keypoints}
% \item	List up to three key points (at least one is required)
% \item	Key Points summarize the main points and conclusions of the article
% \item	Each must be 140 characters or fewer with no special characters or punctuation and must be complete sentences
% \end{keypoints}

\begin{keypoints}
\item Saturn's northern summertime hemisphere was mapped by JWST/MIRI to study seasonal evolution of temperatures, aerosols, and composition.
\item The data show evidence for changing temperatures and winds in the equatorial oscillation, polar vortices, and interhemispheric stratospheric circulation.
\item MIRI spectral coverage and sensitivity enables mapping of several gases for the first time, particularly in ranges inaccessible to Cassini.
\end{keypoints}

%%%%%%%%%%%%%%%%%%%%%%%%%%%%%%%%%%%%%%%%%%%%%%%
%
%  ABSTRACT and PLAIN LANGUAGE SUMMARY
%
% A good Abstract will begin with a short description of the problem
% being addressed, briefly describe the new data or analyses, then
% briefly states the main conclusion(s) and how they are supported and
% uncertainties.

% The Plain Language Summary should be written for a broad audience,
% including journalists and the science-interested public, that will not have 
% a background in your field.
%
% A Plain Language Summary is required in GRL, JGR: Planets, JGR: Biogeosciences,
% JGR: Oceans, G-Cubed, Reviews of Geophysics, and JAMES.
% see http://sharingscience.agu.org/creating-plain-language-summary/)
%
%%%%%%%%%%%%%%%%%%%%%%%%%%%%%%%%%%%%%%%%%%%%%%%

%% \begin{abstract} starts the second page
% Abstract · Be set as a single paragraph. · Be less than 250 words for all journals except GRL, for which the limit is 150 words.

\begin{abstract}

Saturn’s northern summertime hemisphere was mapped by JWST/MIRI (4.9-27.9 $\mu$m) in November 2022, tracing the seasonal evolution of temperatures, aerosols, and chemical species in the five years since the end of the Cassini mission.  We provide algorithms to clean MRS spectral cubes, revealing Saturn’s banded structure and discrete meteorological features.  The transitional spectral region between reflected sunlight and thermal emission (5.1-6.8 $\mu$m) is mapped for the first time, enabling retrievals of phosphine, ammonia, and water, alongside a stacked system of two aerosol layers (an upper tropospheric haze $p<0.3$ bars, and a deeper cloud layer at 1-2 bars).  Ammonia displays substantial equatorial enrichment, suggesting similar dynamical processes to those found in Jupiter’s equatorial zone.  Saturn’s North Polar Stratospheric Vortex has warmed since 2017, is entrained by westward winds at $p<10$ mbar, and exhibits localised enhancements in several hydrocarbons.  The strongest latitudinal temperature gradients are co-located with the peaks of the zonal winds, implying wind decay with altitude.  Reflectivity contrasts at 5-6 $\mu$m compare favourably with albedo contrasts observed by Hubble, and several discrete vortices are observed.  A warm equatorial stratospheric band in 2022 is not consistent with a 15-year repeatability for the equatorial oscillation.  A stacked system of windshear zones dominates Saturn’s equatorial stratosphere, and implies a westward equatorial jet near 1-5 mbar at this epoch.  Lower stratospheric temperatures, and local minima in the distributions of several hydrocarbons, imply low-latitude upwelling and a reversal of Saturn’s interhemispheric circulation since equinox.  Latitudinal distributions of stratospheric ethylene, benzene, methyl and carbon dioxide are presented for the first time, and we report the first detection of new propane bands in the 8-11 $\mu$m region.

\end{abstract}

\section*{Plain Language Summary}
The Saturn system, with its seasonally-varying atmosphere, delicate rings, and myriad satellites, presented an ideal early target for JWST.  Saturn's extended disc, rapid rotation, and infrared brightness provided a challenge for the small fields-of-view of the Mid-Infrared Instrument (MIRI), requiring a mosaic to map Saturn's northern summertime hemisphere.  This exquisite dataset reveals Saturn's banded structure, discrete vortices, the warm polar vortices, and the continued evolution of an oscillatory pattern of warm and cool anomalies over Saturn's equator.  We show evidence that a stratospheric circulation pattern detected by Cassini during northern winter has now fully reversed in northern summer, with the low-latitude stratosphere being cool and depleted in gases due to summertime upwelling.  MIRI provides access to spectral regions that were not possible with the Cassini spacecraft, particularly in the 5-7 $\mu$m region where reflected sunlight and thermal emission blend together.  Our measurements reveal a stacked system of aerosol layers modulating the infrared brightness.  Ammonia and phosphine are enriched at Saturn's equator, suggesting strong mixing from the deeper troposphere.  The high sensitivity of MIRI enables the first identification of previously unseen emission propane bands, along with the first measurements of the distribution of several gaseous species:  tropospheric water, and stratospheric ethylene, benzene, methyl, and carbon dioxide, all of which provide insights into the circulation and chemistry shaping Saturn's seasonal atmosphere.

\section{Introduction}
\label{intro}

Spectroscopic mid-infrared observations of the Saturn system by JWST \cite{23gardner} were designed to build on the legacy of discoveries of the Cassini-Huygens mission (2004-2017), exploiting the unprecedented spectral coverage and sensitivity of the MIRI/MRS \cite<Medium Resolution Spectrometer, 4.9-27.9 $\mu$m,>[]{23wright} integral field units.  As part of a Guaranteed-Time programme for giant planet observations during JWST's first cycle of operations \cite{21fletcher_epsc}, Saturn provided an ideal test of the capabilities of this new facility.  For example, Saturn's large angular size compared to the small fields-of-view of MIRI/MRS presented a challenge for mapping extended, rotating, and moving sources.  Saturn's spectrum has a large dynamic range, with some regions (e.g., near 6 $\mu$m) sufficiently dark as to require long integrations, but others (e.g., near 25 $\mu$m) so bright that they are close to the saturation limit of the sensitive MIRI/MRS detectors.  Observations of Saturn's small satellites are challenged by scattered light from Saturn's atmosphere and rings.  And given that Saturn's forest of molecular emission and absorption features were previously characterised in detail by Cassini, the Saturn observations provided a sensitive check on the calibration of JWST (e.g., wavelength and flux calibration, and the presence of instrumental artefacts).  

In this work, we provide a comprehensive first assessment of JWST mid-infrared observations of Saturn in November 2022 as a baseline for a long-term seasonal legacy for MIRI.  The Cassini record of Saturn's seasonal evolution came to an end in 2017 \cite{20fletcher_saturn}, shortly after Saturn passed northern summer solstice in May 2017 (planetocentric solar longitude of $L_s=90^\circ$).  Five years later, the northern pole was receding from view during Saturn's mid-summer ($L_s=150^\circ$) as the planet approached northern autumn equinox in June 2025 ($L_s=180^\circ$).  Cassini-Huygens did not have the opportunity to observe this particular northern-summer season on Saturn, although it was captured one Saturnian year earlier by ground-based imaging observations \cite<e.g.,>[]{01stam, 08orton_qxo, 22blake} and Hubble observations \cite{92westphal, 93caldwell, 93karkoschka} in the early 1990s.  The mid-IR observations revealed the cooling of the seasonal North Polar Stratospheric Vortex (NPSV), a region of elevated tropospheric and stratospheric temperatures poleward of $75^\circ$N during summer \cite{15guerlet, 18fletcher_poles}.  Observations in 1993-1995 also captured Saturn's equatorial stratospheric oscillation during its cool equatorial phase \cite{08orton_qxo, 22blake}, which should be replicated in JWST/MIRI observations exactly one Saturnian year later if the oscillation were semi-annual \cite<i.e., a 15-year period,>[]{08fouchet, 08orton_qxo}. 

The MIRI/MRS instrument brings exceptional new capabilities for infrared science that eluded even the Cassini-Huygens spacecraft. Saturn's 5-$\mu$m `window,' where deep thermal emission reveals the dynamics and morphology of the cloud-forming region, was extensively mapped by Cassini/VIMS ($R\sim300$ at 5 $\mu$m), but MRS provides an order of magnitude improvement in spectral resolution ($R\sim3500$) required to separate gaseous absorption features of phosphine, ammonia, and water; plus it provides spatial mapping of the 5.1-6.9 $\mu$m range for the first time (reliable Cassini/CIRS spectroscopy started near 7.0 $\mu$m with $R\sim2800$).  Although this region was previously observed in the disc-averaged sense by ISO \cite<the Short-Wave Spectrometer, with $R\sim1000-2000$,>[]{03encrenaz}, the MRS data presented here are the first to spatially map this transitional region between reflected sunlight and thermal emission, revealing discrete features in the cloud deck, and providing access to the spatial distribution of tropospheric water and aerosols, as well as the stratospheric distributions of methane, ethane and other hydrocarbons using previously inaccessible emission bands.  Finally, MIRI/MRS provides a dramatic improvement in signal-to-noise in Saturn's 10-$\mu$m region, which was plagued by noise in Cassini/CIRS observations (this was the overlap between the two mid-infrared focal plane detectors near 9 $\mu$m).

The MIRI/MRS observations provide an exceptionally rich dataset that has revealed how Saturn's atmosphere evolved in the five years since the end of the Cassini mission.  Section \ref{methods} describes the processing steps applied to MIRI/MRS data, including algorithms to address saturation and instrument artefacts.  Section \ref{overview} provides an overview of the spatial structure and spectral features observable by MIRI/MRS, and compares the observations to Cassini data in 2017 to reveal seasonal variability.  Section \ref{spectra} then describes our spectral modelling approach, including both reflected sunlight and thermal emission, which is used in Section \ref{results} to assess Saturn's northern-summer temperatures and zonal wind shears; the spatial distribution of aerosols, tropospheric condensables and disequilibrium species; and a broad range of stratospheric species to trace stratospheric chemistry and circulation.

% Observations of Saturn's rings and small satellites, acquired in the same programme, will be reported elsewhere \cite{23hedman}.  

%%%%%%%%%%%%%%%%%%%%%%%%%%%%%%%%%%%%%%%%%%%%%%%%%%%%%%%%%%
%%%%%%%%%%%%%%%%%%%%%%%%%%%%%%%%%%%%%%%%%%%%%%%%%%%%%%%%%%
%%%%%%%%%%%%%%%%%%%%%%%%%%%%%%%%%%%%%%%%%%%%%%%%%%%%%%%%%%
\section{JWST MIRI Data Processing}
\label{methods}

\subsection{MIRI/MRS Observations}

\begin{figure}[p]
\centering
\includegraphics[width=1.2\textwidth,center]{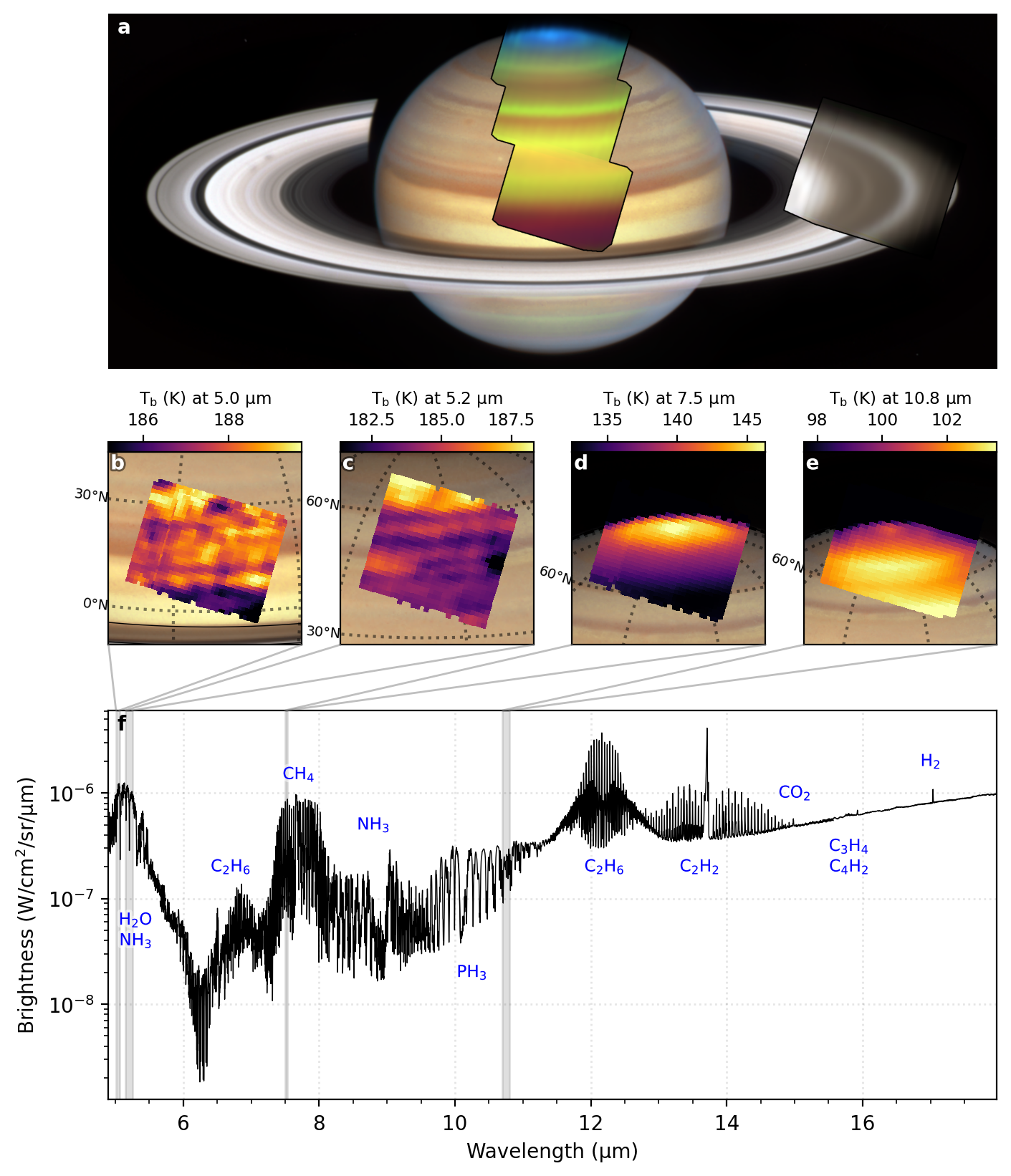}
\caption{Montage of JWST MIRI/MRS observations of Saturn. Panel (a) shows RGB composites of the JWST observations (Saturn: R=\SI{10.3}{\micro\m}, G=\SI{10.1}{\micro\m}, B=\SI{11.6}{\micro\m}; rings: R=\SI{15.5}{\micro\m}, G=\SI{14.6}{\micro\m}, B=\SI{13.5}{\micro\m}) with an HST observation of Saturn in the background \cite{23simon}. Panels (b-e) show spatial structure on Saturn at a range of wavelengths as indicated by the grey shaded regions in panel (f). (f) shows the average spectrum of Saturn with specific spectral features labelled.  Hubble image credit: NASA, ESA, and Amy Simon (NASA-GSFC); Image Processing: Alyssa Pagan (STScI).}
\label{fig:saturn_montage}
\end{figure}

\begin{figure}[t]
\centering
\includegraphics[width=\textwidth]{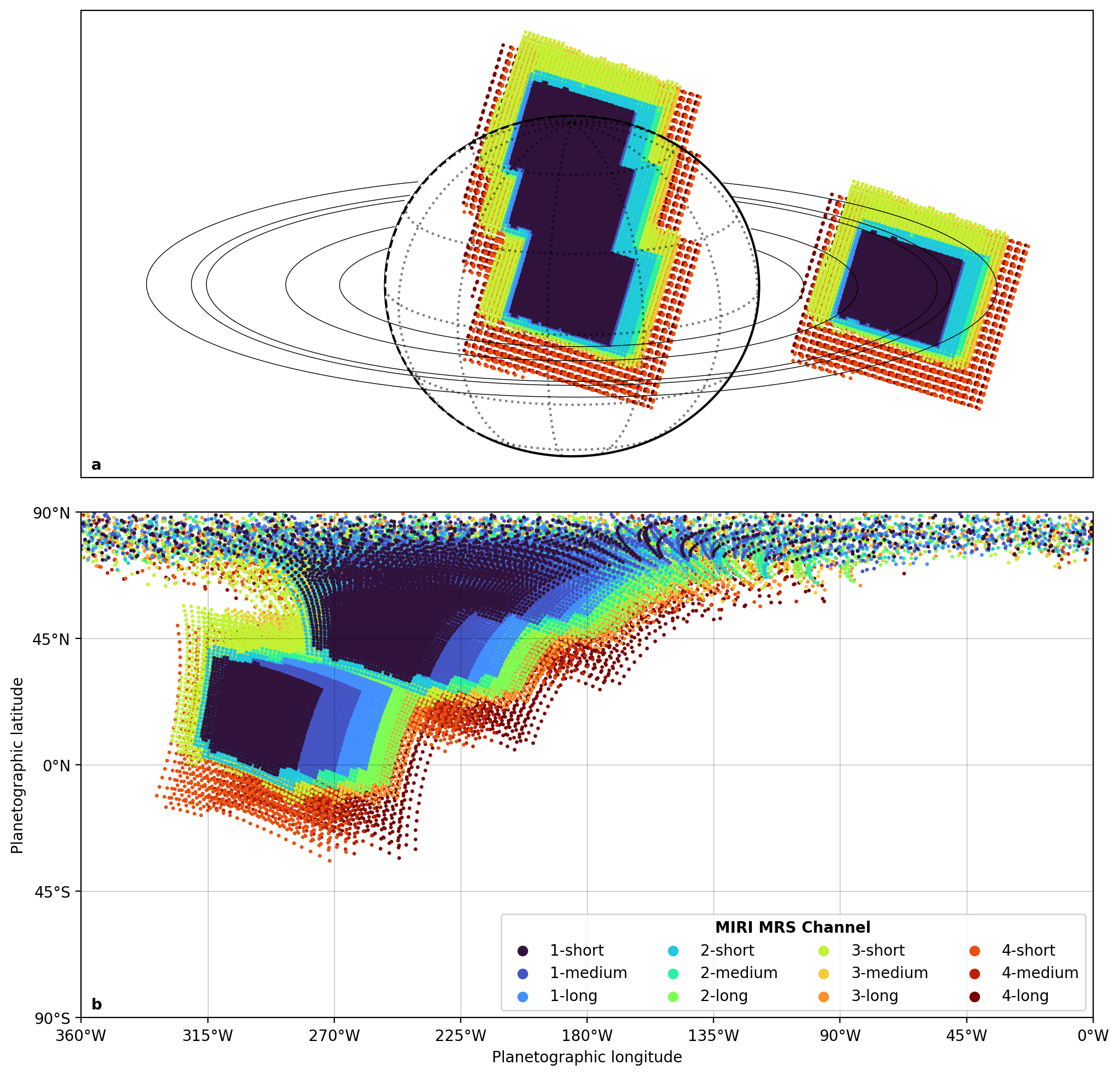}
\caption{Spatial coverage of MIRI observations relative to Saturn's disc (a) and mapped to Saturn's surface (b). Each dot represents the location of a single spaxel with the colour indicating the MIRI channel. The background observation was located 90" to the north of Saturn, so is not shown here (Saturn's disc has a diameter of $\sim17$").}
\label{fig:coverage}
\end{figure}

The MIRI/MRS instrument \cite{15wells} consists of four integral field units (IFUs, channels 1-4) spanning the 4.9-27.9 $\mu$m range with spectral resolutions from $R\sim1330$ at 27.9 $\mu$m to $R\sim3710$ at 4.9 $\mu$m \cite{21labiano}.  Each IFU has a different slice width (from 0.176-0.645") and pixel size (from 0.196-0.273"), and thus field of view, such that they provide different spatial coverage and sampling on Saturn's disc in Fig. \ref{fig:coverage}. Although all four IFUs observe simultaneously (channels 1 and 2 on a \verb|SHORT| detector, channels 3 and 4 on a \verb|LONG| detector), a grating wheel with three different settings (\verb|A, B, C|) is needed for full coverage, leading to a short delay between observations of adjacent portions of the spectra.  The four IFUs and three grating positions provide 12 individual sub-bands, each with its own wavelength coverage and spectral resolution.

The Saturn system observations were part of the Solar System Guaranteed Time Observations (GTO) awarded to H. Hammel, and collated as programme \verb|1247|.  The MRS observations were the first science observations to be executed after a brief hiatus of operations (2022-Aug-24 to 2022-Nov-12), when the MRS grating wheel was found to be experiencing increased friction when moving between short (A), medium (B) and long (C) wavelength settings.  The Saturn observations were redesigned to be executed in reverse-wavelength order (C, to B, to A), with no detriment to the science, and were executed shortly before Saturn left JWST's field of regard in 2022 (i.e., moving below the $85^\circ$ elongation angle from the Sun, as Saturn opposition was earlier in the year on 2022-08-14).  

MIRI/MRS provided a full latitude scan from Saturn's equator to the north pole using three separate mosaic tiles, with a final tile capturing the western ring ansa, as shown in Fig. \ref{fig:saturn_montage}.  Each tile covered the full 4.9-27.9 $\mu$m spectrum, with saturation encountered in the brightest hydrocarbon emission features.  Saturn's angular diameter was $16.9$" at the time of the observations (9.8 AU from JWST, moving away from the observer at 29 km/s), and the spatial coverage of each tile varies with wavelength, from  $3.2\times3.7$" at the shortest wavelength (4.9 $\mu$m) to $6.6\times7.7$" at the longest wavelength (27.9 $\mu$m).  

The MIRI/MRS observations required five separate pointings on 2022-Nov-13 and 2022-Nov-14, as shown in Figure \ref{fig:coverage}.  The western ring ansa was observed first (03:00-04:06UT), followed by an offset 90" north of Saturn (04:11-04:27UT) to determine instrumental artefacts in the MRS observations.  JWST then pointed to Saturn's northern hemisphere, targeting $45^\circ$N (05:40-06:44UT), $15^\circ$N (06:50-07:55UT), but failed to re-acquire a guidestar to complete the final pointing towards Saturn's north pole.  Given that Saturn was about to depart from JWST's field of regard, rapid instructions to re-execute the failed MRS footprint were uploaded to the observatory, enabling the final tile at $75^\circ$N on 2022-Nov-14 (21:58-23:05UT), around 36 hours after the skipped observation was reported.  

With the exception of the $15^\circ$N observation, all MRS tiles used 5 groups (i.e., individual 2.8-second frames) with the \verb|FASTR1| readout pattern, 8 integrations, and a 4-point extended-source dither pattern (no dithers were used for the offset `background' frame).  The equatorial footprint used 4 groups and 10 integrations, to test the impact of using a smaller number of groups on ability to radiometrically calibrate MRS (no problems were identified).  

% \subsubsection{NIRSpec/IFU Observations} 

% NIRSpec/IFU observations of the rings and small satellites used the prism ($R\sim100$) mode to capture the full 0.6-5.3 $\mu$m spectrum, targeting Pandora (2022-Nov-08, 21:55-22:11UT), Epimetheus (2022-Nov-08, 22:15-22:29UT), and Telesto (2022-Nov-10, 12:04-12:25UT).  The Epimetheus observations were also a repeat of an earlier, skipped observation.  Similar observations of Pallene are awaiting execution at the time of writing.  Given the expected brightness of Pallene and Epimetheus, the observations were designed with 2 groups and 3 integrations with the `NRSRAPID' readout mode (10.7-s frames), and a 2-point nod.  Telesto used 15 groups and 1 integration (Pallene was increased to 44 groups), with the `NRSIRS2RAPID' readout mode (14.6-s frames).  The IRS$^2$ mode was intended to improve performance and sensitivity in the longer exposures.

% The NIRSpec $3.0\times3.0$" IFU was used to observe Pandora, Epimetheus and Telesto in prism mode (0.6-5.3 $\mu$m with a spectral resolution $R\sim100$) without spatial resolution, but the first two observations also serendipitously captured the same ring ansa as the MIRI/MRS observations.

\subsection{MIRI/MRS Data Processing}

The MIRI/MRS observations were reduced using the JWST pipeline version \verb|1.9.4| and calibration reference files \verb|1046|\footnote{Available under calibration reference context jwst\_1046.pmap at \url{https://jwst-crds.stsci.edu/context_table/jwst_1046.pmap}.}.  These were applied to the stage-0 \verb|UNCAL| raw data cubes downloaded from MAST.  The final output of the pipeline (stage-3), when run automatically for the archive, are spectral images cubes for each MRS channel with individual dithers combined then rotated and interpolated into a sky reference frame.  However, the Saturn data needed a significant amount of post-processing before being usable, such that the pipeline was run locally, applying all three data reduction stages separately to each dither position and tile.  Stage 1 generates `slope images' (count rates) from the raw \verb|UNCAL| data; stage 2 applies wavelength calibration and absolute flux calibration for each exposure; and stage 3 produces spectral image cubes from the input calibrated slope images - all steps are described in the JWST data processing manual\footnote{\url{https://jwst-docs.stsci.edu/jwst-science-calibration-pipeline-overview/stages-of-jwst-data-processing}}.  The pipeline's \verb|ResidualFringeStep| was used to minimise the effect of spectral fringing, although this remained a challenge at longer wavelengths in Channels 3 and 4 \cite{23wright}.  The default pipeline rotates and interpolates the final cubes into a sky reference frame, but given the significant artefacts from slice to slice (discussed below), we retained the final stage-3 data in the coordinate system of the IFUs (\verb|cube_build.coord_system = 'ifualign'|) to allow later correction of flat field effects.  These corrections must be performed before any attempts to combine individual MRS dithers, otherwise artefacts are blended together in the final products.  Finally, our pipeline included improved wavelength calibration solutions\footnote{Known as FLT-5, now available as the `specwcs' files under calibration reference context jwst\_1082.pmap at \url{https://jwst-crds.stsci.edu/context_table/jwst_1082.pmap}.}  \cite{23argyriou}, which were generated by fitting Jupiter and Saturn spectral models (this work) to the MRS data for each and every spaxel (the wavelength solution varied across the IFU), and estimating the required wavelength shift for each spaxel to align models and MRS data in the rest frame.  These wavelength solutions were only possible below 15 $\mu$m where strong and well-resolved spectral lines were evident, but enabled a significant improvement in spectral fits over the original pipeline.

% \cite{23harkett} 

\begin{figure}[p]
\centering
\includegraphics[width=\textwidth]{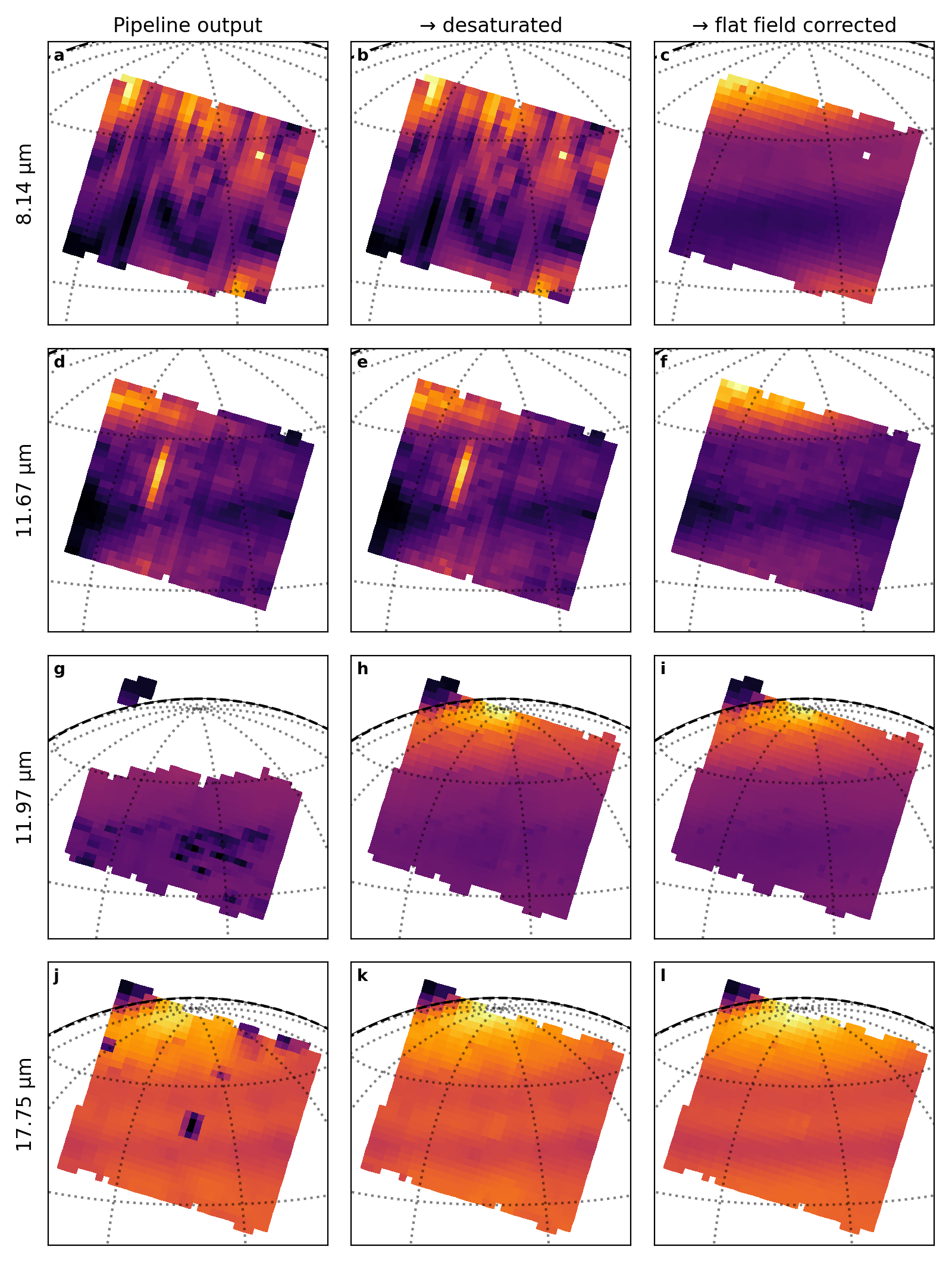}
\caption{Example cube slices at different stages of our custom MIRI data reduction process. The first column shows the output of the standard JWST pipeline, which still contains significant flat field effects (a \& d), saturation (g), and partial saturation (dark pixels in g \& j). The second column shows the data after the desaturation step is applied, and the third column shows the data after the flat field correction is applied.}
\label{fig:pipeline_summary}
\end{figure}

As shown in Figure \ref{fig:pipeline_summary}, significant flat field artefacts and saturation remained in the pipeline output cubes in the IFU-aligned frame. Therefore, we developed custom desaturation and flat field correction routines (discussed below) to correct these effects and produce our final science cubes.

\subsection{Desaturation}
\begin{figure}[t]
\centering
\includegraphics[width=\textwidth]{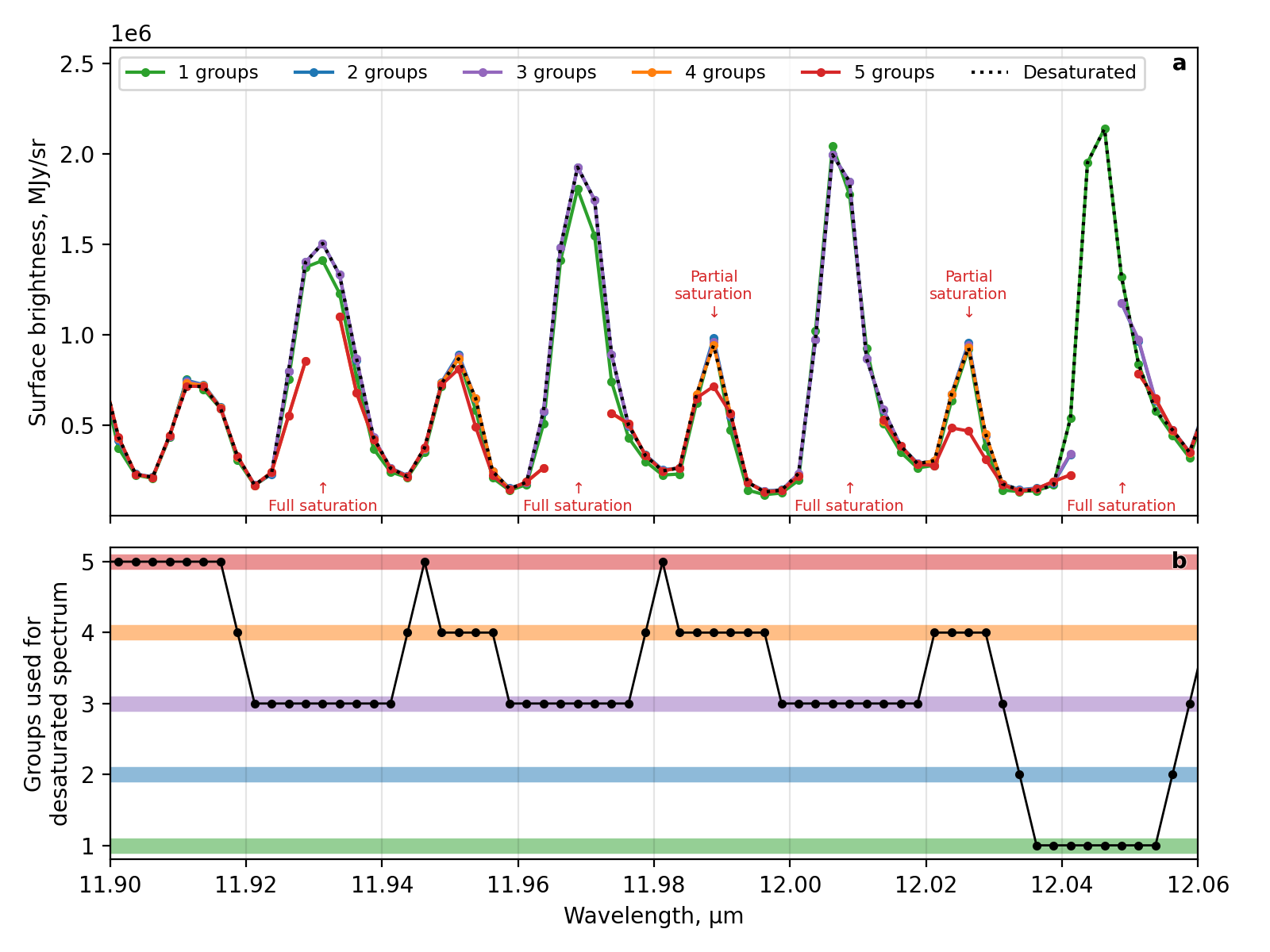}
\caption{Example spectra showing the desaturation routine. (a) shows the spectra using different numbers of groups, where the 5 group data (in red) is the `standard' pipeline output, which shows both saturation and partial saturation. The black dotted line in (a) shows the desaturated spectrum which is constructed using a varying number of groups, as shown in (b), to minimise saturation while maximising SNR. The majority of the spectral range uses the full 5 group data, and only narrow regions (such as this one, in the $\nu_9$ ethane emission band) require desaturation.}
\label{fig:desaturation}
\end{figure}

In the brightest parts of Saturn's spectrum, the MRS detector becomes saturated, leading to a loss or corruption of data (e.g. Figure \ref{fig:pipeline_summary}g). MIRI observations are split into a series of `groups,' which each record the measured flux for a part of the whole exposure. The detector saturates when the integrated flux reaches a certain threshold, meaning that even if the full exposure is saturated, the first few groups in the exposure may still have useful data that was recorded before the detector saturated. Therefore, data reduced using different numbers of groups can be used to desaturate (i.e. `fill in') saturated parts of the spectrum.

Our data processing routine modifies the \verb|UNCAL| raw data cubes to create versions containing the full range of groups (i.e., for the observations which have 5 groups in total: 1, 2, 3, 4, and 5 group versions are created). These different versions are all run through the standard JWST pipeline, creating 5 different versions of each science cube, each of which effectively has a different integration time. These 5 cubes are then merged into a single desaturated cube by dynamically selecting the highest number of groups possible to maximise SNR while minimising saturation, as shown in Figure \ref{fig:desaturation}.

The desaturation routine operates by comparing the different versions of each spectrum (the coloured lines in Figure \ref{fig:desaturation}a). The routine works iteratively, starting with the largest number of groups $n$, then replacing bad regions of the spectrum with the $n-1$ group spectrum. This is repeated until none of the spectrum contains `bad' data, or the 1-group spectrum is reached. The following regions of the spectrum are treated as bad data and replaced:
\begin{itemize}
    \item Regions flagged as saturated by the JWST pipeline (marked as `full saturation' in Figure \ref{fig:desaturation}a).
    \item Regions that appear partially saturated, but have not been flagged as saturated by the JWST pipeline. Regions are classed as partially saturated where $n$ group data is $<90\%$ of the brightness of the $n-1$ group data.
    \item Regions where the  $n$ group data is $>120\%$ of the brightness of the $n-1$ group data are also flagged as outliers, likely caused by cosmic ray hits.
\end{itemize}
The desaturation routine is skipped for regions of the spectrum with a SNR $<300$, as these would not be expected to experience saturation, and high noise levels could lead to false positives when flagging bad regions of the spectrum.

Initial testing of this routine found residual effects at the edge of spectral regions flagged as bad data, so any regions of the spectrum flagged as bad data are expanded by two spectral points. Additionally, the number of groups used for neighbouring spectral points is only allowed to change by 1 (i.e. a spectrum cannot immediately jump from 1 group at one wavelength to 5 groups at the next wavelength). A similar filter is applied in image space, so at a specific wavelength, neighbouring pixels can only vary by 1 group.

The parameters of the desaturation routine were selected by inspecting spectra (e.g. Figure \ref{fig:desaturation}) and images (e.g. Figure \ref{fig:pipeline_summary}) at a wide range of wavelengths, including regions of the spectrum that do and do not experience saturation. Almost all the spectral range uses the full 5 groups (or 4 groups for the equatorial tile), with only the specific spectral and spatial regions ($<10\%$ of all spectral points) that experience saturation replaced with fewer groups.

\subsection{Flat Field Correction}
\begin{figure}[t]
\centering
\includegraphics[width=1.2\textwidth,center]{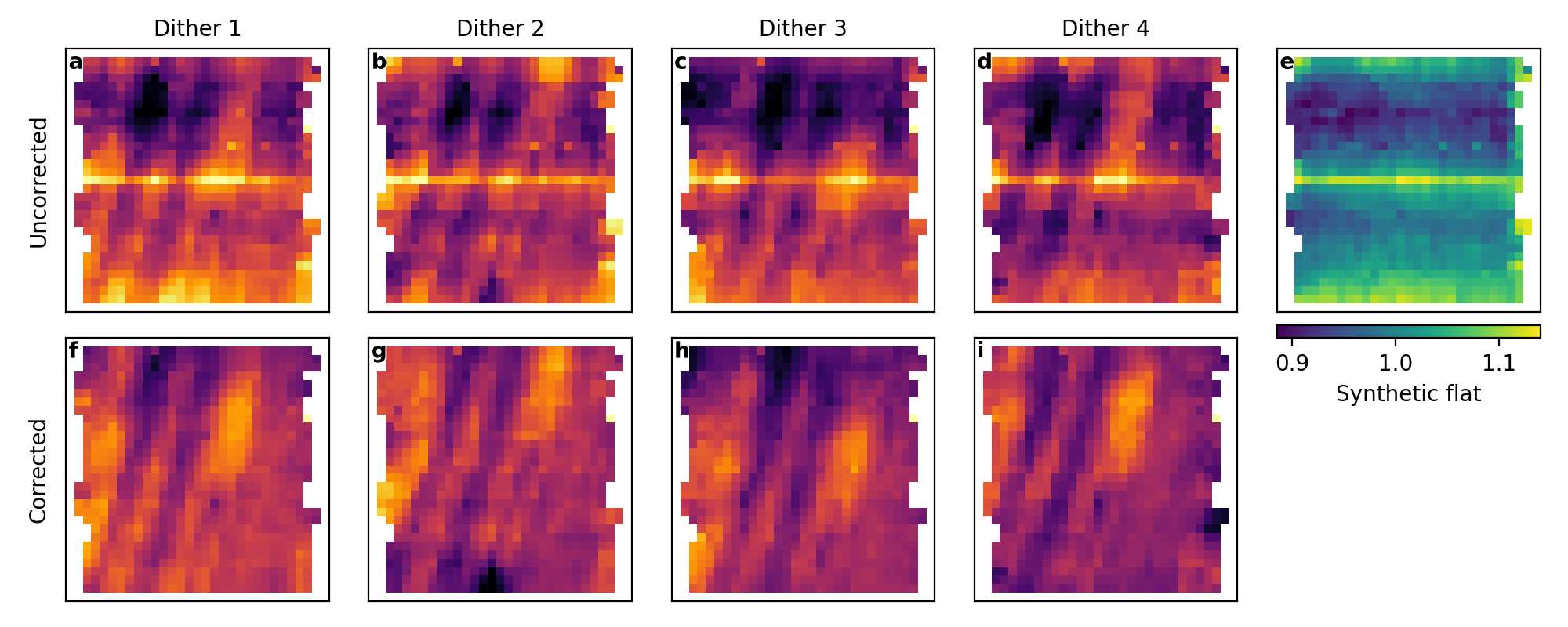}
\caption{Example images at 5.07 $\mu$m (channel 1-SHORT) before (a-d) and after (f-i) the synthetic flat field (e) is applied. The images are shown in the IFU frame, where north is to the top left. The flat field effects in a-e (such as the bright line across the centre of each image) are fixed in the IFU frame, whereas the observed spatial structure on Saturn varies in position on the detector with the different dithers, allowing flat field structure and real spatial structure to be differentiated.}
\label{fig:flat_field}
\end{figure}

After desaturation, significant flat field effects remained visible in the cubes, particularly at shorter wavelengths (e.g. Figure \ref{fig:pipeline_summary}b,e). These effects typically appeared as regular banding patterns aligned along the IFU slices, stripes, and swirls that varied with wavelength, and in many cases completely obscured any detail on Saturn. As shown in Figure \ref{fig:flat_field}, these patterns remained fixed in location on the detector for different dithers (and tiles), demonstrating that they are clearly an instrumental artefact.

These observed flat field patterns may be caused by small discrepancies between the reference flat field images used in the pipeline and the `true' flat field response of the detector. It is also possible that part of the observed apparent flat field is caused by residual artefacts remaining after the pipeline's stray light correction step. For simplicity, we refer to the entire observed pattern as the `flat field effect', regardless of its origin.

To correct for these flat field effects, we used the Saturn observations themselves to create a flat field for each channel and band. We assumed that the flat field can be treated as a purely multiplicative effect, with a corrected cube created by dividing the observed cube by the synthetic flat cube. Note that the background observation, 90" north of Saturn, did not show the same artefacts, suggesting that the flat field is sensitive to how the target illuminates the detectors and IFU slices.

\begin{figure}[t]
\centering
\includegraphics[width=\textwidth]{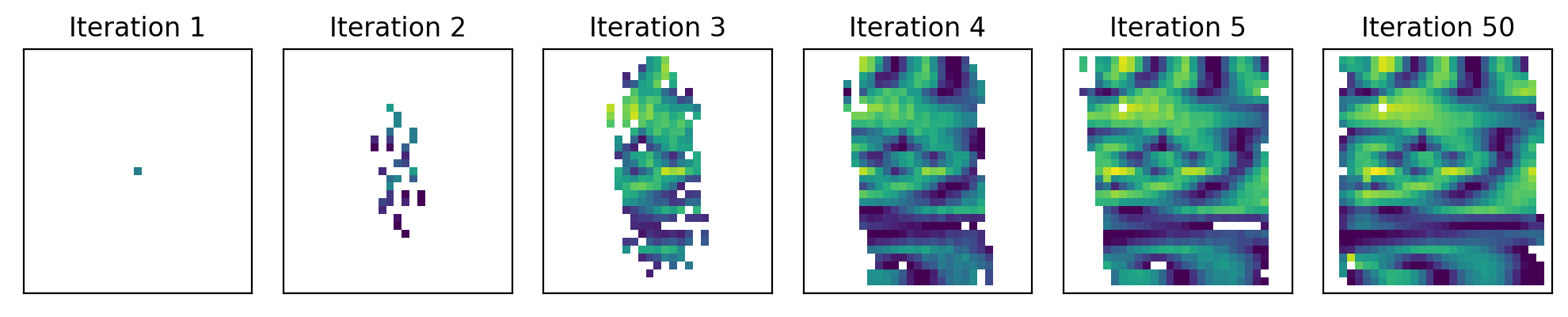}
\caption{Example flat field at 8.14 $\mu$m after the first 5 iterations of the flat field generation routine, and after the final 50th iteration. The flat is seeded with an initial value for the first iteration, which then uses the corresponding pixel ratios (derived from the dithered observations) to propagate the flat values and fill the image. White pixels have a value of NaN, and the NaN values in the final flat field are regions of the detector that do not contain any data.}
\label{fig:flat_generation}
\end{figure}

Our flat generation routine uses a set of four dithered observations to create a synthetic flat field image for each wavelength. We match pairs of pixels that observe (approximately) the same location on the surface of Saturn in different dithers, and assume that any variation in brightness between these pixels is caused by differences in the flat field for these pixels. The ratio of brightness values of all of the pairs of pixels can then be used to construct a flat field image at each wavelength:
\begin{enumerate}
    \item The input data files for each dither are `navigated' to calculate the latitude/longitude coordinates and illumination angles for each pixel. This navigation uses the WCS metadata in the FITS headers (to convert from pixel to celestial coordinates), which is derived from JWST's pointing information. To validate the navigation, we compared the navigated and observed positions of Saturn's limb and rings, and found that no additional manual adjustments were needed.
    \item The four dither input images are filtered to set extreme values (outlier pixels and pixels with emission angles \ang{>75}) to NaN to prevent any outliers contaminating the flat. Wavelengths where the average SNR $<10$ are skipped as the constructed flat would be too contaminated with noise.
    \item Corresponding pixels are identified by matching valid pixels that observe a similar location on Saturn in any of the four input dithers, using the navigated latitude/longitude coordinates for each pixel from step 1. An oval footprint is used around each pixel (in latitude/longitude space), and any pixels that fall within this footprint are treated as observing the same part of Saturn. The height (North-South direction) of the oval $h$ is set at half of the average difference in latitude between neighbouring pixels, and the width (East-West direction) of the oval is set as $w=4h$. This elongated oval footprint is used as Saturn has much more variation in the North-South direction than the East-West direction, so it matches more similar pixels than a simple circular footprint would.
    \item For each pair of corresponding pixels, $A$ and $B$, we calculate the ratio of observed pixel fluxes $R_{AB} = O_A/O_B$. Assuming that the flat field is a multiplicative effect, we can treat each pixel's observed flux $O_i = S_iF_i$ as the `true' flux from Saturn $S_i$ multiplied by the flat field for the given pixel $F_i$. As we define these corresponding pixels to observe (approximately) the same region of Saturn's surface, we assume that the original flux from Saturn is equal for both pixels, $S_A=S_B$, allowing the calculated ratio to be reduced to $R_{AB} = (S_A F_A)/(S_B F_B) = F_A/F_B$, giving the ratio of the flat field values for the two pixels.
    \item The set of calculated pixel ratios are then used to construct the flat field image. A flat image is initialised with the central pixel value set to 1, and all other values set to NaN, and we then iteratively construct the flat using the calculated ratios to propagate values. At each iteration, every pixel value in the flat is updated to $F'_A = \mathrm{median}(R_{Ai}F_i)$ where the $F_i$ are all the non-NaN corresponding pixels. After all the updated values are calculated at each iteration, any pixels outside the range $2/3 < F_i < 3/2$ are set to NaN to protect against outliers. After removing outliers, the flat is then divided by the mean pixel value. The routine is run for 50 iterations to allow the constructed flat to converge on a consistent solution. Figure \ref{fig:flat_generation} shows an example flat field at different generation steps.
    \item This constructed flat image is slightly under-constrained, as all the pixels can be multiplied by an arbitrary scaling factor and still provide a self-consistent result. Therefore, we scale each flat image so that the mean value of the pixels is unity, i.e. $\frac{1}{N}\sum^N_i{F_i} = 1$. This ensures that the application of the flat does not change a spectrum calculated by averaging all the pixels in an entire cube.
\end{enumerate}

Synthetic flat cubes were generated from the four dithers associated with the \ang{15}N and \ang{45}N tiles, and then averaged to produce the final flat cubes used to correct the data (the \ang{75}N tile and ring observation included too much background sky to be useful). The algorithm parameters were refined by studying the flats generated from the different tiles and the quality of the flat corrected data.

Special care was taken to ensure the flats did not contain any features of Saturn's atmosphere, and to ensure that the flats from different tiles produced consistent results. Comparisons of sets of dithered images (e.g. Fig. \ref{fig:flat_field}) allows structure from the flat field (fixed in detector location, identical between tiles) and real structure on Saturn (variable in detector location, different in each tile) to be differentiated. Regions of the spectrum with and without significant spatial structure were studied in detail, as well as the entire spectral range using animations that compared sets of dithers at each wavelength (see supplementary material). The \ang{75}N and rings tiles also provided useful checks, as these tiles were well corrected, even though they were not used in generating the synthetic flats.

As shown in Figure \ref{fig:flat_field}, the synthetic flats are able to correct both small-scale (1-2 pixel) and large scale ($>10$ pixel) variations in the sensitivity of the detector, including in regions with significant spatial structure on Saturn. The application of the flats helped to reveal spatial structure in the Saturn observations that was often completely obscured by the sensitivity variations, and prevented any spurious spatial variation being treated as real spatial variation on Saturn's surface.

\subsection{Zonal Averages}
\label{sec_zonal}
Zonal averages were calculated from the observed data using the following routine:
\begin{enumerate}
    \item All observed pixels, from all tiles and dithers, are binned into \ang{1} latitude bins.
    \item Within each bin, the median spectrum is calculated from all spectra in the bin. The 1/3 of the spectra with the largest RMS relative to this median spectrum are then discarded. This ensures the final zonal averages are protected from the effect of outlier pixels. Median averaging is used here (rather than mean) to ensure any extreme outlier pixels do not cause `good' spectra to be discarded.
    \item The mean spectrum for each bin is calculated from the remaining 2/3 `good' spectra. This mean spectrum is used as the zonal average for each latitude bin, which is then used for spectral modelling in subsequent sections.
\end{enumerate}

% \subsection{NIRSpec Data Processing}

%%%%%%%%%%%%%%%%%%%%%%%%%%%%%%%%%%%%%%%%%%%%%%%%%%%%%%%%%%
%%%%%%%%%%%%%%%%%%%%%%%%%%%%%%%%%%%%%%%%%%%%%%%%%%%%%%%%%%
%%%%%%%%%%%%%%%%%%%%%%%%%%%%%%%%%%%%%%%%%%%%%%%%%%%%%%%%%%
\section{Dataset Overview}  
\label{overview}

\subsection{Saturn's Spectrum}
\label{sec_spectra}
An average of MIRI/MRS observations of Saturn's atmosphere is shown at the bottom of Fig. \ref{fig:saturn_montage} for 4.9-18.0 $\mu$m, omitting MRS data from the longest channel (17.7-27.9 $\mu$m) due to ongoing challenges with fringe removal and calibration.  Below $\sim7.3$ $\mu$m, the spectrum is shaped by a combination of scattered reflected light from aerosols (notably within the deepest absorption bands and near 6 $\mu$m) and thermal emission.  The 5-$\mu$m window is sculpted by PH$_3$ lines below 5.2 $\mu$m ($\nu_1$ at 4.3 $\mu$m) and NH$_3$ above 5.2 $\mu$m ($2\nu_2$ at 5.32 $\mu$m, $\nu_4$ at 6.15 $\mu$m), along with narrow absorption bands of H$_2$O (5.1-5.4 $\mu$m) and AsH$_3$ (4.9-5.0 $\mu$m).  Bright emission from the 5-$\mu$m window implies low aerosol opacity, so cloud bands and small discrete features appear in silhouette in Fig. \ref{fig:saturn_montage}b-c against the bright background glow from Saturn's 4-6 bar region.  Bright reflection near 6 $\mu$m provides a means of constraining upper tropospheric aerosols.  Hydrocarbon emission from CH$_4$ ($\nu_2$ at 6.5 $\mu$m) and C$_2$H$_6$ ($\nu_8$ at 6.8 $\mu$m; $\nu_6$ at 7.3 $\mu$m) also contribute to this 4.9-7.3 $\mu$m range.  This range is particularly noteworthy as it has only been previously observed by ISO/SWS in the disc-average \cite{03encrenaz}, and neither Cassini/VIMS ($R\sim300$ at 5 $\mu$m) nor Cassini/CIRS ($R\sim2800$ at 7.0 $\mu$m) could observe in the 5.1-6.9 $\mu$m range.  Thus MIRI channel-1 provides access to tropospheric NH$_3$ and H$_2$O, along with the properties of Saturnian aerosols, in this range for the first time, with spectral resolutions of $R\sim3100-3750$.  

Aerosol contributions diminish at longer wavelengths in channels 2 and 3 (7.3-18.0 $\mu$m), which are dominated by the collision-induced absorption due to H$_2$ and He, with emission and absorption features superimposed.  PH$_3$ ($\nu_2$ at 10.08 $\mu$m, $\nu_4$ at 8.94 $\mu$m) and NH$_3$ ($\nu_2$ at 10.5 $\mu$m) provide absorption features that dominate the 8-12 $\mu$m range; with strong emission features from methane (CH$_4$ $\nu_4$ at 7.7 $\mu$m), acetylene (C$_2$H$_2$ $\nu_5$ at 13.7 $\mu$m) and ethane (C$_2$H$_6$ $\nu_9$ at 12.2 $\mu$m); and weaker emission features from CO$_2$ ($\nu_2$ at 14.9 $\mu$m), diacetylene (C$_4$H$_2$ $\nu_8$ at 15.9 $\mu$m), methylacetylene (C$_3$H$_4$ $\nu_9$ at 15.8 $\mu$m), ethylene (C$_2$H$_4$ $\nu_7$ at 10.5 $\mu$m), propane (C$_3$H$_8$ $\nu_{26}$ at 13.4 $\mu$m) and benzene (C$_6$H$_6$ $\nu_4$ at 14.83 $\mu$m).  The H$_2$ S(1) quadrupole at 17.03 $\mu$m and its associated dimer absorption can be seen in Channel 3-Long, but the S(0) quadrupole at 28.2 $\mu$m is just outside the MRS range.  

The spectral database used in MRS modelling was initially based on that used for Cassini retrievals \cite{18fletcher_poles}, with updates to AsH$_3$ \cite{19coles} and CH$_3$ \cite{19ahmad_ch3} from the ExoMol database \cite{16tennyson}, and GeH$_4$ from HITRAN \cite{22gordon}.  Voigt broadening was used for all bands - the sub-Lorentzian lineshape of \citeA{04bailly} did not have an impact on the quality of the spectral fits. The line database was used to calculate $k$-distributions for each gas within each of the 12 MRS subbands, using the wavelength grid from stage 3 of the standard pipeline, and the wavelength-dependent resolving power of each channel determined from ground-based measurements \cite{21labiano}.  Note that in-flight commissioning updates to the spectral resolution \cite{23jones} have not been incorporated into our spectral models at this stage.

Collision-induced absorption of H$_2$ and He was included based on their dimer absorptions \cite{18fletcher_cia}.  During the course of the spectral fitting, residuals between model and data were used to identify missing bands of known species, and to search for any new species. Multiple bands of propane are observed on Saturn for the first time - only $\nu_{26}$ at 13.4 $\mu$m had been included in our line database based in GEISA \cite{20geisa}, and had been previously used to study the propane distribution \cite{09guerlet, 18fletcher_poles}. Residuals at high latitudes revealed the presence of the $\nu_7$, $\nu_{20}$, $\nu_{21}$ and $\nu_8$ emission bands at 8.63, 9.49, 10.85 and 11.51 $\mu$m, respectively, for the first time.  These were introduced into our line database using the pseudo-linelist of \citeA{13sung}, and the improvement in spectral residuals are shown in Supplemental Fig. \ref{{fig:residc3h8}}.
% the Appendix (Fig. \ref{fig:residc3h8}).

% $\nu_6+\nu_8$ band of diacetylene at 8.06 $\mu$m \cite{92mcnaughton} detected for the first time and included in HITRAN database \cite{22rothman}.

The GeH$_4$ $\nu_2$ band at 10.74 $\mu$m is too weak to be seen, lost amongst numerous NH$_3$ absorption features.  The AsH$_3$ $\nu_4$ band at 9.97 $\mu$m does have a detectable signature on the edge of a PH$_3$ absorption line, but in a region of the spectrum that is affected by MRS fringing at the longward end of channel 2-medium.  The small spectral feature is reproduced by an abundance of $\sim0.4$ ppb, with a decline from equator to pole that cannot be explained by fringing.  Nevertheless, precise constraints must await more robust defringing strategies.  Finally, we see no evidence of emission from the HCN $\nu_2$ at 14.05 $\mu$m, discussed in Section \ref{strat_chem}.

%- this is perhaps unsurprising, given that nitrile-producing chemistry is unexpected on Saturn, as the domains of NH$_3$ photolysis (troposphere) and CH$_4$ photolysis (stratosphere) are separated from each other in altitude \cite{16fletcher}.

% C2H6
% https://lweb.cfa.harvard.edu/hitran/vibrational.html
% 823 12.15 nu_9 
% 995 10.05 nu_3 (not in model - do we see it?)
% 1388 7.20 nu_2
% 1379 7.25 nu_6 (Guerlet)
% 1468 6.81 nu_8 (Guerlet)

% C2H2 
% 674 14.83 nu4 (not in model)
% 729 13.71 nu_5
% 1974 5.07 nu2 (not in model)

% CO2
% 1388 7.2 nu1 (not in model)

% C2H4
% 3026 1
% 1623 2
% 1342 3
% 1023 4
% 3103 5
% 1236 6
% 949 10.53 7
% 943 8
% 3106 9
% 826 10
% 2989 11
% 1444 12 - this is in the database 6.92 µm

% C3H8
% 748,	n26
% 869,	n8 11.51
% 922,	n21 10.85
% 1054,	n20 9.49
% 1158,	n7 8.63
% 1192,	n25 (buried in nu7) 8.38
% 1339, n19 7.47
% 1376, n18 7.26
% 1472, n24 6.79
% 1477, n4 (buried in nu24) 6.77

%%%%%%%%%%%%%%%%%%%%%%%%%%%%%%%%%%%%%%%%%%%%%%%%%%%%%%%%%%
%%%%%%%%%%%%%%%%%%%%%%%%%%%%%%%%%%%%%%%%%%%%%%%%%%%%%%%%%%
%%%%%%%%%%%%%%%%%%%%%%%%%%%%%%%%%%%%%%%%%%%%%%%%%%%%%%%%%%
\subsection{Spatial Structure}
\label{spatial}

\begin{figure}[h]
\centering
\includegraphics[width=1.3\textwidth,center]{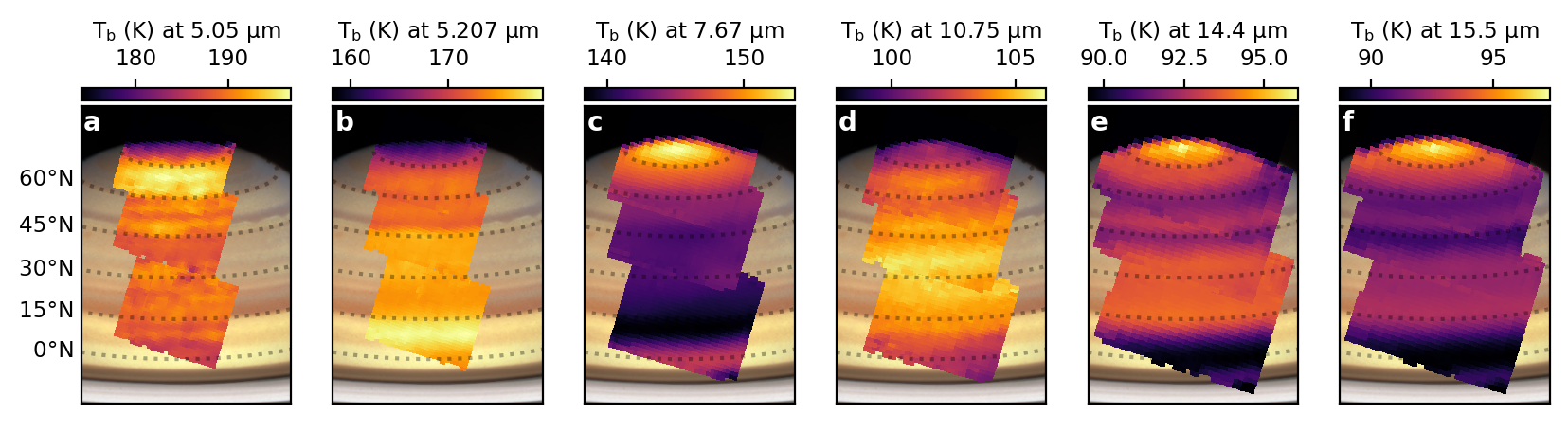}
\caption{Composite images created by combining all three Saturn tiles to show the equator-to-pole variation in brightness temperature at different wavelengths.  5.05 $\mu$m senses aerosol opacity in the deep troposphere (3-6 bars), whereas 5.207 $\mu$m is in a strong NH$_3$ absorption and senses a blend of thermal emission and reflected sunlight from upper tropospheric aerosols.  7.67 $\mu$m senses stratospheric temperatures (0.1-5 mbar) via CH$_4$ emission; whereas 10.75 $\mu$m senses a blend of tropospheric temperature and ammonia opacity near 400-600 mbars.  14.4 and 15.5 $\mu$m are primarily sensitive to the H$_2$-He continuum, sounding tropospheric temperatures in the 100-to-300-mbar range.  Note that each tile observed a different longitude range, but are shown overlapping here for simplicity - this causes some artificial inconsistencies in brightness at regions of overlap.}
\label{fig:equator_pole}
\end{figure}

\begin{figure}[h]
\centering
\includegraphics[width=1.4\textwidth,center]{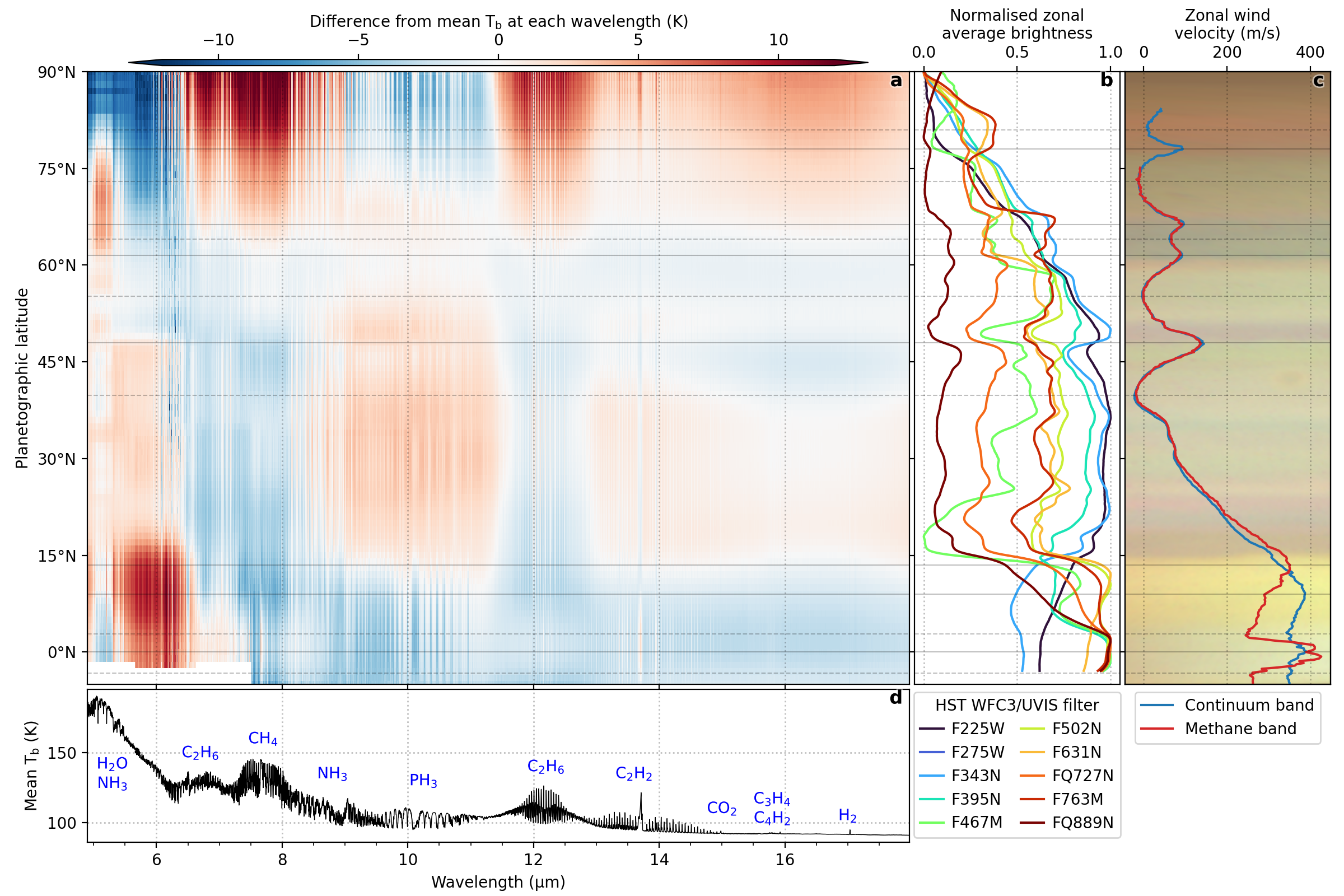}
\caption{Variation in Saturn's zonal average brightness temperature with wavelength and latitude. (a) shows the difference from the mean brightness temperature spectrum (d), where red areas are brighter than the average and blue areas are dimmer. (b) shows the normalised zonal average brightness of HST observations of Saturn in September 2022 \cite{23simon} and (c) shows Saturn's zonal wind profiles \cite{11garcia}. Solid and dashed horizontal lines in (a-c) indicate the peaks and troughs of the zonal wind profiles respectively. The background of (c) shows part of the colour composite map of Saturn created from the HST observations.}
\label{fig:zonal_average}
\end{figure}

Fig. \ref{fig:equator_pole} shows selected wavelengths from the three MIRI/MRS 4.9-27.9 $\mu$m cubes spanning from Saturn's equator to the north pole.  Given Saturn's relative longitudinal homogeneity, these can be taken as a good approximation to a zonal mean, which is calculated as described in Section \ref{methods}.  Fig. \ref{fig:zonal_average} then shows the difference between the zonally-averaged brightness and the mean brightness temperature spectrum, highlighting strong gradients as a function of latitude.  These gradients are compared to the cloud-tracked zonal winds from Cassini in both the continuum and methane bands \cite<Fig. \ref{fig:zonal_average}c,>[]{11garcia}, showing how thermal-infrared brightness is related to the peaks of the eastward and westward jets.  We also compare the MIRI/MRS maps (acquired in November 2022) to visible-light reflectivity scans acquired by Hubble (HST) in September 2022 \cite<Fig. \ref{fig:zonal_average}b>[]{23simon}, to show how brightness temperature and aerosol reflectivity are related.  The zonally-averaged reflectivity in ten HST WFC3/UVIS filters has been normalised for plotting purposes, to highlight the similarities in the location of strong brightness gradients.

Together, the composite images of Fig. \ref{fig:equator_pole} and the zonal-mean brightness in Fig. \ref{fig:zonal_average} reveals a wealth of detail.  The exquisite sensitivity of MRS, even compared to previous spectroscopic maps from Cassini, reveal Saturn's banded structure in both reflected sunlight and thermal emission.  The following three features stand out in the meridional (latitudinal) direction:

\begin{itemize}
    \item \textbf{Belt/Zone Structure:}  The strongest meridional gradients in tropospheric and stratospheric brightness temperatures are co-located with the peaks of the eastward and westward jets, as measured at Saturn's cloud-tops \cite{11garcia}, supporting a geostrophic balance between the winds and temperature gradients via the thermal windshear equation, and the decay of Saturn's tropospheric winds with altitude \cite{81pirraglia, 83conrath}.  At mid-latitudes where Ferrel-like meridional circulation cells are expected to dominate \cite{20fletcher_beltzone}, Saturn's zones are defined as cool, anticyclonic bands equatorward of eastward jets, whereas the belts are warm, cyclonic bands poleward of eastward jets \cite{09delgenio}.  MIRI continuum emission from the troposphere (e.g., Fig. \ref{fig:equator_pole}e-f) reveals subtle cool zones equatorward of eastward jets at $31.5^\circ$ (an inflection in the broad equatorial jet), $47.8^\circ$, $61.5^\circ$, and $78.0^\circ$N, in addition to the broad cool Equatorial Zone at $<9.2^\circ$N where continuum-band cloud tracking reveals a maximum eastward windspeed \cite{11garcia}.  Stratospheric banding is more subtle, but a bright equatorial band is observed in methane emission at 7.67 $\mu$m and in the peak of the acetylene emission at 13.7 $\mu$m \cite<corresponding to the equatorial stratospheric oscillation,>[]{08orton_qxo,22blake}.

    \item \textbf{North Polar Stratospheric Vortex (NPSV):}  The warm NPSV, defined by the strong gradient in stratospheric brightness temperature near $78^\circ$N, is visible throughout the MIRI/MRS dataset, particularly near 7-8 $\mu$m sensing stratospheric CH$_4$, and in regions of tropospheric continuum emission longward of 14 $\mu$m.  This should be contrasted with generally low polar brightness temperatures in the 5-6.5 $\mu$m and 9-11 $\mu$m regions that probe higher pressures.  The NPSV formed during northern spring \cite{18fletcher_poles} and is expected to have reached its maximum contrast with respect to lower latitudes in 2021-22 \cite<confirmed by ground-based thermal imaging>[]{22blake}, and to decline in visibility in the coming years - see below for discussion of seasonal evolution since the end of the Cassini mission in 2017.  Embedded within the NPSV, the central north polar cyclone discovered by Cassini \cite<NPC,>[]{08fletcher_poles} remains visible in the MIRI/MRS maps as a peak in brightness temperature right at the pole (Fig. \ref{fig:equator_pole}c-f).

    \item \textbf{Deep Cloud Structure:}  As has been previously shown by ground-based imaging at 5.1-5.2 $\mu$m \cite{01yanamandra} and Cassini/VIMS spectroscopy \cite<$<5.1$ $\mu$m,>[]{06baines, 11fletcher_vims}, MIRI 5-$\mu$m maps in Fig. \ref{fig:equator_pole}a and the zonal mean in Fig. \ref{fig:zonal_average} reveal cloud banding on a much finer spatial scale than the broad belts and zones defined by the temperature field.  Indeed, the distribution of aerosols revealed by MIRI spectroscopy between 4.9 and 7.3 $\mu$m is a good match for the albedo contrasts in Fig. \ref{fig:zonal_average}c, but are not a good proxy for Saturn's belts and zones \cite{20fletcher_beltzone} - aerosols do not simply condense where it is cold, and sublimate where it is warm, with a more complicated pattern emerging.  Rather, the finescale cloud banding observed at 5 $\mu$m is a closer match to that seen in visible light \cite<e.g.,>[]{05vasavada}.  Deep NH$_3$ absorption features in the 5.1-5.3 $\mu$m range, and in the 6-$\mu$m region, also reveal the contribution of reflected sunlight from upper tropospheric aerosols (Fig. \ref{fig:equator_pole}b) - without this reflection, there would be almost no radiance from these deep bands.  The equatorial region is particularly bright in the 5.5-6.5 $\mu$m range in Fig. \ref{fig:zonal_average}a due to the presence of upper tropospheric aerosols observed in visible light in Fig. \ref{fig:zonal_average}a.
    
\end{itemize}

\begin{figure}[h]
\centering
\includegraphics[width=1.3\textwidth,center]{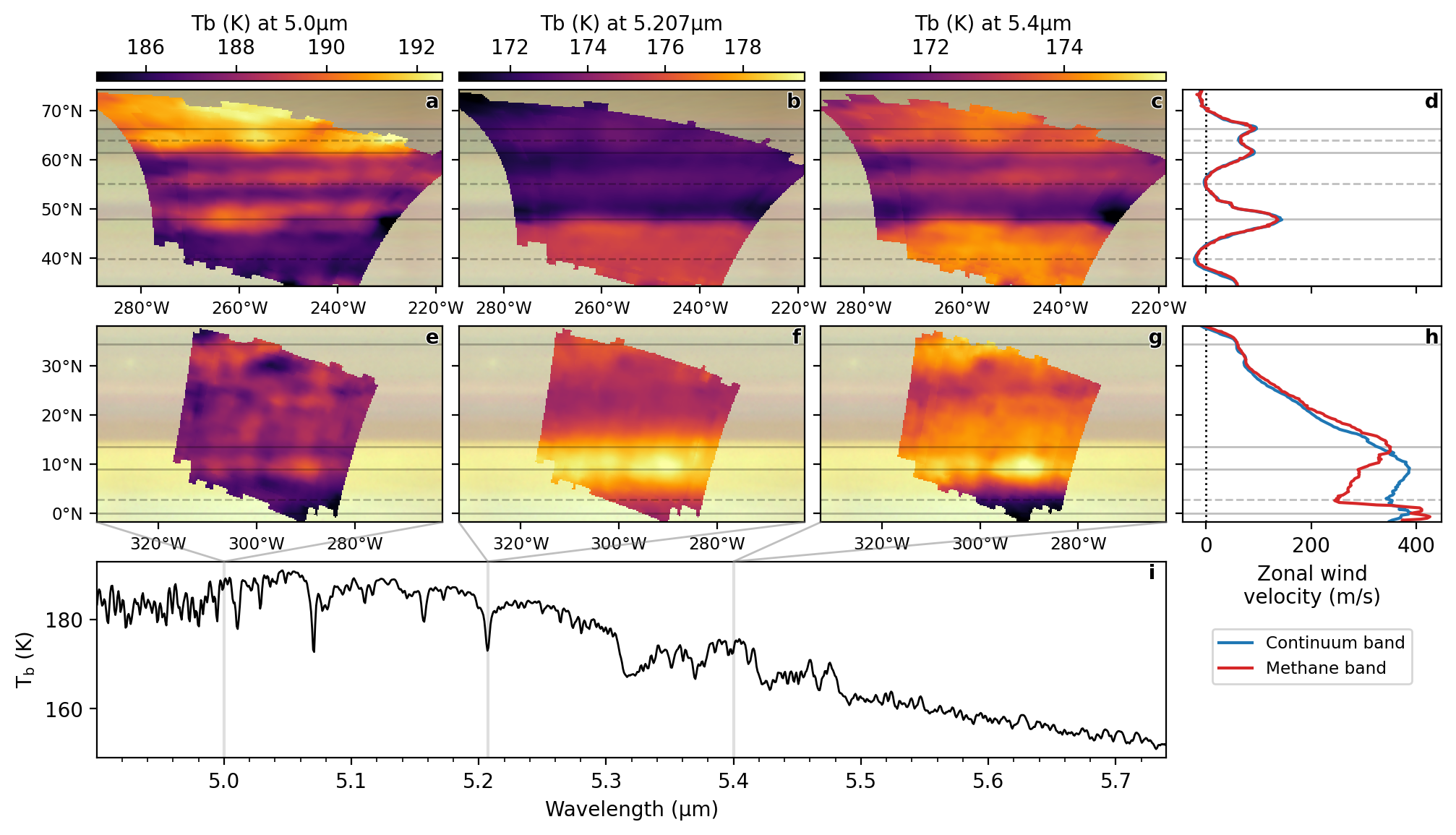}
\caption{Mapped spatial structure in channel 1-short for low- and mid-latitudes. (d) and (h) show the zonal wind velocities \cite{11garcia} and (i) shows an example brightness temperature spectrum.  While all panels sense a blend of thermal emission and reflected sunlight, panels (a) and (e) primarily reveal deep thermal emission modulated by overlying aerosol opacity, whereas panels (b)-(c) and (f)-(g) primarily reveal reflectivity variations from aerosols in the upper troposphere.  Solid and dashed horizontal lines indicate the peaks and troughs of the zonal wind profiles, respectively. The backgrounds of (a-c \& e-g) show the colour composite map of Saturn created from the HST observations.}
\label{fig:ch1a_maps}
\end{figure}

The deep clouds revealed by MIRI/MRS are shown in more detail in Fig. \ref{fig:ch1a_maps}, which uses only the shortest MRS channel (1A).  Animations in the Supporting Material show that Saturn's appearance changes dramatically with wavelength, from inside a deep absorption band (sensing sunlight reflected from upper tropospheric aerosols) to the intervening `continuum' (sensing deeper cloud opacity).  The Equatorial Zone (0-9.2$^\circ$N) is generally dark (e.g., high aerosol opacity), with a brighter band near 10$^\circ$N marking the boundary with the NEB.  At 5.0 $\mu$m small low-contrast features are observed up to the prograde jet at $48^\circ$N, but these cannot be seen at 5.2 $\mu$m where reflected sunlight creates a more homogeneous appearance.  Poleward of the $48^\circ$N jet, the 5-$\mu$m thermal emission increases substantially at the same time as the reflectivity sharply declines.  We no longer see the 5-$\mu$m-bright, aerosol-depleted band near 35-40$^\circ$N that had dominated the appearance of the northern hemisphere after the 2010-11 storm \cite{16sromovsky}, consistent with a re-population of the band by Saturn's seasonal aerosols in the decade since the storm, so that this band no longer stands out.  This fading of the 5-$\mu$m storm emission is consistent with the ground-based record \cite{20bjoraker_dps}.  Fine banding is observed up to $61.5^\circ$N, when the 5-$\mu$m emission again rises substantially as 5.2 $\mu$m reflection falls, with bright emission continuing to the latitude of the hexagon at $78.0^\circ$N.  Interior to the hexagon, the polar domain in Fig. \ref{fig:equator_pole}a-b is 5-$\mu$m dark, and also dark in reflected sunlight at 5.2 $\mu$m, consistent with its appearance in Cassini/VIMS observations in 2016 \cite{21sromovsky}.

Seasonally-generated aerosols are not homogeneous over the whole of Saturn's northern hemisphere, but are confined by the banded structure - the reflected sunlight component in Fig. \ref{fig:ch1a_maps}b and the 5-6 $\mu$m range in Fig. \ref{fig:zonal_average}a decreases in distinct steps from the equator to the pole, with notable boundaries between the EZ and NEB (\ang{9.2}N) and at \ang{48.7}N; the thermal 5-$\mu$m emission component shows a notable increase between \ang{61.5}N and \ang{78}N, consistent with the lowest aerosol opacity there, and with the suggestion of cloud-clearing near $65^\circ$N in 2019-20 \cite{20bjoraker_dps}.  A four-fold increase in aerosol opacity of upper-tropospheric hazes interior to the hexagon between 2013 and 2016 was observed by Cassini \cite{21sromovsky}, possibly accounting for the dark hexagon appearance to JWST in 2022.  MIRI does not observe the hexagon latitude as a bright 5-$\mu$m band, which had been evident in VIMS observations in 2013 and 2016, suggesting continued increases in aerosol opacity in the polar domain, spreading to lower latitudes, but not yet reaching the \ang{60}-\ang{78}N range.

Right at the equator, equatorward of $5^\circ$N, both the 5-$\mu$m brightness and the reflected sunlight drop to create an unusually dark band.  This can be seen in Fig. \ref{fig:zonal_average}, where the dark band corresponds to subtle colour contrasts in Hubble images from September 2022, and to a peak of reflectivity in several of the individual HST filters, notably the strong CH$_4$ band at 889 nm.  Such a dark band was not evident in VIMS maps in 2006-11, although this region was generally bland and diffuse at 5 $\mu$m, with several dark plume-like discrete features bordering it at \ang{6}-\ang{7}N \cite{11fletcher_vims}.  This dark band coincides with the rapid increase in the upper tropospheric winds observed in the 890-nm CH$_4$ band i Fig. \ref{fig:zonal_average}c \cite{11garcia}.  We will return to this unique region, and the challenging spectral fits, in Section \ref{results}.

%%%%%%%%%%%%%%%%%%%%%%%%%%%%%%%%%%%%%%%%%%%%%%%%%%%
%%%%%%%%%%%%%%%%%%%%%%%%%%%%%%%%%%%%%%%%%%%%%%%%%%%
%%%%%%%%%%%%%%%%%%%%%%%%%%%%%%%%%%%%%%%%%%%%%%%%%%%
\subsubsection{Discrete Vortices and Hexagon} 

Although Saturn remains longitudinally homogeneous at most latitudes, we subtracted zonal averages from the MRS images to search for evidence for discrete features.  Such longitudinal contrasts were only definitively observed in the shortest MRS channels in Fig. \ref{fig:ch1a_maps}, where significant structure is observed.  In particular, we see a bright cloud-free region near \ang{48}N, \ang{261}W; at least two dark anticyclonic vortices near \ang{48}N, \ang{230}W and \ang{30}N, \ang{299}W; a patch of high 5-$\mu$m brightness at the edge of the equatorial zone near \ang{10}N, \ang{292}W; and spatial structure in the bright band surrounding the dark polar domain near \ang{62}N, \ang{215}W.  In an effort to determine the history and longevity of these features, we compared the MIRI maps (13-14 November 2022) to HST observations \cite<21 September 2022,>[]{23simon} and observations by amateur astronomers \cite<using the PVOL database throughout November 2022,>[]{18hueso} to search for the presence of these discrete features in multiple datasets. Comparisons of the JWST and HST data are shown in the Supplemental Material. 
% Appendix \ref{app:discrete_features}.  

Unfortunately, there were no conclusive detections of any of the features observed by MIRI in these supporting visible-light datasets.  Given the $\sim7$-week time gap between HST and JWST observations this is perhaps unsurprising, despite attempts to account for zonal drifts during this interval in the Supplemental Material.  In particular, HST observations in September revealed the continued presence of the anticyclonic vortex (AV) that was generated by the 2010-11 storm \cite{13sayanagi} near \ang{42}N, \ang{190}W, and the presence of a pair of vortices near \ang{63}N, \ang{335}W that could be related to the coupled vortex system on the `double jet' near \ang{62}-\ang{67}N studied by \citeA{18gaztelurrutia}.  Whilst the shared latitudes are compelling, the limited longitudinal coverage of the MRS maps, combined with the 7-week time gap since Hubble, makes it unlikely that these are the same features.  Indeed, the long-lived AV at \ang{42}N was expected to be near \ang{176}W on 2022-Nov-13, and was therefore missed by MIRI (Fig. \ref{fig:coverage}).  This strongly argues for contemporaneous MIRI/MRS spectroscopy and NIRCAM (or HST) imaging in future observing programmes.

Finally, when the MIRI observations had originally been designed (assuming a 2018 launch), we had hoped to detect the vertices of Saturn's polar hexagon during northern summer, both in tropospheric and stratospheric thermal emission \cite{18fletcher_poles}.  Unfortunately, the decreasing sub-observer latitude in 2022 reduced the chances of success, and no convincing evidence of hexagon vertices can be observed, despite considerable work to clean up and combine the individual channel-1 dithers.  We will likely need to wait to the 2040s for our next infrared views of the hexagon itself.

%%%%%%%%%%%%%%%%%%%%%%%%%%%%%%%%%%%%%%%%%%%%%%%%%%%
%%%%%%%%%%%%%%%%%%%%%%%%%%%%%%%%%%%%%%%%%%%%%%%%%%%
%%%%%%%%%%%%%%%%%%%%%%%%%%%%%%%%%%%%%%%%%%%%%%%%%%%
\subsection{Seasonal Change Since Cassini}

\begin{figure}[p]
\centering
\includegraphics[width=0.5\textwidth]{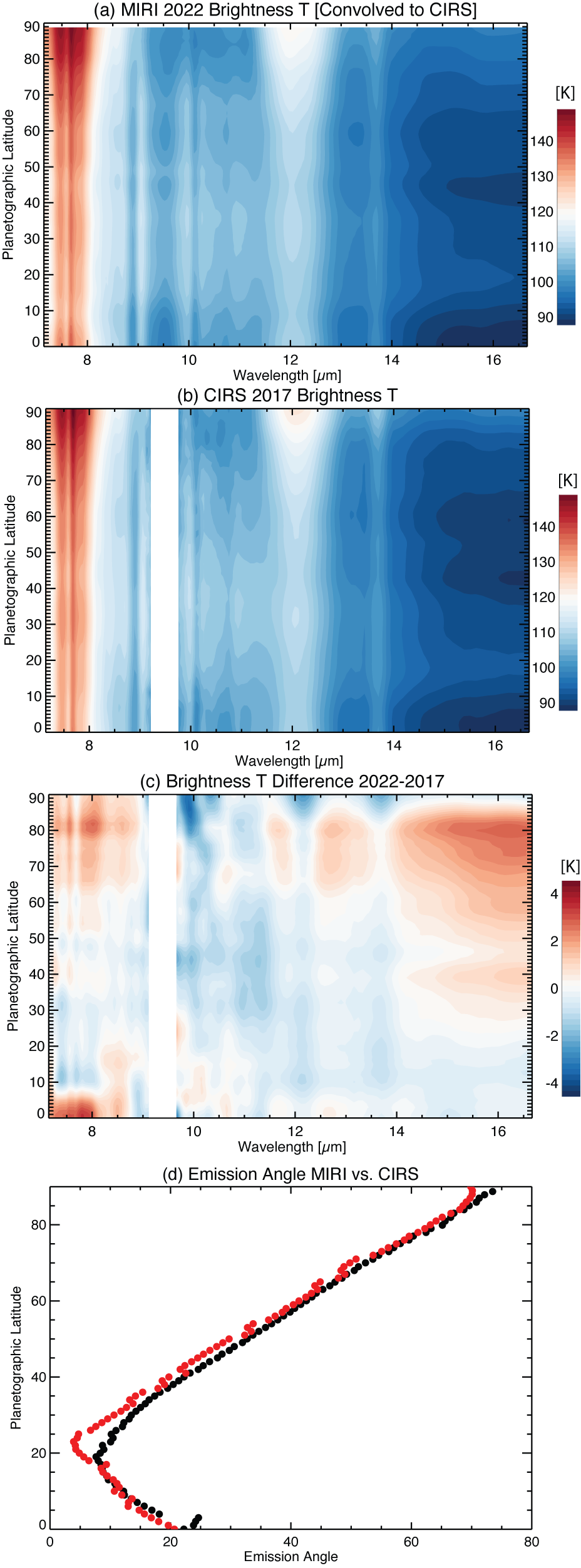}
\caption{Comparison of Cassini/CIRS observations in 2017 to MIRI/MRS observations in 2022. MRS spectra were convolved to the same spectral resolution as CIRS (15 cm$^{-1}$), and three CIRS northern hemisphere maps were zonally averaged onto a latitudinal grid.  Brightness temperatures as a function of wavelength are shown in (a)-(c), showing stratospheric CH$_4$ emission on the left, ethane emission near 12 $\mu$m, and continuum H$_2$-He absorption on the right. In panel (d), the emission angles for MIRI (black) and CIRS (red) for each latitude are approximately the same, so differences are not due to limb brightening/darkening. CIRS measurements between $\sim9$-10 $\mu$m are disregarded (white rectangle) as they had low signal-to-noise.}
\label{fig:cirs_comp}
\end{figure}

Before modelling the MRS spectra, we compare the calibrated JWST data to other observations in the mid-infrared.  Although ground-based studies have been able to monitor morphological changes to Saturn's mid-IR emission since 2017 \cite{22blake}, these have typically been calibrated to match a low-latitude average of Cassini/CIRS radiances due to the difficulties arising from variable telluric contamination, making genuine assessments of global-scale temperature changes rather challenging.  Fig. \ref{fig:cirs_comp} compares the MIRI 7-17 $\mu$m brightness with those from Cassini/CIRS in 2017.  Three CIRS northern hemisphere maps (2017-Jan-19, 2017-Apr-17 and 2017-Aug-26), acquired from a near-equatorial sub-spacecraft latitude at a spectral resolution of 15 cm$^{-1}$, were combined and zonally-averaged onto a $1^\circ$ latitude grid. The MIRI/MRS zonal averages were convolved with the CIRS instrument lineshape to achieve the same low spectral resolution, and the differences are shown in Fig. \ref{fig:cirs_comp}.  Note that the viewing geometry for each latitude was approximately the same in 2022 and 2017, such that limb darkening/brightening should be negligible.  

The 7-8 $\mu$m CH$_4$ emission in Fig. \ref{fig:cirs_comp}(c) shows the change in Saturn's equatorial stratospheric oscillation \cite{08orton_qxo, 08fouchet}, which reveals a brighter equatorial band in 2022 compared to 2017.  This warm band is also visible at 7.65 $\mu$m in Fig. \ref{fig:equator_pole}.  This is consistent with the ground-based record \cite{22blake} which showed continued warming at the equator since the end of the Cassini mission.  It is not, however, consistent with expectations from one Saturnian year earlier, where the equatorial band was in its cool phase in 1993-1995 during the same season.  Thus the semi-annual nature of Saturn's equatorial oscillation remains in doubt \cite{13sinclair, 22blake}.  

The stratosphere from $10^\circ$ to $35^\circ$N appears to be cooler in 2022 than in 2017, consistent with the idea of upwelling and adiabatic cooling in the summer hemisphere as part of an interhemispheric circulation from summer to winter, reminiscent of the Earth's Brewer-Dobson circulation \cite{12friedson, 22bardet}.  During northern winter ($L_s=310^\circ$), enhancements of stratospheric hydrocarbons detected by Cassini near 25$^\circ$N \cite{10guerlet} were consistent with stratospheric subsidence as part of a meridional circulation from summer to winter \cite{12friedson}.  The cooler brightness temperatures measured by MIRI suggest that this circulation has now reversed by $L_s=150^\circ$, having switched direction near equinox in 2009 \cite{22bardet}.  In the following sections, we will attempt to verify this via measurements of trace hydrocarbon species to determine vertical motions in the northern low-latitude stratosphere.  

Poleward of $60^\circ$N, the MIRI/MRS observations reveal warmer temperatures than the Cassini/CIRS observations, consistent with the continued warming of the NPSV in Fig. \ref{fig:equator_pole} as northern summer progressed.  This is also true in the northern troposphere, sampled by H$_2$-He emission longward of 14 $\mu$m, which has warmed since 2017.  Radiative models \cite{14guerlet}, combined with the viewing geometry from Earth \cite{22blake}, predict the visibility of the NPSV will drop considerably in the next 1-2 years as autumn approaches.  Cooler MIRI north-polar temperatures at $\sim10$, 12.2, and 13.7 $\mu$m could be a consequence of differences in spatial and spectral resolution (particularly the spectral convolution in the Q-branches of ethane and acetylene), rather than reflecting a real change between 2017 and 2022.

As a further assessment of seasonal change, we compare zonal-mean scans of ground-based VLT/VISIR observations of Saturn at 7.9, 12.3 and 17.6 $\mu$m from 2016 to 2022 \cite{22blake} to the results from CIRS and MIRI (see Supplemental Fig. \ref{fig:visir_comp}).  All three techniques capture the contrasts associated with Saturn's bands, with the 17.6-$\mu$m observations confirming the tropospheric warming during northern summer; and 12.3 $\mu$m showing the warm equatorial band and increased brightness of the NPSV.  However, at 7.9 $\mu$m, Supplemental Fig. \ref{fig:visir_comp}c shows the problems associated with scaling ground-based images to a low-latitude CIRS average, as it missed the stratospheric cooling between $10^\circ$ and $50^\circ$N, and therefore overestimated the brightness of the NPSV.  So whilst the NPSV has warmed since 2017 (by 3-4 K in brightness temperature at 7.9 $\mu$m), the magnitude is smaller than that presented in ground-based studies \cite{22blake}.  In summary, MIRI/MRS observations in 2022 reveal changes to Saturn's equatorial oscillation, unexpected stratospheric cooling equatorward of $40^\circ$N, tropospheric warming at most latitudes, and the continued warming of the NPSV.  
   % Appendix \ref{app:visir_comp}, Fig. \ref{fig:visir_comp}

%%%%%%%%%%%%%%%%%%%%%%%%%%%%%%%%%%%%%%%%%%%%%%%%%%%%%%%%%%
%%%%%%%%%%%%%%%%%%%%%%%%%%%%%%%%%%%%%%%%%%%%%%%%%%%%%%%%%%
%%%%%%%%%%%%%%%%%%%%%%%%%%%%%%%%%%%%%%%%%%%%%%%%%%%%%%%%%%

\section{Spectral Modelling}
\label{spectra}

Although inspection of the cleaned MIRI/MRS cubes can demonstrate the spatial contrasts in Saturn's mid-IR emission associated with cloud banding and discrete storms, further progress can be made by inverting the MIRI spectra to determine Saturn's temperatures, zonal winds, aerosols, and distributions of gaseous species.

\subsection{MRS Spectral Fitting}
Zonally-averaged MRS spectra (Section \ref{sec_zonal}) were fitted using the NEMESIS optimal estimation retrieval algorithm \cite{08irwin}, which has been previously applied to Cassini/VIMS and Cassini/CIRS spectra of Saturn.  Sources of spectral linedata and the generation of $k$-distributions were described on Section \ref{sec_spectra}.  Spectral uncertainties reported by the MRS pipeline (the \verb|ERR| backplane) were averaged for the pixels used for each latitude, but were found to be unrealistically small - we retain the spectral shape of the uncertainty envelope, but increase the error by factors of 10 to 40 (depending on MIRI channel) to enable spectral fits with a goodness-of-fit of approximately one - this is the equivalent of adding forward-modelling uncertainty during the retrieval process \cite{08irwin}.

\begin{figure}[p]
\centering
\includegraphics[width=1.4\textwidth,center]{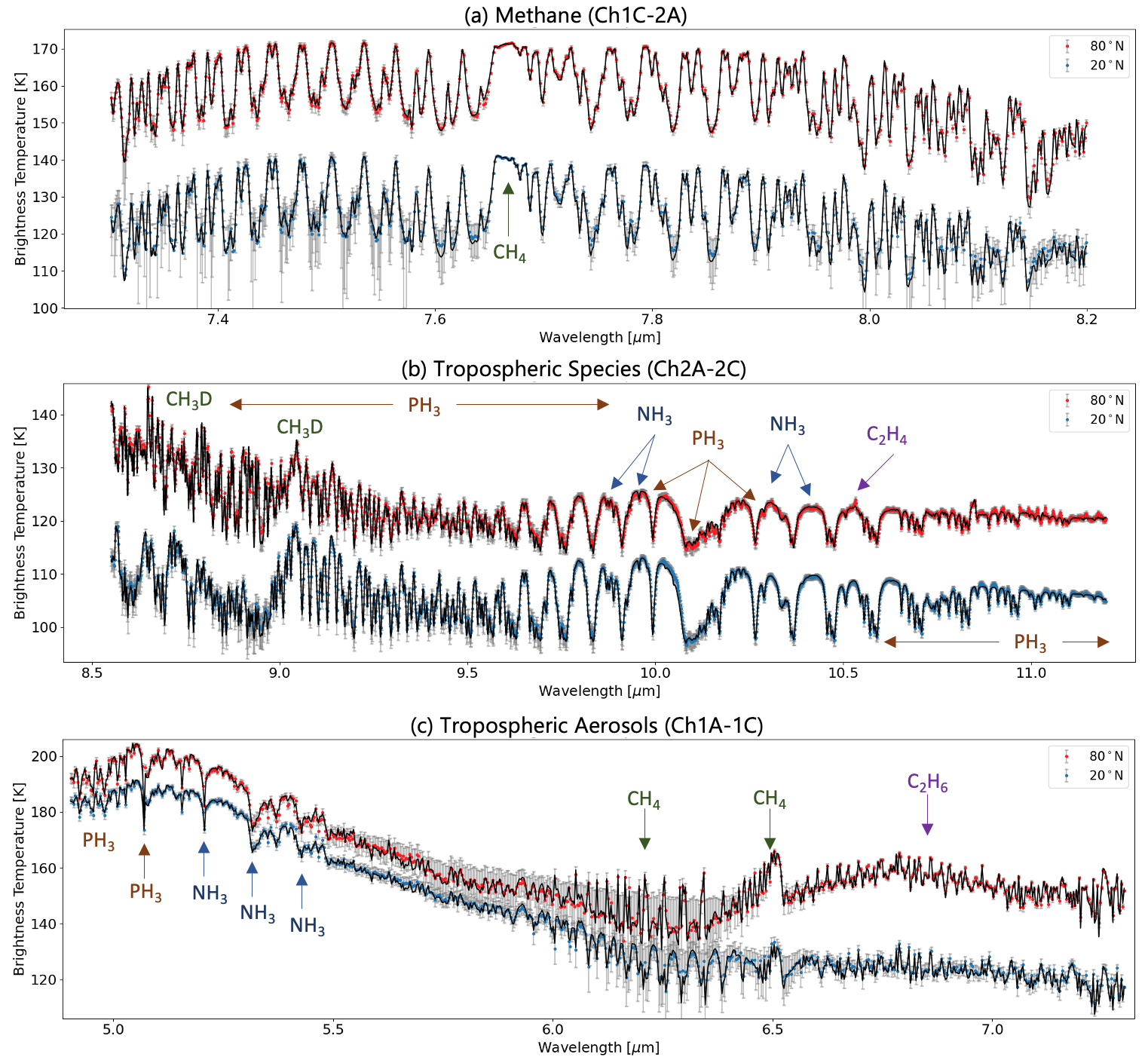}
\caption{Examples of the quality of spectral fits (solid lines) to MIRI/MRS data (points with error bars) at different stages within our multi-stage retrieval.  Panel (a) is dominated by stratospheric CH$_4$ emission, used to determine the vertical temperature structure; panel (b) sounds tropospheric species like PH$_3$, NH$_3$ and CH$_3$D; panel (c) sounds reflection and absorption by tropospheric aerosols, in addition to PH$_3$, NH$_3$, H$_2$O, and emission from stratospheric CH$_4$ and C$_2$H$_2$.  Spectra at \ang{80}N have been offset from the \ang{20}N spectra by 20 K for clarity.  Uncertainties are a scaled version of those reported by the JWST pipeline, as described in Section \ref{spectra}.}
\label{fig:example_fits}
\end{figure}

\begin{figure}[ht]
\centering
\includegraphics[width=1.4\textwidth,center]{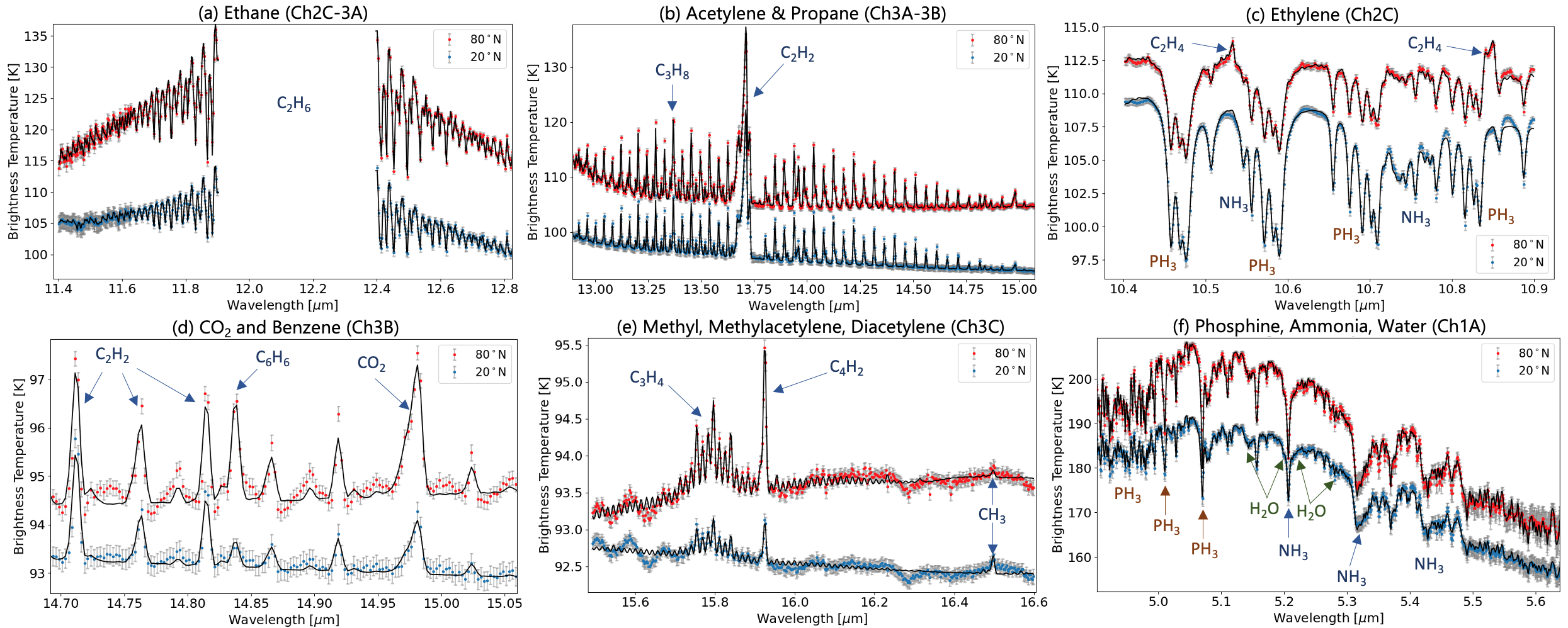}
\caption{Example fits (black line) to observations (points with error bars) in specific regions of Saturn's spectrum, including those regions sensing stratospheric hydrocarbons (ethane in (a) with the central region omitted as described in the text; acetylene and propane in (b); ethylene in (c); CO$_2$ and benzene in (d); methylacetylene, diacetylene, and methyl in (e)) and tropospheric absorptions from ammonia, phosphine and (tentatively) water in (c) and (f).  Spectra for \ang{80}N have been offset from \ang{20}N for clarity by 10 K in (a) and (c); 20 K in (b) and (f) - no offsets were used in (d) and (e).}
\label{fig:cxhy_fit}
\end{figure}

Given the broad spectral coverage, and the high spectral resolution, we adopted a multi-stage approach, first fitting undersampled broadband spectra to constrain atmospheric temperatures and aerosols, before fitting narrowband MRS spectra at their native sampling to study specific gaseous species.  Each MRS subband was fitted simultaneously, but used the correct geometry (emission, incidence, and azimuthal angle) to calculate the atmospheric path, as small differences arise from different pointings between tiles and dithers.  In addition, we divide the data at 7.3 $\mu$m, with longer wavelengths considering only thermal emission and no scattering, but shorter wavelengths considering multiple scattering of both reflected and thermal photons, as has been typical of Cassini studies (see Supplemental Figure \ref{fig:cloud_tests}, which shows that aerosol-free models and cloudy, scattering models converge in the 6.6-7.3 $\mu$m range, although this is somewhat dependent on the choice of refractive indices described below).  The sequential stages were as follows:
% \ref{fig:cloud_tests}

\begin{enumerate}
    \item \textbf{Global Fit:}  Fitting the full 7.3-16.3 $\mu$m region, sampling every 4th point in the spectrum to accelerate the retrieval process, to estimate the $T(p)$ profile as a function of latitude, along with initial assessments of gaseous variability.  Temperatures, continuous profiles of ethane and acetylene, parameterised profiles of NH$_3$ and PH$_3$, and scaled abundances of C$_2$H$_4$, C$_4$H$_2$, C$_3$H$_4$, C$_3$H$_8$, C$_6$H$_6$, CH$_3$, and CO$_2$ were retrieved during step 1 (see below for discussion of priors).  The gaseous abundances would be refined later, but this provided a first zonal-mean $T(p)$ structure.  Tropospheric temperatures and gas abundances were determined by simultaneously fitting the H$_2$-He continuum observed beyond 15 $\mu$m and the absorption of NH$_3$ and PH$_3$ between 8-12 $\mu$m.  Stratospheric temperatures are largely controlled by CH$_4$ emission at 7.8 $\mu$m, but also by the simultaneous fitting of temperature and composition from the ethane and acetylene emission.  We omit wavelengths beyond 16.3 $\mu$m due to challenging fringing/ripples that dominate the long-wavelength spectrum, and the lack of an in-flight MRS calibration for channel 4 at the time of writing.  We omit 11.9-12.3 $\mu$m due to a known artefact in this MRS subband, whereby light leaks through the MRS dichroic filter from 6.1 $\mu$m to 12.2 $\mu$m to create a source-dependent artefact in the data.  This artefact was identified due to difficulties in fitting Saturn's C$_2$H$_6$ emission band at 12.2 $\mu$m simultaneously with CH$_4$ at 7.8 $\mu$m. 

    \item \textbf{Refined Temperatures:}  The initial $T(p)$ from step 1 was then used as a prior for (i) a refined estimate of the stratospheric temperatures, using data from channels 1C and 2A (7.3-8.4 $\mu$m) at their full spectral resolution; and (ii) a refined estimate of tropospheric temperatures, ammonia, and phosphine from 8.0-11.5 $\mu$m (channels 2A to 2C).  Examples of the quality of the spectral fits for latitudes at \ang{20}N (representative of low latitudes) and \ang{80}N (representative of the bright polar emission) are shown in Fig. \ref{fig:example_fits}a and Fig. \ref{fig:example_fits}b.

    \item \textbf{Aerosol Fitting:}  Temperatures were then fixed for aerosol and gaseous retrievals from the 4.9-7.3 $\mu$m range (channels 1A to 1C, sampling every 3rd spectral point).  Unlike the longer wavelengths, this region requires multiple scattering of reflected and thermal light to fit, and full details of the aerosol model are provided below, with example spectral fits shown in Fig. \ref{fig:example_fits}c.  The resulting aerosol profiles were then incorporated back into the initial temperature inversions to check for any changes in the $T(p)$ structure, using only their absorption cross-sections (i.e., without scattering).  Changes to the resulting $T(p)$ were negligible, but would be dependent on the refractive indices of the aerosols, which are not uniquely constrained (see Section \ref{aerosol_fit}).

    \item \textbf{Refined Composition:}  Finally, we adopted the $T(p)$ and aerosol cross-sections, alongside the derived NH$_3$ and PH$_3$ distributions, as priors for focused retrievals of specific spectral features such as the hydrocarbons, HCN, CO$_2$ and H$_2$O.  Examples for a range of gaseous species are shown in Fig. \ref{fig:cxhy_fit}, with key features labelled.
    
\end{enumerate}

\subsection{Fitting Temperatures and Gases}
Saturn's prior $T(p)$ at $p>50$ mbar, as well as PH$_3$ and NH$_3$ profiles, are based on a low-latitude ($\pm30^\circ$ latitude) average from Cassini/CIRS nadir observations \cite{09fletcher_ph3}.  The $T(p)$ is extended deeper than 0.8 bar using a dry adiabatic lapse rate \cite{11fletcher_vims}, and higher ($p<50$ mbar) using a global average of Cassini/CIRS limb observations \cite{09guerlet}, resulting in a prior defined from 1 $\mu$bar to 10 bars.  We adopt the He/H$_2$ ratio from Voyager \cite{00conrath}, CH$_4$ and its isotopologues from Cassini/CIRS \cite{09fletcher_ch4}; C$_2$H$_2$, C$_2$H$_6$ and C$_3$H$_8$ from an average of CIRS limb measurements \cite{09guerlet}; all other hydrocarbons (C$_2$H$_4$, C$_4$H$_2$, C$_3$H$_4$, C$_6$H$_6$, CH$_3$), and CO$_2$ come from averages of the seasonal photochemical model of \citeA{05moses_sat}, updated to a finer latitude grid with zero meridional mixing \cite<$K_{yy}=0$ m$^2$/s,>[]{07moses}.  Prior deep abundances for CO \cite<1.0 ppb,>[]{90noll}, GeH$_4$ \cite<0.4 ppb,>[]{90noll}, H$_2$O \cite<0.176 ppb,>[]{97degraauw} and AsH$_3$ \cite<2.2 ppb,>[]{11fletcher_vims} come from previous investigations at 5 $\mu$m.  HCN uses the stratospheric upper limit from Herschel \cite<22 ppb at $p<1$ mbar,>[]{12fletcher_spire}.

\begin{figure}[ht]
\centering
\includegraphics[width=1.1\textwidth,center]{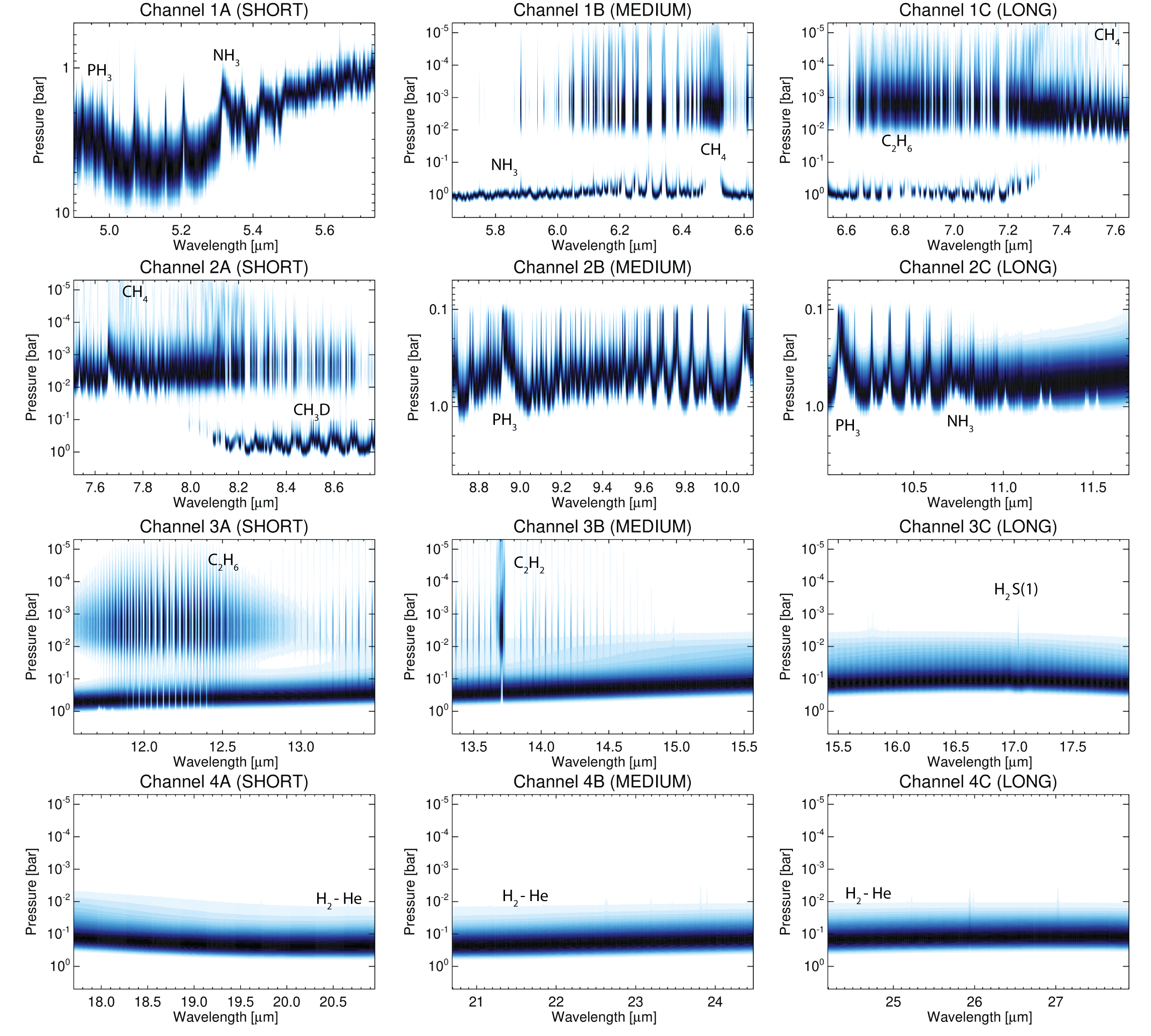}
\caption{Cloud-free contribution functions (Jacobians showing the rate of change of spectral radiance with temperature) calculated for nadir viewing and a typical mid-latitude composition on Saturn.  These contribution functions have been normalised at each wavelength, and darker shading indicates the greatest contribution. No aerosols are included in this model. Key gaseous features have been labelled, note that minor stratospheric species become apparent with higher emission angles (not shown).  The y-axis changes in pressure range to emphasise dominant features.}
\label{fig:contribfns}
\end{figure}

With the exception of CO, AsH$_3$, and GeH$_4$, all species were allowed to vary from their priors during the MIRI/MRS retrievals.  Depending on the size and strength of their spectral contributions in Fig. \ref{fig:contribfns}, gases were either (i) retrieved as full, continuous profiles with height (C$_2$H$_2$, C$_2$H$_6$); (ii) parameterised in terms of a deep mole fraction, transition pressure, fractional scale height (compared to the atmospheric scale height) up to the tropopause (NH$_3$, H$_2$O and PH$_3$); or (iii) simply scaled versions of the prior profiles (i.e., which implicitly assumes that the vertical profile is an accurate representation of Saturn's atmosphere).  Contribution functions for these spectral ranges are shown in Fig. \ref{fig:contribfns}, indicating the approximate sensitivity levels to which the different MRS channels are sensitive.

\subsection{Fitting Aerosols}
\label{aerosol_fit}
Saturn's aerosol distribution is best constrained via remote sensing at visible and near-infrared wavelengths, but the opacity, location, and wavelength-dependent scattering/absorption properties of aerosols contributes significantly in the MIRI/MRS range, particularly at wavelengths below 7.3 $\mu$m.  Studies in this range remain somewhat limited, and are dominated by investigations of aerosol changes in discrete regions such as the polar domain \cite{21sromovsky, 20sanchez_hex} and the northern storm band \cite{13sromovsky, 16sromovsky}.  Attempts to study the latitudinal distribution of aerosols have used Hubble \cite{01stam, 16perez-hoyos}, Cassini/ISS \cite{13roman}, and Cassini/VIMS \cite{11fletcher_vims}, with the latter using only nightside 5-$\mu$m spectra to avoid the complications of reflected sunlight, whereas JWST/MIRI spectra are a blend of thermal emission and reflected sunlight.  To our knowledge, there have been no aerosol retrieval studies that utilise the 5.1-6.8 $\mu$m domain inaccessible to Cassini.

Given the degeneracies inherent in fitting reflected sunlight spectra, as evidenced by the range of different results in the literature \cite{20fletcher_saturn}, we initially adopted an Occam's-razor approach to fitting the 4.9-7.3 $\mu$m range.  Fig. \ref{fig:contribfns} shows that the 5-$\mu$m region senses high pressures (3-7 bars) in the absence of aerosols, so we initially considered a single aerosol population near $\sim1.5$ bars (i.e., above the primary contribution functions near 5.0-5.2 $\mu$m), rather than multiple different compact cloud decks, and allowed the base pressure, top pressure, total opacity and vertical extension to vary freely during the fitting process.  We also avoided imposing any particular spectral shape to the aerosol cross-section, single-scattering albedo, and phase functions (modelled via two-term Henyey Greenstein functions fitting the results of Mie scattering calculations).  Values for particle radius (from a standard gamma distribution with a 5\% variance) and the spectrally uniform real and imaginary refractive indices were chosen after tests at several latitudes where these values were allowed to vary freely.  The radius had a small effect, with $r=1.0\pm0.05$ $\mu$m selected as a best fit. The spectral fits were largely insensitive to the real refractive index, with values representative of NH$_3$ ice $n_r\sim1.4$ \cite{84martonchik} and NH$_4$SH solid $n_r\sim2.3$ \cite{07howett_nh4sh} fitting equally well, so a mean $n_r=1.8$ was selected.  The most important parameter was the imaginary refractive index $n_i$, which varies considerably in the infrared depending on the assumed composition of the aerosols.  Fits were significantly improved with smaller values of $n_i$ in the 5-$\mu$m region than would be typically expected for `pure' NH$_3$ and NH$4$SH aerosols - the final selected value was $n_i=1\times10^{-3}$ over the whole range, resulting in weakly absorbing particles (single scattering albedoes $>0.95$ in this spectral range).

We attempted to fit the 4.9-7.3 $\mu$m range (step 3, above) using this single-cloud layer and the choices of optical properties described above.  Retrievals with multiple scattering of reflected and thermal photons are numerically intensive, requiring numerical evaluation of Jacobians at each step of the inversion, so MIRI/MRS spectra from channels 1A, 1B and 1C were undersampled by a factor of 4 to ensure a good fit across the whole 4.9-7.3 $\mu$m range.  The latitudinally-resolved $T(p)$ derived from steps 1 and 2 of our retrieval scheme were required to correctly reproduce emission from the CH$_4$ $\nu_2$ band at 6.5 $\mu$m and the C$_2$H$_6$ emission band at 6.8 $\mu$m, seen in Fig. \ref{fig:example_fits}c, particularly at high-latitudes where emission from the warm NPSV dominates.  We varied the vertical location and extent of the cloud simultaneously with parametric profiles of NH$_3$, PH$_3$ and H$_2$O.  

% Adding double clouds
This single-cloud model was remarkably successful in fitting most of the 4.9-7.3 $\mu$m range, allowing us to then search for discrepancies which might hint at more complicated cloud structure, such as the multi-layer clouds of \citeA{21sromovsky}, or the potential wavelength-dependent absorptions of photochemical hazes such as those of \citeA{15guerlet}.  Poor fits in the dark equatorial band and polar latitudes led us to experiment with a second cloud layer in the upper troposphere sitting at higher altitudes than the original layer, providing further degrees of freedom but informed by the observation of hazes in Saturn's upper troposphere \cite{13roman}.  A real refractive index of $n_r=1.74$ was selected, similar to the value for diphosphine at 195 K (P$_2$H$_4$) that had been adopted in Cassini/VIMS studies of upper-tropospheric hazes \cite{21sromovsky}, based on the expected photochemical production from PH$_3$.  The resulting fits were weakly sensitive to the aerosol radius ($1.0\pm0.05$ $\mu$m was selected) and imaginary refractive index ($n_i=5\times10^{-3}$ was selected).  This double-cloud scheme can be thought of as representing the upper tropospheric haze and the top-most condensate cloud, with their base pressures, opacities, and vertical extensions all freely varying during the 4.9-7.3 $\mu$m fitting, and producing the high-quality fits shown in Fig. \ref{fig:example_fits}(c).  The results will be discussed in Section \ref{results}.

Finally, although Cassini/CIRS observed aromatic and aliphatic hydrocarbon aerosols in the polar stratosphere ($p<8$ mbar) in limb observations \cite{15guerlet}, there is limited need to include them in the nadir MIRI/MRS spectral fitting.  The CIRS results suggested a peak in opacity near $6.9\pm0.3$ $\mu$m, and there is a subtle but compelling residual in the spectral fits in the same location (see Supplemental Fig. \ref{fig:resid}), which will be the topic of future investigations.
% \ref{fig:resid}
%%%%%%%%%%%%%%%%%%%%%%%%%%%%%%%%%%%%%%%%%%%%%%%%%%%%%%%%%%
%%%%%%%%%%%%%%%%%%%%%%%%%%%%%%%%%%%%%%%%%%%%%%%%%%%%%%%%%%
%%%%%%%%%%%%%%%%%%%%%%%%%%%%%%%%%%%%%%%%%%%%%%%%%%%%%%%%%%
\section{Saturn's Temperatures, Aerosols and Composition in 2022}
\label{results}

The results of the multi-stage retrievals of zonally-averaged temperatures, aerosols, and gaseous species are described in the following subsections.

\subsection{Temperatures and Winds}
\label{sec:temp}

\begin{figure}[p]
\centering
\includegraphics[width=0.6\textwidth]{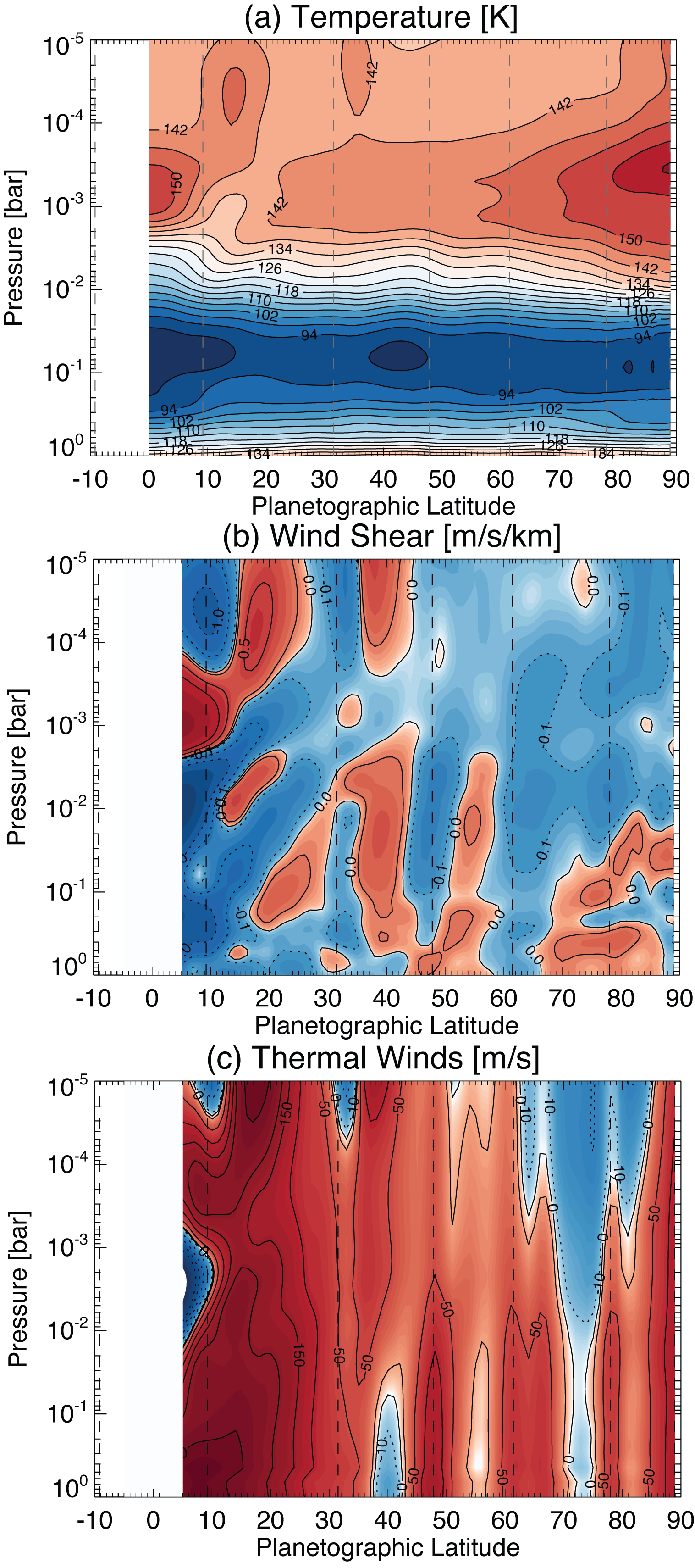}
\caption{Temperatures, windshear, and thermal winds as a function of latitude derived from MIRI/MRS spectroscopy.  The peaks of cloud-tracked eastward winds are shown as vertical dashed lines for context.  In panels b and c, regions of negative windshear/westward winds are shown as blue with dotted contours; regions of positive windshear/eastward winds are shown as red with solid contours.  Temperature contours are every 4 K; windshear and wind contours are logarithmic to show structure, but windshears are labelled at $\pm0.1$, $\pm0.5$ and $\pm1.0$ m/s/km; winds are labelled at 10, 50, 100, 150, and 200 m/s.  Absolute values for derived thermal winds are subject to significant uncertainties related to the integration of the thermal wind equation \cite{16fletcher}, so should be used as a guideline to trends only. }
\label{fig:temp-winds}
\end{figure}

Saturn's zonal-mean temperatures during northern summer are shown in Fig. \ref{fig:temp-winds}, as derived from step two (i.e., refined spectral fitting at full spectral resolution after a `global fit' to the 7.3-16.4 $\mu$m spectrum).  The temperature inversion confirms many of the conclusions available from the brightness temperature maps alone.  The troposphere is characterised by a cool EZ; temperature gradients between mid-latitude belts and zones that correlate with the peaks of the cloud-tracked zonal winds; and a warm polar domain.  The cool EZ is coincident with the highest aerosol opacity (Section \ref{sec:aerosols}), and it is therefore possible that our aerosol-free assumption for wavelengths beyond 7.3 $\mu$m is inadequate, despite tests in Section \ref{spectra} suggesting that the derived aerosols had minimal absorption at these wavelengths.  Further refinement of the aerosol refractive indices, incorporating wavelengths longer than 7.3 $\mu$m, would be needed to fully resolve this potential degeneracy between aerosols and temperatures.

The stratosphere exhibits warming within the polar domain, reaching maximum temperatures of $154\pm1$ K within the NPSV (poleward of \ang{78}N) and $160\pm2$ K within the NPC (poleward of \ang{87}N).  These peak temperatures are only slightly cooler than those observed within the southern stratosphere in 2004-05 - the SPSV and SPC - by Cassini/CIRS \cite{18fletcher_poles}, a seasonal asymmetry possibly due to Saturn's orbital eccentricity.  Radiative models demonstrate that the bulk of this warming is driven by seasonal radiative heating \cite{14guerlet, 16hue, 22blake}, but the sharp boundaries are related to dynamics (e.g., stratospheric winds encircling the NPSV and NPC).  Temperature inversions indicate a moderate stratospheric cooling at pressures below 300 $\mu$bar, with temperatures approaching a $140\pm3$ K quasi-isothermal structure up to the base of the thermosphere.  While retrieved low-pressure temperatures are somewhat influenced by the choice of prior, this isothermal structure is consistent with radiative modelling \cite<e.g.,>[]{14guerlet}.  

The most prominent feature of the MIRI temperature field is the vertical structure of Saturn's Equatorial Stratospheric Oscillation (we refer to this as the SESO, as the semi-annual nature of the oscillation is questionable).  During the November-2022 phase, a prominent warm anomaly ($153\pm0.8$ K) is observed between the equator and $\sim$\ang{10}N centred near 0.7 mbar.  This is some 12-14 K warmer than temperatures at 0.1 mbar, and is responsible for the warm band of bright stratospheric emission observed in Fig. \ref{fig:equator_pole} and Fig. \ref{fig:zonal_average}.  This is accompanied by off-equatorial temperature maxima near \ang{13}N, a strong warm anomaly near 0.05 mbar, and a weaker cool anomaly near 1 mbar.  Similar vertical patterns were observed $\sim16-17$ years earlier by Cassini/CIRS limb spectroscopy \cite{08fouchet} in 2005-06, and are revealed here due to the high spectral resolution of MIRI/MRS.  

Latitudinal temperature gradients $dT/dy$ (where $y$ is the north-south distance in km) are converted to vertical windshears $du/dz$ (where $z$ is altitude) via the thermal wind equation \cite<e.g.>[]{04holton}, omitting the equatorial region where the Coriolis parameter tends to zero.  These are shown in Fig. \ref{fig:temp-winds}b, and show intriguing shear structure equatorward of \ang{30}N (broadly the domain occupied by Saturn's equatorial jet).  The positive equatorial shearzone in near 1 mbar in 2022 is likely to be the one that was seen near 0.1-0.3 mbar in 2017 by Cassini \cite{17fletcher_QPO}, having descended over a decade in pressure in five years.  Positive and negative shear zones associated with the SESO and its off-equatorial structures appear to move diagonally to higher pressures with decreasing latitude.  For example, the negative shear zone in the upper troposphere appears continuously connected to the negative shear zone at 1 mbar and \ang{30}N.  The same connection is seen for the negative shear zone at 10 mbar (\ang{5}N) and 0.05 mbar (\ang{30}N).  Such a system of stacked shear zones connecting the equator and off-equatorial features was nicely captured by the model of \citeA{22bardet}.  

Windshears derived from nadir infrared data provide a good picture of the shearzones as a function of altitude and latitude, but using them to integrate winds with altitude and across regions of low information content (e.g., the tropopause and lower stratosphere) can generate significant uncertainties \cite<e.g.,>[]{08fouchet, 16fletcher}.  Nevertheless, we display the estimated thermal winds in Fig. \ref{fig:temp-winds}c, assuming that the continuum-band winds measured by \citeA{11garcia} are placed at 500 mbar.  Treating these thermal winds with appropriate caution, we infer a localised equatorial westward jet (exceeding -200 m/s) near 1-5 mbar and equatorward of \ang{10}N, embedded within a larger region of eastward flow that spans the tropics equatorward of \ang{30}N.  The westward jet is below an eastward equatorial jet ($\sim200$ m/s) near 0.1-0.5 mbar, with the peak-to-peak contrast of $\sim400$ m/s between eastward and westward maxima being comparable to that derived from Cassini limb observations \cite{08fouchet}.  Direct observations of Saturn's stratospheric winds by ALMA \cite{22benmahi} four years earlier (2018) observed a $\sim300$ m/s eastward jet between \ang{20}S and \ang{20}N but with a coarse vertical resolution, covering 0.01-20 mbar.  Although this is qualitatively consistent with the broad eastward flow inferred from MIRI in Fig. \ref{fig:temp-winds}c, the coarse resolution of the ALMA data may average over any oscillatory wind patterns over a decade of pressure.  Alternatively, differences between ALMA and MIRI might simply reflect the downward propagation of these stacked zonal jets over four years, and future joint campaigns between ALMA and JWST would be welcome to confirm this, alongside ground-based spectroscopic monitoring of the equatorial CH$_4$ emission at high spatial and spectral resolution.

% We find a strong equatorial westward jet (exceeding -200 m/s) near 1-5 mbar% similar to those observed by Cassini \cite{08fouchet}
% , below a strong eastward equatorial jet ($\sim200$ m/s) near 0.1-0.5 mbar. 
%, and a narrower westward $\sim50$ m/s jet near 0.3 mbar.
%Whilst these are opposite to that predicted by the thermal wind equation applied to MIRI, differences might simply reflect the downward propagation of these stacked zonal jets.  For example, movement of the westward jet from 0.3 mbar (2018) to near 3 mbar (2022) is consistent with the rate of descent of shear zones observed by Cassini \cite{17fletcher_QPO}, and future joint campaigns between ALMA and JWST would be welcome to confirm this.  

Beyond the equator, the correlation between $dT/dy$ and the cloud-tracked zonal winds causes the decay of the cloud-top winds with height (e.g., the westward jet near \ang{42}N becomes eastward at $p<80$ mbar), as previously observed by Cassini \cite{09read}.  Finally, the strong $dT/dy$ at the edge of the NPSV implies negative windshear and the inference of westward flow around the edge of the vortex for $p<10$ mbar \cite{18fletcher_poles}, at a latitude consistent with the westward winds directly observed by ALMA near \ang{74}N planetographic \cite{22benmahi}.  This westward wind is zonally symmetric and in balance with the seasonal temperature gradients from radiative heating within the NPSV.

%, with limited relationships to the auroras.

%%%%%%%%%%%%%%%%%%%%%%%%%%%%%%%%%%%%%%%%%%%%%%%%%%%
%%%%%%%%%%%%%%%%%%%%%%%%%%%%%%%%%%%%%%%%%%%%%%%%%%%
%%%%%%%%%%%%%%%%%%%%%%%%%%%%%%%%%%%%%%%%%%%%%%%%%%%
\subsection{Aerosols}
\label{sec:aerosols}
The zonal-mean distribution of Saturn's aerosols are shown in Fig. \ref{fig:clouds-ph3}(a), derived from fits to the 4.9-7.3 $\mu$m region using the double-cloud scheme and multiple scattering of reflected and thermal photons as outlined in Section \ref{aerosol_fit}.  Although significant degeneracy exists in the choices of aerosol refractive indices and size distributions, the need for two clouds was evident from the difficulty fitting equatorial latitudes (where clouds are most reflective).  The base pressures, vertical extensions, top-most pressures, and total opacity were then varied as free parameters for the two aerosol populations, resulting in the stacked cloud decks observed here.  

% Tropospheric haze
The upper cloud, which may comprise a photochemically-produced haze potentially associated with diphosphine \cite{80ferris}, is optically thickest, highest (a base near 200 mbar), and most extended at the equator, reaching into the lower stratosphere near 60-70 mbar. This upper haze is found deeper (near 300 mbar) and more compact from \ang{18}-\ang{48}N, and then declines considerably at higher latitudes beyond \ang{60}N.  The latitudinal dependence and altitude of this haze layer is well-matched to the small inflection in the $T(p)$ profile observed by Cassini/CIRS and known as the `temperature knee' \cite{20fletcher_saturn}, suggesting that seasonal heating of this aerosol population is responsible for the change in the curvature of the vertical temperature profile.  The absence of this aerosol at high latitudes may be partially responsible for the bright 5-$\mu$m emission between \ang{60}-\ang{80}N in Figs. \ref{fig:equator_pole}, \ref{fig:zonal_average} and \ref{fig:ch1a_maps}.  The latitudinal distribution of this upper haze also matches that derived from Palomar visible-light observations acquired in 1995 \cite{01stam}, and Cassini/ISS determinations of aerosols during 2004-2007 \cite{13roman}, which identified haze-top pressures ranging from 40 to 150 mbar, with the thickest and highest at the equator, becoming deeper and thinner at mid-latitudes.  

% Deeper cloud
The deeper cloud, which may be associated with condensates of NH$_3$ \cite<potentially coated in other material, e.g.,>[]{21sromovsky}, resides between 1-2 bars, but with an extension to lower pressures that maybe merges with the upper haze.  The base pressure of this cloud varies with latitude, reaching the lowest pressures (1.2 bars) near \ang{10}N, which was the site of the highest reflectivity in Fig. \ref{fig:ch1a_maps}, and the deepest pressure (2.6 bars) within the polar domain.  Indeed, the inversions poleward of \ang{78}N suggest that this deep cloud resides at $p>2$ bars and is vertically compact, responsible for both the dark thermal emission at 5 $\mu$m and the absence of reflectivity at 5.2 $\mu$m - the dark north pole is therefore caused by this deep aerosol layer.  The compact nature of this cloud deck appears to be consistent with Cassini observations \cite{11fletcher_vims, 21sromovsky}, although we caution that the vertical extension is a rather poorly constrained parameter in these retrievals.  The mean pressure of the cloud base is again consistent with that found by Cassini/ISS \cite<$1.75\pm0.4$ bars,>[]{13roman}, even though optical measurements only reveal this deep cloud when there are gaps in the overlying haze. 

% What isn't included?
This simple two-cloud scheme does not provide constraints on two other proposed cloud layers:  neither a deep cloud \cite<2.7-4.5 bars, potentially due to NH$_4$SH>[]{21sromovsky}; nor the stratospheric hazes \cite<near 50 mbar,>[]{13roman, 15guerlet, 21sromovsky}.  The deep cloud does not seem to be required to reproduce the MIRI data, but this may be due to lack of constraint from reflected sunlight at shorter wavelengths (e.g., from NIRSpec).  The very low opacities ($0.08\pm0.05$ at 619 nm) and small particle sizes ($<0.3$ $\mu$m) of the stratospheric aerosols inferred by \citeA{13roman} make it very unlikely that they would contribute significant opacity at longer mid-IR wavelengths.  Nevertheless, very subtle residuals in our fits to north polar latitudes near $6.8\pm0.2$ $\mu$m could be related to the stratospheric aerosols (see Supplemental Fig. \ref{fig:resid}c), and will be the topic of future studies, given their photochemical nature and potential importance in the balance of radiative heating and cooling in the polar domain \cite{15guerlet}.

%%%%%%%%%%%%%%%%%%%%%%%%%%%%%%%%%%%%%%%%%%%%%%%%%%%
%%%%%%%%%%%%%%%%%%%%%%%%%%%%%%%%%%%%%%%%%%%%%%%%%%%
%%%%%%%%%%%%%%%%%%%%%%%%%%%%%%%%%%%%%%%%%%%%%%%%%%%
\subsection{Tropospheric Gases}

Saturn's tropospheric gases - namely NH$_3$, PH$_3$, and H$_2$O - are accessible in the 4.9-6.0 $\mu$m region (primarily channel 1A) and the 8-11 $\mu$m region (spanning channels 2A to 2C).  The latter sounds higher altitudes in the upper troposphere, whereas the former provides access to the deeper cloud-forming layers.

\subsubsection{Phosphine}
Phosphine (PH$_3$) is retrieved parametrically from both spectral regions (i.e., varying the deep mole fraction and scale height for a fixed transition pressure of 1 bar), and the results from 5 $\mu$m are shown in Fig. \ref{fig:clouds-ph3}b.  For $p>1$ bar, the deep abundance varies between 4.0-5.0 ppm over most of the northern hemisphere, consistent with results from Cassini/VIMS \cite{11fletcher_vims, 21sromovsky}, but with no notable contrasts at the equator, and a higher abundance $6.5\pm1.0$ ppm poleward of \ang{80}N.  Much of the latitudinal structure is therefore found in the upper troposphere, driven by the fractional scale height of the gas.  For $p<1$ bar, PH$_3$ is enriched equatorward of \ang{15}N with evidence for a slight depletion right at the equator.  Further enriched bands are found at \ang{33}N, \ang{46}N and between \ang{60}-\ang{80}N, with a general decline in abundance towards the north pole.   The equatorial peak and presence of bands of elevated PH$_3$ were also observed by Cassini/CIRS \cite{09fletcher_ph3}, but the precise locations differ - in particular, there is no good correspondence between mid-latitude PH$_3$ bands and zones of cooler temperatures, as we might expect if only dynamics (i.e., upwelling) were controlling the mid-latitude distribution.  Conversely, there is a good correspondence between higher cloud bases and the elevated PH$_3$, reinforcing links between PH$_3$ and aerosols shielding the gas from UV photolysis.  This correspondence breaks down in the 5-$\mu$m-bright band near \ang{60}-\ang{80}N, where we have thin clouds but enriched PH$_3$, suggesting a more complex balance between aerosol shielding and vertical mixing.  Where clouds are deepest and most compact poleward of \ang{80}N, PH$_3$ could be depleted either by polar subsidence \cite{08fletcher_poles} or by diminished aerosol shielding.

Inversions from the 10-$\mu$m region show a similar morphological structure but primarily sense $p<1$ bar (see Fig. \ref{fig:contribfns}).  However, the mid-latitude peaks are at different locations - \ang{29}N and \ang{44}N, and the deep abundances vary between 7-10 ppm, a factor of $\sim2$ higher than those derived from the 5-$\mu$m region.  This is a known discrepancy between PH$_3$ abundances derived from the two regions and occurs on both Saturn \cite{11fletcher_vims} and Jupiter \cite{15giles}, and reconciliation will require joint fitting of both spectral domains with multiply-scattering aerosols.

\subsubsection{Ammonia}
The latitudinal distribution of NH$_3$ was also retrieved parametrically (with a transition pressure at 1.75 bars), with the 5-$\mu$m results shown in Fig. \ref{fig:clouds-ph3}c.  As previously observed by Cassini/RADAR \cite{13laraia} and VIMS \cite{11fletcher_vims}, MIRI reveals a strong equatorial enhancement within \ang{5} of the equator, with deep abundances below $p>1.75$ bars of 450-650 ppm, compared to a minimum of $\sim150$ ppm at \ang{10}N.  This equatorial enhancement on Saturn shares similarities with that observed on Jupiter \cite{16depater, 17li}, and suggests shared dynamics driving the unique compositions of their cold Equatorial Zones, compared to strong NH$_3$ depletion at other latitudes.  Despite the strong equatorial maximum at depth, the abundance falls steeply with height equatorward of \ang{20}N, such that NH$_3$ at the 200-mbar level displays a local equatorial minimum, perhaps due to the enhanced condensation through cold equatorial temperatures to form the thicker clouds observed here. 

Further local maxima in the deep abundance occur at \ang{26}N and \ang{49}N, which are similar in size (but at different latitudes) compared to peaks observed by Cassini/VIMS in 2006 \cite{11fletcher_vims}, suggesting temporal variability in the abundance of NH$_3$ in the mid-latitude bands.  The region near \ang{40}N is notable as displaying the shallowest gradient in NH$_3$ for $p<1$ bar, coinciding with the coldest tropospheric band in Fig. \ref{fig:temp-winds}a, but is actually a local minimum ($\sim100$ ppm) in the deep NH$_3$ abundance for $p>1.75$ bars.  Similarly, there is a general increase in deep abundance towards the polar domain ($\sim600$ ppm for $p>1.75$ bars), but also a shallower upper tropospheric gradient implying a local polar minimum at 200 mbar.  These MIRI results imply that ammonia displays different latitudinal distributions above and below the condensed aerosols near 1.75 bars, suggestive of differing circulation patterns at different heights \cite{20fletcher_beltzone}.  To reinforce this, NH$_3$ was also retrieved from the 10-$\mu$m region, sensing higher altitudes of the upper troposphere, and resulted in a relatively meridionally uniform distribution similar to that derived by Cassini/CIRS \cite{12hurley}.  The strongest NH$_3$ contrasts are therefore only seen in the cloud-forming region sensed near 5 $\mu$m, with only small latitudinal gradients in the stably-stratified upper troposphere sensed near 10 $\mu$m.

\subsubsection{Water}

Perhaps the most tantalising prospect offered by MIRI/MRS is the possibility of mapping Saturn's tropospheric water in the 5.1-5.5 $\mu$m region, which remained out of reach to Cassini/VIMS because of the lack of spectral coverage beyond 5.1 $\mu$m.   \citeA{97degraauw} detected Saturn's tropospheric water with disc-averaged observations from ISO/SWS, fitting it with 0.2 ppm at $p>3$ bars and generally sub-saturated conditions.  H$_2$O was retrieved parametrically, with initial testing at mid-latitudes suggesting the transition pressure should lie at $p>4$ bars (5 bars was chosen), a weak sensitivity to the chosen deep abundance (10 ppm was selected), but a stronger sensitivity to the fractional scale height, so only the latter parameter was allowed to vary.  

\begin{figure}[ht]
\centering
\includegraphics[width=1.1\textwidth,center]{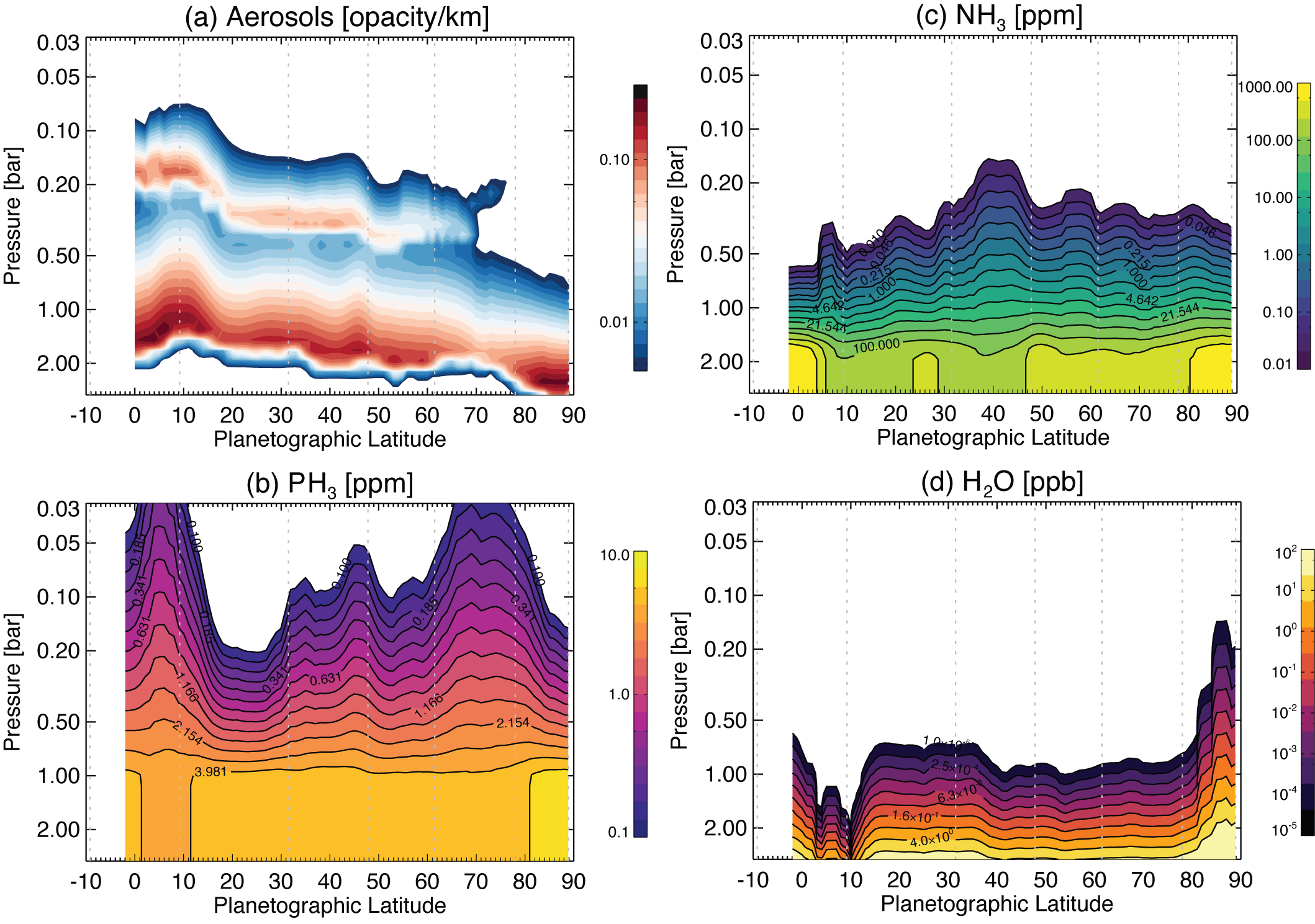}
\caption{Aerosols, phosphine, ammonia and water derived from the 4.9-7.3 $\mu$m region.  Aerosols in panel (a) are plotted in opacity/km at a reference wavelength of 5 $\mu$m, calculated following the scheme in Appendix C of \citeA{22irwin}.  A logarithmic colour bar is used to show structure within the aerosol cross-section.  For the gases in panels (b)-(d), the contours are also logarithmic to allow for the rapid decline of the abundance with altitude.  PH$_3$ and NH$_3$ are provided in ppm, H$_2$O is given in ppb.  Vertical dotted lines show the latitudes of tropospheric eastward jets. }
\label{fig:clouds-ph3}
\end{figure}

Tropospheric water lines are difficult to observe between the forest of PH$_3$ features in Fig. \ref{fig:example_fits}f, and are most visible as small notches either side of the narrow NH$_3$ feature near 5.2 $\mu$m.  These H$_2$O features are present at all latitudes, but are close to the level of the spectral residuals seen elsewhere in the channel-1 data (see Supplemental Fig. \ref{fig:resid}c).  Nevertheless, the MIRI/MRS data provide tentative evidence of latitudinal variability of H$_2$O in Fig. \ref{fig:clouds-ph3}d.   At 3.3 bars, the water abundance varies from a maximum of $\sim200$ ppb right at the equator \cite<consistent with ISO estimates from>[]{97degraauw}, to $\sim10$ ppb between \ang{5}-\ang{10}N, to $\sim145$ ppb from \ang{15}-\ang{35}N, then declines to $\sim80$ ppb from \ang{40}-\ang{80}N.  Poleward of \ang{15}N, the changes in abundance with latitude mirror those in the aerosol distribution, with a tendency for enhanced H$_2$O where the aerosols have a higher optical depth, although we caution that this could reflect a model-dependent degeneracy given the weakness of the H$_2$O features.  The low values near \ang{5}N are very poorly constrained due to the thicker aerosol opacity, but are possibly due to enhanced condensation in the cool EZ.  The shallower gradient in the polar domain are possibly due to the warmer upper-tropospheric temperatures there and the absence of upper tropospheric aerosols, but the same is not seen for NH$_3$.  Conversely, the increase in H$_2$O right at the equator mimics the equatorial column of NH$_3$, suggesting a volatile-rich domain at Saturn's equator.

We note that the challenge of calibrating MRS channel 4 (i.e., beyond 18 $\mu$m) limits our sensitivity to the distribution of stratospheric H$_2$O, which will be the topic of future investigations.

\subsection{Stratospheric Chemistry}
\label{strat_chem}
The spatial distribution of chemical species provides a means to trace Saturn's stratospheric circulation, to understand the photochemical lifetimes of different products, the exogenic supply of oxygenated species, and the potential influence of ionisation on the chemistry of the polar domains.  As described in Section \ref{sec_spectra}, MIRI/MRS provides access to a host of stratospheric chemicals with a higher spectral resolution and sensitivity than Cassini/CIRS.  However, fringing artefacts that still plague wavelengths beyond 10 $\mu$m make precise quantitative measurements challenging, particular for minor species.  In the following sections, we present an initial assay of Saturn's stratospheric composition based on MIRI/MRS data.

\subsubsection{Acetylene and Ethane}

\begin{figure}[ht]
\centering
\includegraphics[width=1.0\textwidth,center]{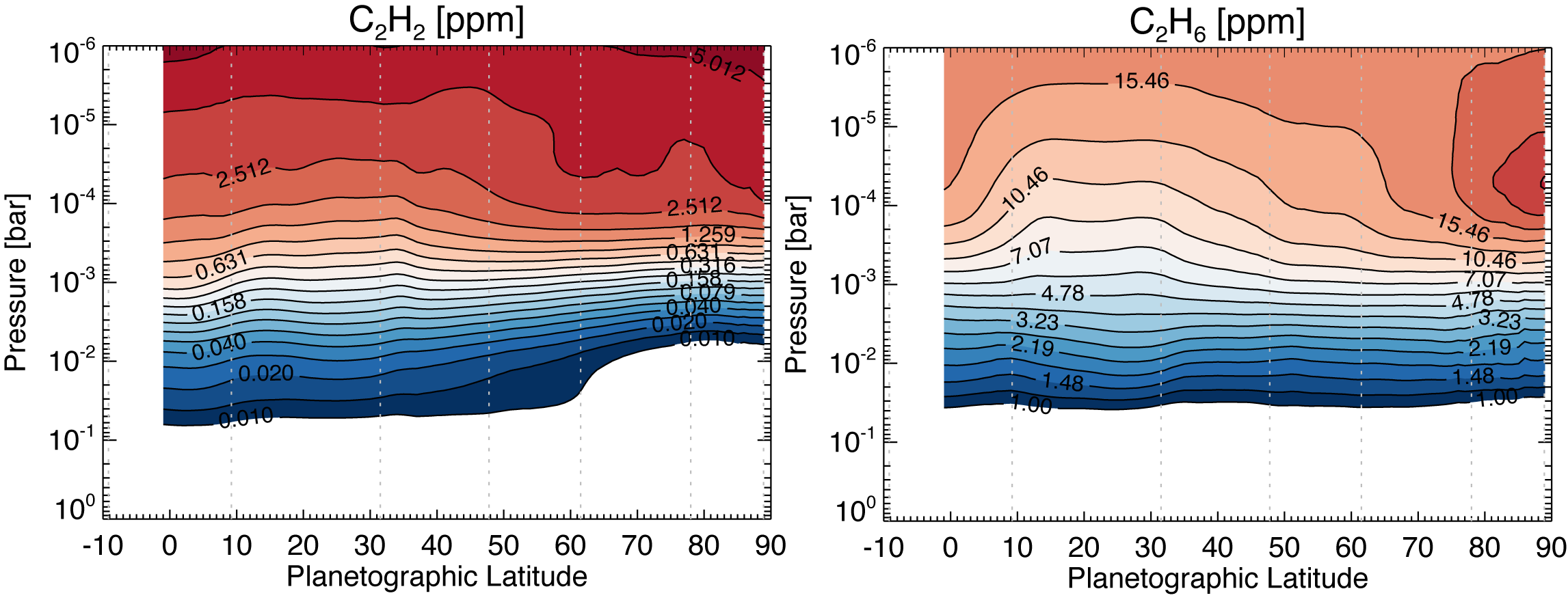}
\caption{Distributions of acetylene and ethane, derived as continuous profiles to capture changes in the vertical gradients of these species.  Contours are logarithmic, and labelled in ppmv.  Vertical dotted lines show the latitudes of tropospheric eastward jets. A cross-section at 0.5 mbar is shown in Fig. \ref{fig:cxhy_all}.}
\label{fig:cxhy_distn}
\end{figure}

\begin{figure}[p]
\centering
\includegraphics[width=1.4\textwidth,center]{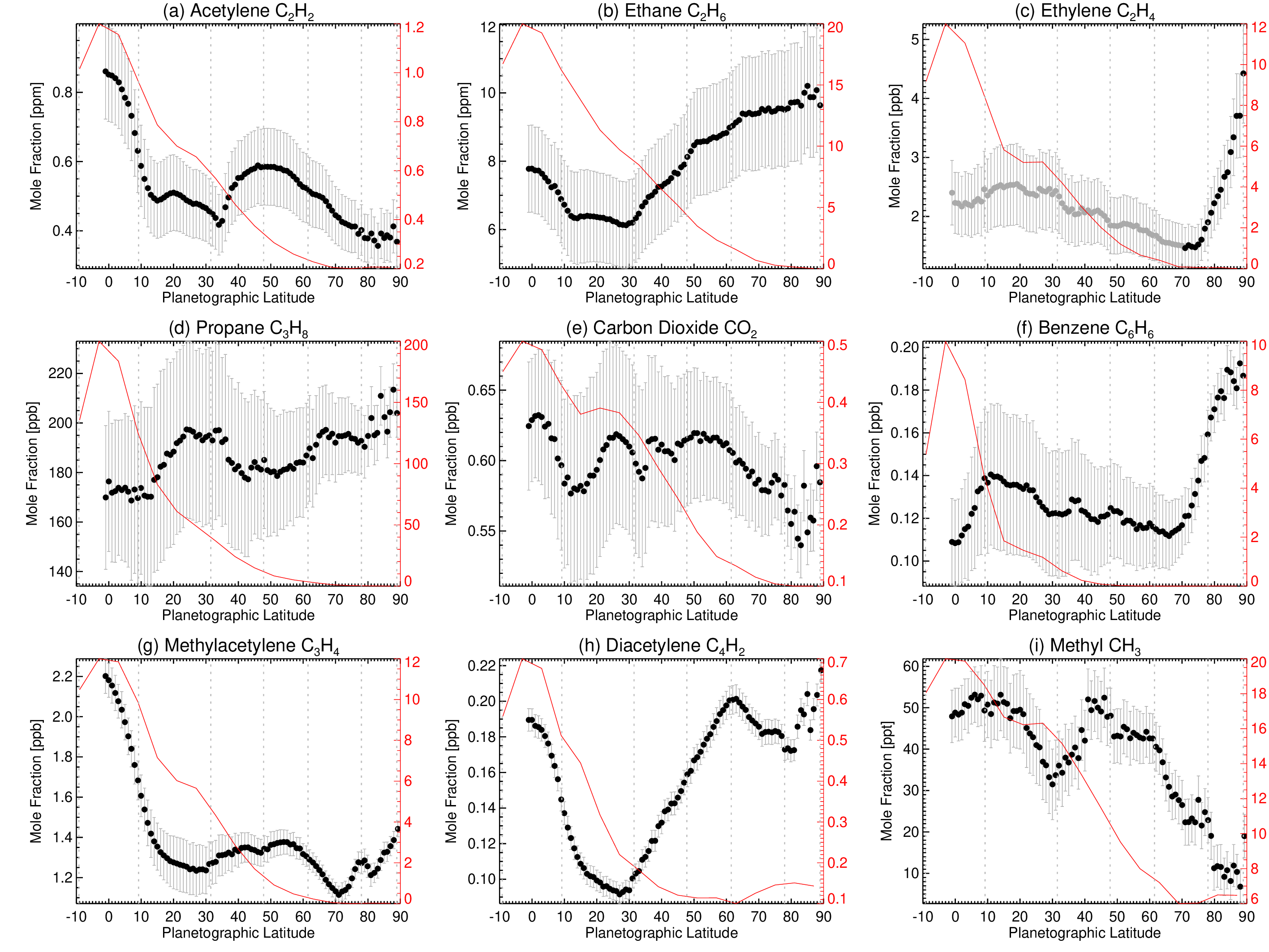}
\caption{Distributions of stratospheric species at 0.5 mbar: acetylene, ethane, ethylene, propane, CO$_2$, benzene, methyacetylene (propyne), diacetylene (1,3-butadiyne), and methyl in Saturn's northern summer, compared to the predictions of the neutral photochemistry model for $L_s=150^\circ$ \cite{07moses, 05moses_sat}.  The photochemical model is shown in red, and is referenced to the right-hand axis so that both the shape of the distribution and differences in absolute abundances can be compared.  Error bars are shown in grey, but these do not include uncertainties due to scaling a latitudinally-uniform \textit{a priori} profile. If the shape of the profile changes significantly with latitude (as is expected for benzene and C$_4$H$_2$ from previous studies), this could influence the retrieved values.  Vertical dotted lines show the latitudes of tropospheric eastward jets.  Gases in panels (a)-(c) were derived as full vertical profiles, other gases are derived as scale factors for our \textit{a priori} profiles.  Grey points in panel (c) signify a lack of obvious detection of C$_2$H$_4$ emission by eye.}
\label{fig:cxhy_all}
\end{figure}

Latitudinal cross-sections of ethane (C$_2$H$_6$) and acetylene (C$_2$H$_2$) are shown in Fig. \ref{fig:cxhy_distn}, and at the 0.5-mbar level in Fig. \ref{fig:cxhy_all}a,b, based on spectral fits shown in Fig. \ref{fig:cxhy_fit}.  Cassini observations revealed that both species are time-variable, responding to Saturn's seasonally-evolving circulation and solar flux.  The equator-to-pole gradient of C$_2$H$_2$ changes with height, with the upper stratosphere showing polar enrichment (poleward of \ang{60}N and $p<0.1$ mbar), but the lower stratosphere showing polar depletion ($p>0.1$ mbar) due to the stronger gradient of C$_2$H$_2$ near the poles compared to at other latitudes.  C$_2$H$_6$ also shows strong polar enrichment (poleward of \ang{78}N and $p<0.1$ mbar, within the NPSV).  Both species show enhancements at the equator, as previously observed by Cassini \cite{09guerlet, 13sinclair, 15sylvestre} and predicted by photochemical models based on the annual-average insolation \cite{05moses_sat, 15hue}, but this too is time variable, with evidence that the equatorial C$_2$H$_6$ peak has strengthened with time whereas C$_2$H$_2$ has remained reasonably constant \cite{20fletcher_saturn}.  

The different latitudinal trends in the lower stratosphere  (C$_2$H$_2$ declining towards high latitudes, C$_2$H$_6$ increasing) have been observed previously, with short-lived C$_2$H$_2$ more closely following photochemical predictions of \citeA{05moses_sat} (Fig. \ref{fig:cxhy_all}a), whereas long-lived C$_2$H$_6$ is more sensitive to stratospheric circulation.  Intriguingly, local maxima between \ang{10}-\ang{30}N observed in both species during northern winter by Cassini \cite<2005-2012,>[]{09guerlet, 15sylvestre} have now been replaced by local minima between \ang{10} and \ang{35}N during northern summer observed by JWST.  This supports the idea that wintertime subsidence has been replaced by summertime upwelling in this latitude range (i.e., upwelling of hydrocarbon-depleted air from the lower stratosphere), associated with the seasonal reversal of the inter-hemispheric Hadley cell \cite{12friedson, 22bardet}.  This upwelling may provide an explanation for why MIRI observed colder stratospheric temperatures in 2022 (Section \ref{sec:temp}) compared to Cassini in 2017.

\subsubsection{Ethylene}

Ethylene was previously only detected within Saturn's storm-perturbed stratosphere \cite{12hesman, 15moses}, due to the elevated temperatures and a photochemical increase in the C$_2$H$_4$ abundance in 2011.  Other than the storm and a reported ground-based detection \cite{01bezard_dps}, C$_2$H$_4$ had proven elusive until the high sensitivity of MIRI/MRS, which reveals 10.5-$\mu$m C$_2$H$_4$ emission within the NPSV for the first time (Fig. \ref{fig:cxhy_all}c, 0.5-mbar abundances of $4.0\pm0.5$ ppb poleward of \ang{80}N).  This abundance is a factor of 2-3$\times$ lower than that expected (but not seen) at the equator due to neutral photochemistry \cite{05moses_sat}.  Note that the emission features are only readily detectable in spectra poleward of \ang{70}N - at lower latitudes, our model provides the highest C$_2$H$_4$ abundance that would be consistent with the non-detection of emission (within uncertainties).  The seasonal model overpredicts the estimated low-latitude abundance ($2.3\pm0.5$ ppb at 0.5 mbar compared to $\sim12$ ppb from the model) by a factor of 4, but the data are more consistent with updated models for Saturn's stratosphere \cite<1-2 ppb at 0.5 mbar from Fig. 6 of>[]{15moses}.  

Nevertheless, the polar maximum in C$_2$H$_4$ in Fig. \ref{fig:cxhy_all}c is unexpected on the grounds of neutral photochemistry, and is suggestive of either subsidence of hydrocarbon-rich air from higher altitudes (leading to the maxima in ethane and acetylene), or due to an enhanced contribution from ion-neutral chemistry at the highest latitudes.  Distinguishing these scenarios will require future modelling work for the chemistry of the NPSV.

\subsubsection{$C_3$, C$_4$ and C$_6$ Hydrocarbons}

\textbf{Propane} is detected as a perturbation to the stronger C$_2$H$_2$ lines near 13.37 $\mu$m (indicated in Fig. \ref{fig:cxhy_fit}), which results in the large uncertainties on the distribution in Fig. \ref{fig:cxhy_all}d.  Propane bands $\nu_{20}$ and $\nu_{21}$ are observed near 9.4 and 10.7 $\mu$m, respectively (see Supplemental Fig. \ref{fig:residc3h8}), but primarily at high latitudes.   The strongest detections are within the warm NPSV, with abundances near $200\pm10$ ppb, and a relatively uniform distribution with latitude, consistent with that found by \citeA{15sylvestre}.  It is clear that C$_3$H$_8$ does not follow the predictions of seasonal photochemistry, suggesting the influence of meridional circulation on the distribution of long-lived propane \cite<e.g., similar to the argument for C$_2$H$_6$,>[]{05moses_sat}.  

Unlike propane and ethane, the unsaturated hydrocarbons \textbf{methylacetylene} (C$_3$H$_4$) and \textbf{diacetylene} (C$_4$H$_2$) are relatively short-lived species that track the photochemical model predictions at latitudes equatorward of \ang{45}N, with C$_3$H$_4$ in particular showing the expected equator-to-pole contrast in Fig. \ref{fig:cxhy_all}g.  C$_4$H$_2$ shows a local minimum near \ang{20}-\ang{40}N that has evolved from the local maximum observed by Cassini in 2005-06 \cite{10guerlet}, possibly related to the onset of low-latitude stratospheric upwelling during northern summer.  Surprisingly, C$_4$H$_2$ then increases towards the NPSV where some of the highest abundances are obtained in Fig. \ref{fig:cxhy_all}h.  This is unlikely to be a circulation effect, as C$_4$H$_2$ has only a slightly shorter loss timescale than C$_3$H$_4$ \cite{10guerlet}, so we might expect to see the same structure in C$_3$H$_4$.  Photolysis of C$_2$H$_2$ is the dominant production mechanism for C$_4$H$_2$ \cite{05moses}, so the polar C$_4$H$_2$ may simply be a result of the excess C$_2$H$_2$ within the NPSV at high altitudes in Fig. \ref{fig:cxhy_distn}, which is not captured by the photochemical models.  Conversely, C$_3$H$_4$ is primarily formed from interconversion of other C$_3$ hydrocarbons \cite{05moses}, which do not display the same enrichment as C$_2$H$_2$.  Furthermore, the model of \citeA{05moses_sat} overestimates equatorial C$_3$H$_4$ and C$_4$H$_2$ by factors of $\sim5$ and $\sim4$, respectively, similar to that found by \citeA{10guerlet}.

\textbf{Methyl} \cite<CH$_3$, first detected by ISO,>[]{98bezard} is also a short-lived species produced directly from methane photolysis \cite{00moses}.  MIRI/MRS provides the first latitudinally-resolved measurements of CH$_3$, showing the same equator-to-pole decline as C$_2$H$_2$ and C$_3$H$_4$, along with a local minimum near \ang{30}N that may be related to the seasonal upwelling.  The seasonal model underpredicts the equatorial abundances by $\sim2.5$.  Unlike most other hydrocarbon species, CH$_3$ appears to be most depleted within the NPSV, by a factor of $\sim5$ compared to equatorial abundances.  This may reflect the annual-average insolation, with less CH$_4$ photolysis at the highest latitudes, although we note that methyl-methyl recombination reactions are dominant producer of C$_2$H$_6$, so it may have been mostly converted into the enriched ethane of the NPSV.  Alternatively, we note that the CH$_3$ abundance is very sensitive to the methane homopause pressure \cite{98bezard}, so the CH$_3$ depletion in the NPSV could potentially be caused by subsidence through the upper stratosphere and lower thermosphere that pushes the methane homopause to deeper pressures, providing a consistent picture with the enhanced C$_2$H$_4$, C$_2$H$_6$, and some other hydrocarbon abundances.

\textbf{Benzene} \cite<C$_6$H$_6$, first detected by>[]{01bezard} also differs substantially from neutral photochemical predictions, which would expect a distribution similar to that of C$_2$H$_2$.  Instead, we see the C$_6$H$_6$ emission at all latitudes (e.g., between strong C$_2$H$_2$ lines in Fig. \ref{fig:cxhy_fit}), with enhancement by 1.5-2.0$\times$ within the NPSV compared to mid-latitudes.  A similar latitudinal gradient was observed by Cassini/CIRS in the southern hemisphere between 2007 and 2012, with a slight enhancement within the SPSV \cite{15guerlet}.  Nevertheless, the peak abundance at 0.5 mbar remains $50\times$ smaller than the photochemical model prediction of \citeA{05moses_sat}, as previously found by \citeA{15guerlet}, and some $10\times$ smaller than the coupled ion-neutral chemistry of \citeA{23moses}.  As discussed by \citeA{16koskinen} and \citeA{23moses}, benzene on Saturn is greatly enhanced by ion chemistry, and increased production due to auroral-induced ion chemistry may play a role at high latitudes, but finding a specific match between models and data remains a challenge.

It is interesting to note that, of all the hydrocarbon species observed by MIRI in 2022, none of them show the substantial chemical consequences predicted by neutral photochemistry \cite{23moses} if the large influx of organic-rich ring material detected by Cassini during its Grand Finale in 2017 \cite{22serigano} were vapourised/ablated upon entry into the equatorial stratosphere.  Furthermore, no new nitriles were observed in 2022 (see below).  We thus favour the hypothesis of \citeA{23moses} that the ring material enters as small dust particles that do not ablate and affect the stratospheric composition.

\subsubsection{Exogenic Species}

Carbon dioxide \cite<CO$_2$, first detected by ISO,>[]{97feuchtgruber} is detected at all latitudes by MIRI/MRS, with a relatively uniform abundance with latitude ($0.60\pm0.05$ ppb at 0.5 mbar, Fig. \ref{fig:cxhy_all}e) and hints of a slightly lower abundance ($0.56\pm0.03$ ppb) within the NPSV.  The distribution is approximately consistent with the few measurements available from Cassini \cite{13abbas}, but is inconsistent with photochemical models that assume a globally constant flux of incoming oxygen species and predict CO$_2$ abundances that are greatest near the equator and tail off toward high latitudes \cite{05moses_sat}.  There are no strong latitudinal gradients that might imply a spatially-localised source \cite<e.g., recent comets,>[]{10cavalie}, or Enceladus plume and ring material entering at specific latitudes, but icy grain ablation remains too small a source to explain the measurements \cite{17moses}.  Indeed, the uniform CO$_2$ distribution derived here does not match the exogenic H$_2$O distribution derived from Herschel observations, which showed enhancements at low latitudes \cite{19cavalie}.  More work is required to robustly compare the CO$_2$ and H$_2$O distributions to elucidate their sources.

Unless Saturn suffers a large-scale impact event, or the equatorial ring influx does provide excess nitrile production \cite{23moses}, HCN is not an expected photochemical product on Saturn, as the regions of photolytic destruction of CH$_4$ and NH$_3$ are separated by hundreds of kilometres in the vertical.  Any detection of the 14.04 $\mu$m ($\nu_2$) HCN line is challenging due to blending with lines of C$_2$H$_2$.  We also note that the MRS wavelength calibration \cite{23argyriou} and spectral resolution \cite{23jones} creates fitting artefacts in channel 3B, preventing a rigorous study of upper limits in this preliminary study.  Nevertheless, we find that the MRS data support HCN abundances no larger than 1 ppb at $p<1$ mbar, an improvement over the previous upper limit of 22 ppb at $p<1$ mbar in the sub-millimetre from Herschel/SPIRE \cite{12fletcher_spire}.  We do not see any evidence of the $\nu_5$ band of HC$_3$N at 15.08 $\mu$m, again confirming an absence of nitriles related to ring influx \cite{23moses}.

%%%%%%%%%%%%%%%%%%%%%%%%%%%%%%%%%%%%%%%%%%%%%%%%%%%
%%%%%%%%%%%%%%%%%%%%%%%%%%%%%%%%%%%%%%%%%%%%%%%%%%%
%%%%%%%%%%%%%%%%%%%%%%%%%%%%%%%%%%%%%%%%%%%%%%%%%%%
\section{Conclusions}

This initial survey of JWST/MIRI observations of Saturn has revealed a wealth of new insights into the evolution of the seasonal atmosphere during northern summer (November 2022, $L_s=150^\circ$); demonstrated MIRI/MRS capabilities to observe extended, bright, rotating and moving planetary objects that are much larger than the fields-of-view; and provided a means to evaluate and mitigate challenges related to wavelength calibration, detector saturation, and instrumental artefacts for MIRI/MRS.  Spatially-resolved 4.9-27.9 $\mu$m maps of Saturn (three tiles spanning from the equator to the north pole) have been inverted to study the zonal-mean temperatures, windshears, aerosols, and gaseous composition from the cloud-forming region of the troposphere into the mid-stratosphere.  This includes the first maps of the transitional region of Saturn's spectrum between 5.1-6.9 $\mu$m, where both thermal emission and scattered sunlight shape the spectrum, which were inaccessible to the VIMS and CIRS instruments on Cassini.  

Although the JWST data reduction pipeline continues to evolve, we have presented algorithms for correcting artefacts such as wavelength calibration offsets, correction of partially saturated spectral regions, and the development of `flat-field' corrections by exploiting multiple dithers on the same target.  As the MIRI/MRS spectral cubes required significant processing outside of the pipeline, we perform customised cleaning on each tile and dither independently, to prevent the blending of artefacts that, in the most extreme cases, can completely obscure the spatial structure of Saturn itself.  With these artefacts removed, we summarise the initial survey of Saturn as follows:

\begin{enumerate}
    \item \textbf{Saturn's banded structure:}  Latitudinal gradients in temperatures and windshears show strong correlations with the locations of Saturn's mid-latitude eastward and westward jets \cite{11garcia}, with the contrasts between cool zones and warmer belts confirming the decay of the zonal winds with altitude via the thermal windshear equation.  Conversely, gradients in reflectivity (and the derived aerosol structure from 4.9-7.3 $\mu$m) show similarities to albedo contrasts in observations acquired by Hubble in September 2022 \cite{23simon}, with a decrease in aerosols in distinct steps from equator to pole, with the highest cloud base and thickest tropospheric hazes within the broad reflective band equatorward of \ang{15}N.  A narrow equatorial band ($<$\ang{5}N) is dark at 5 $\mu$m, coinciding with the location where the narrow upper-tropospheric jet was previously detected \cite{11garcia}.  The 5-$\mu$m brightness has evolved with time, such that a pole-encircling band from \ang{62}-\ang{78}N is now the brightest on the planet, coinciding with a dearth of aerosol opacity, whereas the polar domain (interior to the hexagon) is the darkest at 5 $\mu$m due to an absence of upper-tropospheric aerosols and an optically thick and compact cloud at higher pressures. We no longer see the 5-$\mu$m-bright, aerosol-depleted band near 35-40$^\circ$N that had dominated the appearance of the northern hemisphere after the 2010-11 storm,  consistent with a re-population of the band by Saturn's seasonal aerosols in the decade since the storm. 

    \item \textbf{North Polar Stratospheric Vortex (NPSV):}  The seasonal stratospheric vortex that developed poleward of \ang{78}N during northern spring \cite{18fletcher_poles} remained present in 2022, with warmer temperatures than those measured in 2017, now approaching the same temperatures as those observed in the SPSV during southern summer (2004-05).  Radiative models \cite<e.g.,>[]{14guerlet}, combined with our Earth-based vantage point \cite{22blake}, suggest that the visibility of the warm NPSV will decline substantially in the next 1-2 years before autumn equinox.  The sharp thermal gradient at the edge of the NPSV promotes negative windshear, and thus westward stratospheric winds entraining the NPSV for $p<10$ mbar. MIRI/MRS did not detect evidence of the vertices of Saturn's hexagon, meaning that we will need to wait until the 2040s for our next infrared views of the hexagon.  However, the bright North Polar Cyclone (NPC) was still visible, embedded within the NPSV.  The NPSV was enriched in several stratospheric hydrocarbon species (ethane, acetylene, ethylene, benzene, diacetylene) due to a combination of polar subsidence over the summer pole, and potential ion-neutral chemistry at high latitudes. 

    \item \textbf{Cyclones and anticyclones: }  Channel 1-short of MIRI/MRS offers the best opportunity to probe the fine-scale cloud banding, as well as discrete features.  We observe contrasts associated with cyclones (bright and aerosol-free) and anticyclones (dark and cloudy) near \ang{48}N, \ang{10}N and \ang{62}N, but these could not be identified in contemporaneous amateur observations, nor Hubble observations $\sim7$ weeks earlier.  This strongly argues for near-simultaneous MIRI/MRS spectroscopy and NIRCAM (or Hubble) imaging in future observing programmes, alongside long-term records of variability from ground-based facilities.

    \item \textbf{Saturn's Equatorial Stratospheric Oscillation:} Windshears derived from retrieved temperature gradients reveal diagonally-stacked shear zones that rise to higher altitudes with higher latitudes, connecting the stratosphere throughout the low-latitudes equatorward of \ang{30}N.  MIRI/MRS reveals warm and cool temperature anomalies at the equator (and off-equatorial anomalies at \ang{13}N) that are consistent with the downward propagation of the oscillatory pattern over a decade of pressure in the five years since the 2017 Cassini observations.  A warm equatorial anomaly centred near 0.7 mbar is responsible for the bright equatorial band of emission observed in CH$_4$ and C$_2$H$_2$ emission, but this is opposite to the dark band observed from the ground one Saturnian year earlier \cite<1993-1995,>[]{08orton_qxo, 22blake}, raising doubts about the semi-annual nature (i.e., 15-year period) of the equatorial oscillation. Thermal wind calculations imply the presence of an equatorial westward jet near 1-5 mbar in 2022 superimposed onto a broader region of eastward flow, but future joint campaigns between JWST and ALMA are necessary to confirm the validity of the thermal winds derived here.

    \item \textbf{Reversal of Saturn's interhemispheric stratospheric circulation:}  MIRI/MRS observations in 2022 reveal cooler temperatures in the \ang{10}-\ang{40}N domain compared to Cassini in 2017.  This coincides with local minima in several hydrocarbon species in 2022 (notably C$_2$H$_2$, C$_2$H$_6$, C$_4$H$_2$ and CH$_3$), opposite to the local maxima detected by Cassini \cite{09guerlet, 10guerlet, 15sylvestre}.  This adiabatic cooling and hydrocarbon-depleted air implies a transition from wintertime subsidence to summertime upwelling in the northern hemisphere, as part of the seasonal reversal of Saturn's interhemispheric stratospheric circulation \cite{12friedson, 22bardet}.

    \item \textbf{Stacked aerosol layers:}  We present the first assessment of aerosol opacity in 5.1-6.8 $\mu$m range, requiring two aerosol layers to reproduce the thermal emission and reflected sunlight components.  The \textbf{upper aerosol layer}, possibly a photochemically-produced haze related to diphosphine \cite{80ferris}, is thickest, highest (near 200 mbar) and most extended at the equator, but deeper (300 mbar) and more compact at mid-latitudes, before becoming negligible poleward of \ang{60}N.  This layer is likely to be the same as that detected by Cassini/ISS \cite{13roman}, and coincides with a region of localised radiative heating detected by Cassini/CIRS \cite{07fletcher_temp}.  The \textbf{deeper aerosol layer}, possibly associated with condensates of NH$_3$ and other species, resides at 1-2 bars, being shallowest (1.2 bars) near \ang{10}N and deepest (2.6 bars) and most compact in the polar domain, responsible for both the dark thermal emission at 5 $\mu$m and the absence of reflectivity at 5.2 $\mu$m.  Aerosol fitting for MIRI/MRS is somewhat degenerate, and further constraints could be provided in future by NIRSpec observations of reflected sunlight. 

    \item \textbf{Tropospheric species:}  Latitudinal variability of \textbf{phosphine} appears to be most confined to $p<1$ bar, displaying an equatorial maximum \cite{09fletcher_ph3} and general decline in abundance towards the polar domain.  Regions of higher aerosol opacity generally correspond to regions of elevated PH$_3$, suggesting this disequilibrium species is most abundant when aerosols shield the molecule from photolysis.  \textbf{Ammonia} displays a strong equatorial enrichment $<5^\circ$ of the equator (450-650 ppm), similar to the NH$_3$-enriched column in Jupiter's equatorial zone \cite{06achterberg, 17li, 16depater} and suggesting similar dynamical processes at the equators of both gas giants.  NH$_3$ also displays different latitudinal distributions above and below the condensed aerosols near 1.7 bars, suggestive of differing circulation patterns at different heights.  Neither PH$_3$ nor NH$_3$ display consistent belt/zone contrasts at mid-latitudes, suggesting secondary circulation patterns associated with mid-latitude Ferrel cells may be weak.  Tropospheric \textbf{water} is mapped for the first time in its condensation region (i.e., above the expected H$_2$O cloud), with estimates at 3.3 bars varying from 200 ppb at the equator, to 10 ppb near \ang{5}N, and a distinct step in the abundance near $\sim$\ang{40}N from 145 ppb at low latitudes to 80 ppb at higher latitudes.

    \item \textbf{Stratospheric chemicals:}  Short-lived hydrocarbons, such as \textbf{acetylene, diacetylene, methylacetylene, ethylene, and methyl}, tend to track the annual-average insolation \cite<i.e., peaking at low latitudes, consistent with the predictions of seasonal photochemical models,>[]{05moses_sat}, whereas longer-lived species (\textbf{ethane, propane, and possibly benzene}) appear to be more influenced by long-term stratospheric circulation redistributing gases to higher latitudes.  Acetylene has a steeper vertical gradient within the NPSV than elsewhere, and several species (ethane, ethylene, benzene, diacetylene) show strong enrichments within the NPSV.  Methyl is mapped with latitude for the first time, showing a factor of $\sim5$ decrease from equator to pole.  Ethylene is detected in the NPSV for the first time.  The benzene abundance remains an order of magnitude smaller than the predictions of photochemical models.  \textbf{Carbon dioxide} is detected at all latitudes with a relatively uniform distribution, lacking strong latitudinal gradients that might imply a spatially-localised source of exogenic oxygen species.

    \item \textbf{Absence of nitriles:}  None of the hydrocarbon distributions support the substantial chemical consequences predicted by neutral photochemistry \cite{23moses} if the large influx of ring material detected by Cassini in 2017 \cite{22serigano} were vapourised upon entry into the equatorial stratosphere.  Neither HCN nor HC$_3$N, both predicted by nitrile chemistry related to ring influx, are detected by MIRI. This suggests that ring material influx is not strongly influencing Saturn's equatorial stratosphere.
   
\end{enumerate}

We caution the reader that the MIRI/MRS calibration continues to evolve, and that refined cleaning techniques will enable future in depth studies - in particular, removal of instrument artefacts from the 17-28 $\mu$m domain will enable studies of temperature and para-H$_2$ at longer wavelengths; plus a more sensitive search for stratospheric emission lines throughout the MIRI spectrum.  Future joint campaigns with other JWST instruments (NIRSpec and NIRCAM) would aid in breaking degeneracies in aerosol retrievals, and collaboration with ALMA would allow a direct comparison of stratospheric winds and thermal winds associated with equatorial and polar jets.  

JWST observations of the Saturn system in November 2022 ($L_s=150^\circ$) have revealed the wealth of possibilities offered by IFU spectroscopy from space in the mid-infrared.  But Saturn's atmosphere will continue to change with the onset of northern autumn (May 2025, $L_s=180^\circ$), and we hope that these Cycle-1 observations will mark the starting point for a long-term MIRI legacy programme to track the seasonal evolution of Saturn's circulation and chemistry through to the next southern summer solstice (April 2032, $L_s=270^\circ$), completing the seasonal assessment begun by Cassini.

%%%%%%%%%%%%%%%%%%%%%%%%%%%%%%%%%%%%%%%%%%%%%%%
%
% DATA SECTION and ACKNOWLEDGMENTS
%
%%%%%%%%%%%%%%%%%%%%%%%%%%%%%%%%%%%%%%%%%%%%%%%

\section*{Open Research Section}
%Level-3 calibrated Saturn MIRI/MRS data from the standard pipeline are available directly from the MAST archive \cite{MAST1247_MIRI}.  Hubble observations used for comparison were acquired by the Outer Planets Legacy Program \cite{MAST_OPAL}.

%The NEMESIS radiative transfer and retrieval code \cite{08irwin} used in this study is open-access and is available for download \cite{22irwin_nemesis}.

%The JWST calibration pipeline is available via \citeA{23bushouse}, this work used version \verb|1.9.4|.

%The custom pipeline and data processing code developed in this study is available at \citeA{23king_pipeline}.

%The data products produced in this study, including synthetic flat fields, zonal average spectra and quick-look visualisations, are available from \citeA{23fletcher_data}.

%\section*{Open Research Section}
Level-3 calibrated Saturn MIRI/MRS data from the standard pipeline are available directly from the MAST archive \cite{MAST1247_MIRI} via \url{http://dx.doi.org/10.17909/wjpz-7383}.  Hubble observations used for comparison were acquired by the Outer Planets Legacy Program \cite{MAST_OPAL} via \url{https://archive.stsci.edu/prepds/opal/}, also available here: \url{http://dx.doi.org/10.17909/T9G593}.

The NEMESIS radiative transfer and retrieval code \cite{08irwin} used in this study is open-access and is available for download \cite{22irwin_nemesis} from GitHub or Zenodo (\url{https://doi.org/10.5281/zenodo.5816724}).

The JWST calibration pipeline \cite{23bushouse} is available at \url{https://github.com/spacetelescope/jwst}, this work used version \verb|1.9.4|.

The custom pipeline and data processing code \cite{23king_pipeline} developed in this study is available at \url{https://github.com/JWSTGiantPlanets/pipelines/} (\url{https://doi.org/10.5281/zenodo.7891560}).

The data products produced in this study \cite{23fletcher_data}, including synthetic flat fields, zonal average spectra and quick-look visualisations, are available at \url{https://github.com/JWSTGiantPlanets/saturn-atmosphere-miri} (\url{https://doi.org/10.5281/zenodo.7891588}).

\acknowledgments
Fletcher, King, and Roman were supported by a European Research Council Consolidator Grant (under the European Union’s Horizon 2020 research and innovation programme, grant agreement No 723890) at the University of Leicester.  Harkett was supported by an STFC studentship; Melin was supported by an STFC James Webb Fellowship (ST/W001527/1).  Hammel and Milam acknowledge support from NASA JWST Interdisciplinary Scientist grant 21-SMDSS21-0013.

We wish to express our gratitude to the JWST support team for their patience and perseverance as we designed these MRS observations - in particular Beth Perriello, Bryan Holler, Misty Cracraft, Tony Roman, and John Stansberry for their aid in setting up the observations in APT, and David Law for his tireless support as we developed codes to interpret MIRI/MRS data.  We are grateful to Conor Nixon, Imke de Pater, Patrick Irwin, Pablo Rodriguez-Ovalle and Thierry Fouchet for their helpful discussions during the development of this work.  We thank amateur observers Chris Go, Trevor Barry, Anthony Wesley, and Tiziano Olivetti for their efforts to identify discrete features during the JWST MIRI observation epoch.  We thank two reviewers, Glenn Orton and Bruno B\'{e}zard, for their thorough critiques of this article.   This research used the ALICE High Performance Computing Facility at the University of Leicester. 

This work is based on observations made with the NASA/ESA/CSA James Webb Space Telescope. The data were obtained from the Mikulski Archive for Space Telescopes at the Space Telescope Science Institute, which is operated by the Association of Universities for Research in Astronomy, Inc., under NASA contract NAS 5-03127 for JWST. These observations are associated with program 1247 (PI: Fletcher).  JWST observations were compared to data acquired from the NASA/ESA HST Space Telescope, associated with OPAL program (PI: Simon, GO13937), and archived by the Space Telescope Science Institute, which is operated by the Association of Universities for Research in Astronomy, Inc., under NASA contract NAS 5-26555. 

For the purpose of Open Access, the corresponding author has applied a CC-BY public copyright licence to any Author Accepted Manuscript (AAM) version arising from this submission.

%%%%%%%%%%%%%%%%%%%%%%%%%%%%%%%%%%%%%%%%%%%%%%%
% REFERENCES and BIBLIOGRAPHY
%
% \bibliography{<name of your .bib file>} don't specify the file extension
% don't specify bibliographystyle
%
%%%%%%%%%%%%%%%%%%%%%%%%%%%%%%%%%%%%%%%%%%%%%%%

\bibliography{references.bib}

\section*{Supplemental Material}
\label{supplement}

\begin{figure}
% \centering
\noindent\includegraphics[width=\textwidth]{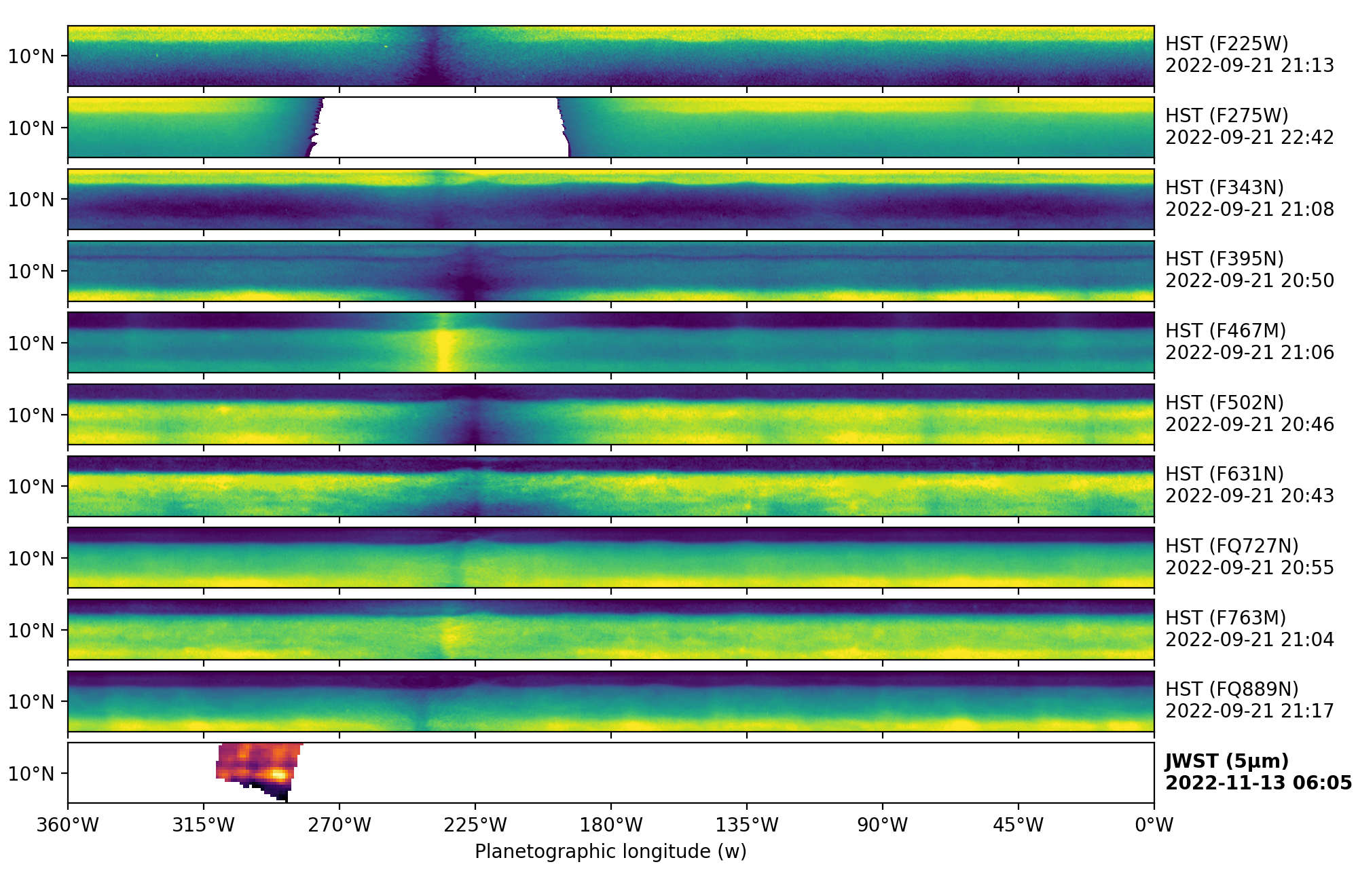}
\caption{Comparison of the \ang{10}N, \ang{290}W warm feature on Saturn with HST observations using a range of different filters. Each subplot shows a planetographic cylindrical map projection, with a \ang{20} latitude range (i.e. from \ang{0}N to \ang{20}N for this figure). The variability in zonal wind speed with latitude and time difference between the HST and JWST observations makes it difficult to match specific discrete features between the observations.}
\label{fig:feature_10n}
\end{figure}

\begin{figure}[h]
\centering
\includegraphics[width=\textwidth]{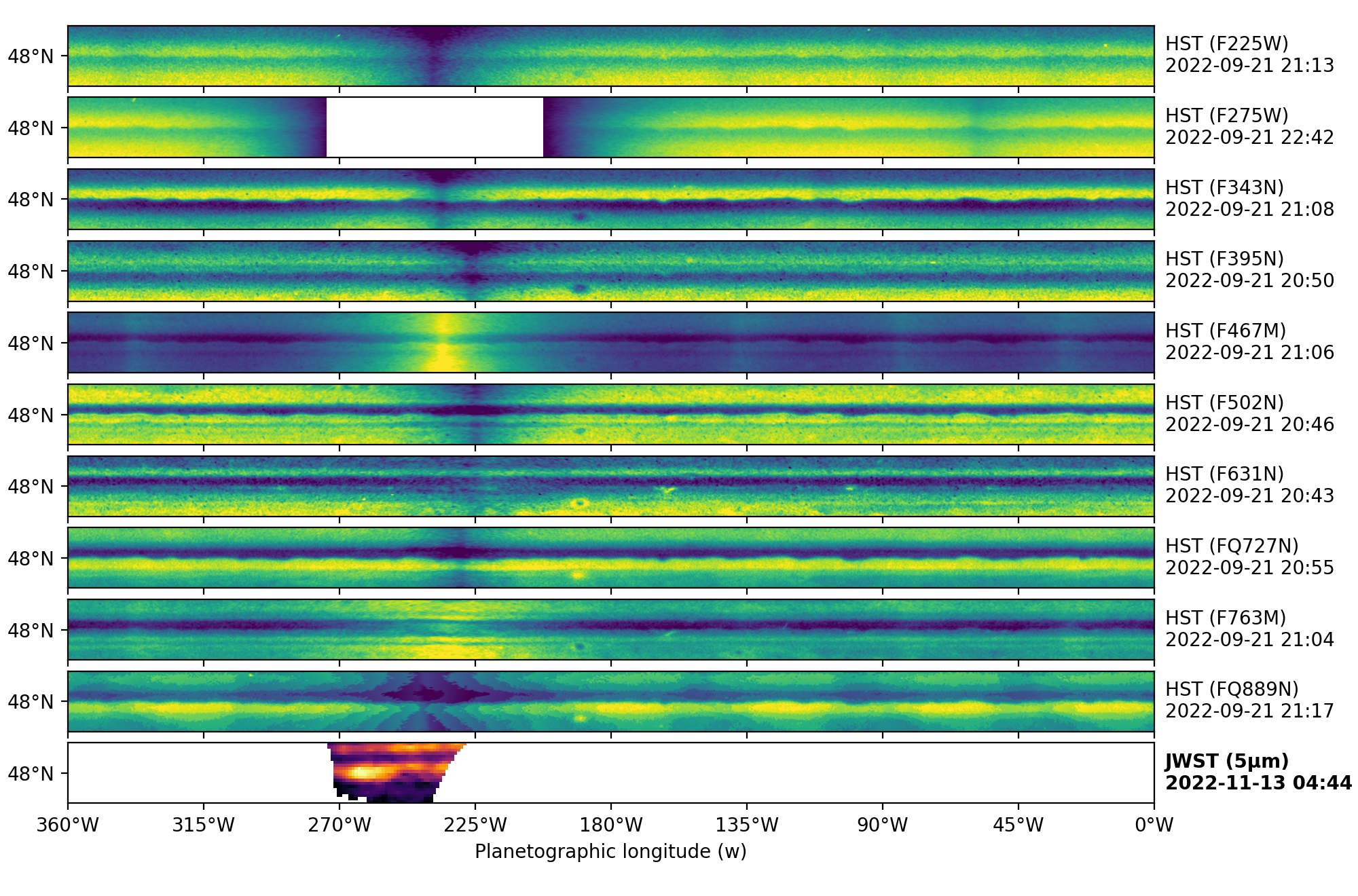}
\caption{Comparison of the \ang{48}N, \ang{261}W 5-$\mu$m warm feature with HST observations. See Fig \ref{fig:feature_10n} for more detail.   }
\label{fig:feature_48n_warm}
\end{figure}

\begin{figure}[h]
\centering
\includegraphics[width=\textwidth]{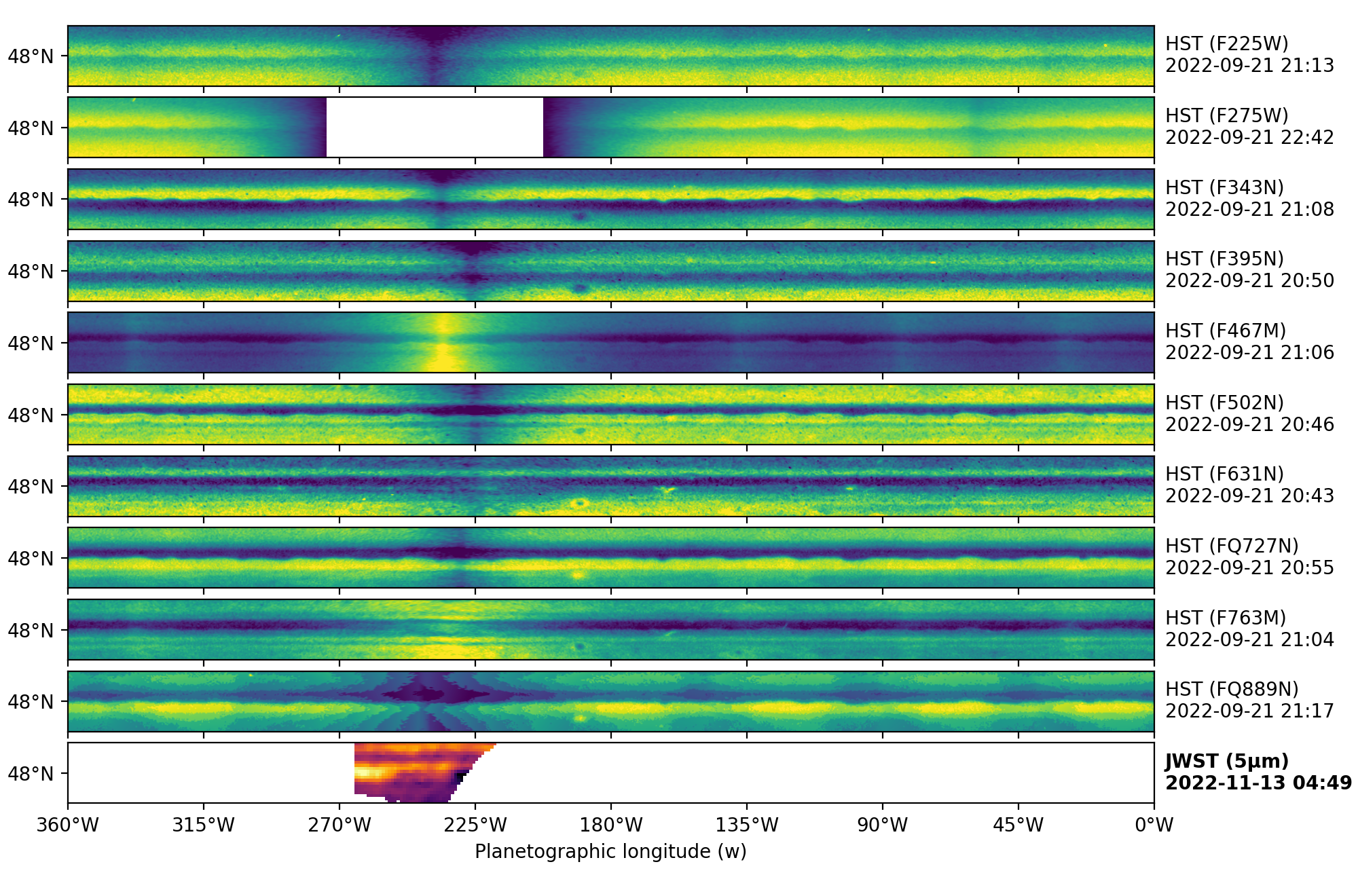}
\caption{Comparison of the \ang{48}N, \ang{230}W 5 $\mu$m cool feature with HST  observations. See Fig \ref{fig:feature_10n} for more detail. Note that the anticyclonic vortex at \ang{42}N observed in the HST data is not seen in the MIRI observation.}
\label{fig:feature_48n_cool}
\end{figure}

\begin{figure}[h]
\centering
\includegraphics[width=\textwidth]{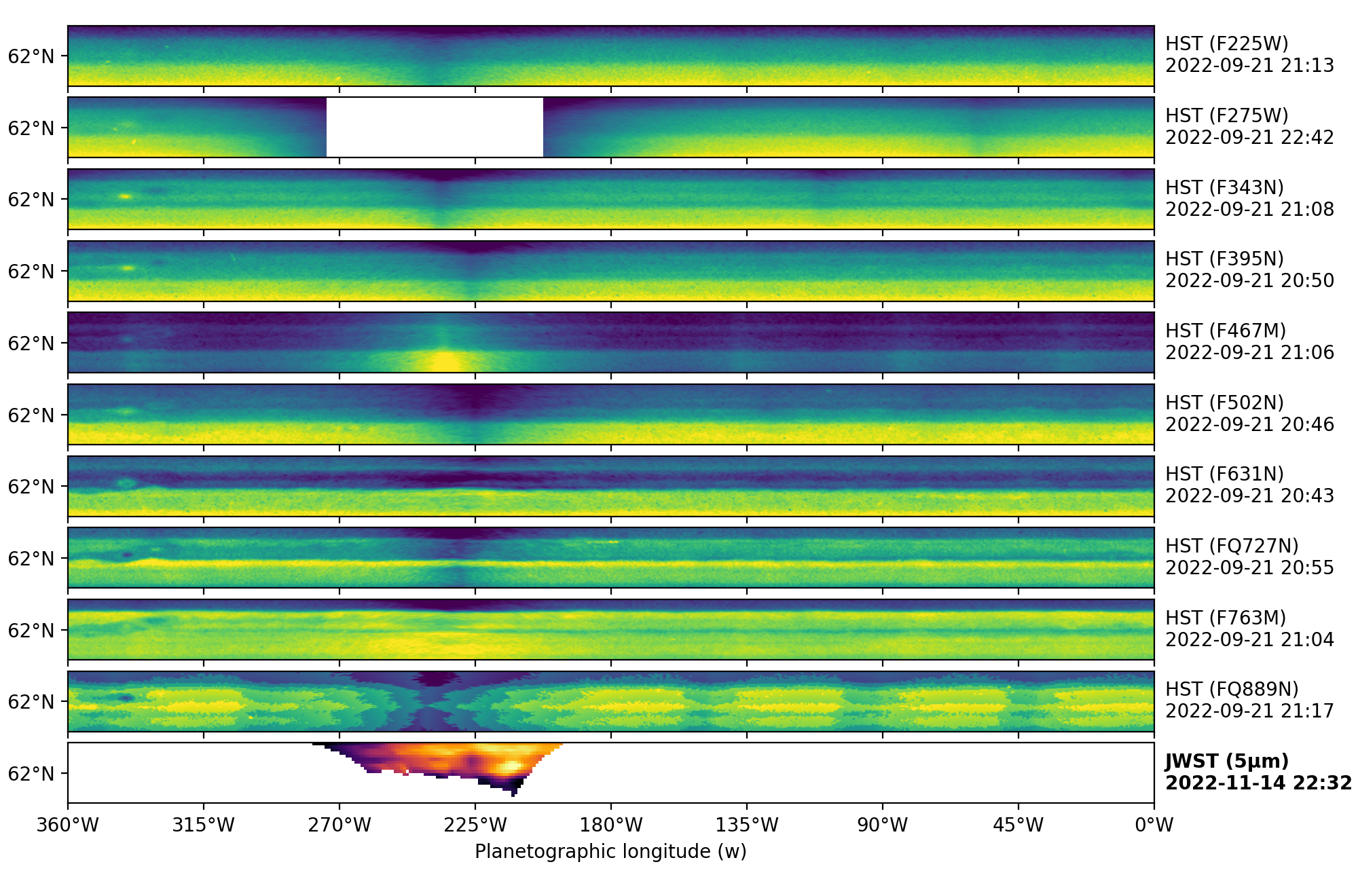}
\caption{Comparison of the \ang{62}N, \ang{215}W 5 $\mu$m warm feature with HST  observations. See Fig \ref{fig:feature_10n} for more detail.}
\label{fig:feature_62n}
\end{figure}

Figures \ref{fig:feature_10n}-\ref{fig:feature_62n} show comparisons of the discrete features observed in the November-2022 MIRI/MRS observations to HST observations of the corresponding latitudes in September 2022.  Discrete features will have drifted substantially in the $\sim7$ weeks between observations, so these four figures are meant to show the longitudinal context for the more focussed MRS maps.  They allow the reader to observe whether any similar vortices or other structures are seen in the same latitude band as those observed by MRS - see the main text for a full discussion of these features.

\begin{figure}[h]
\centering
\includegraphics[width=0.8\textwidth]{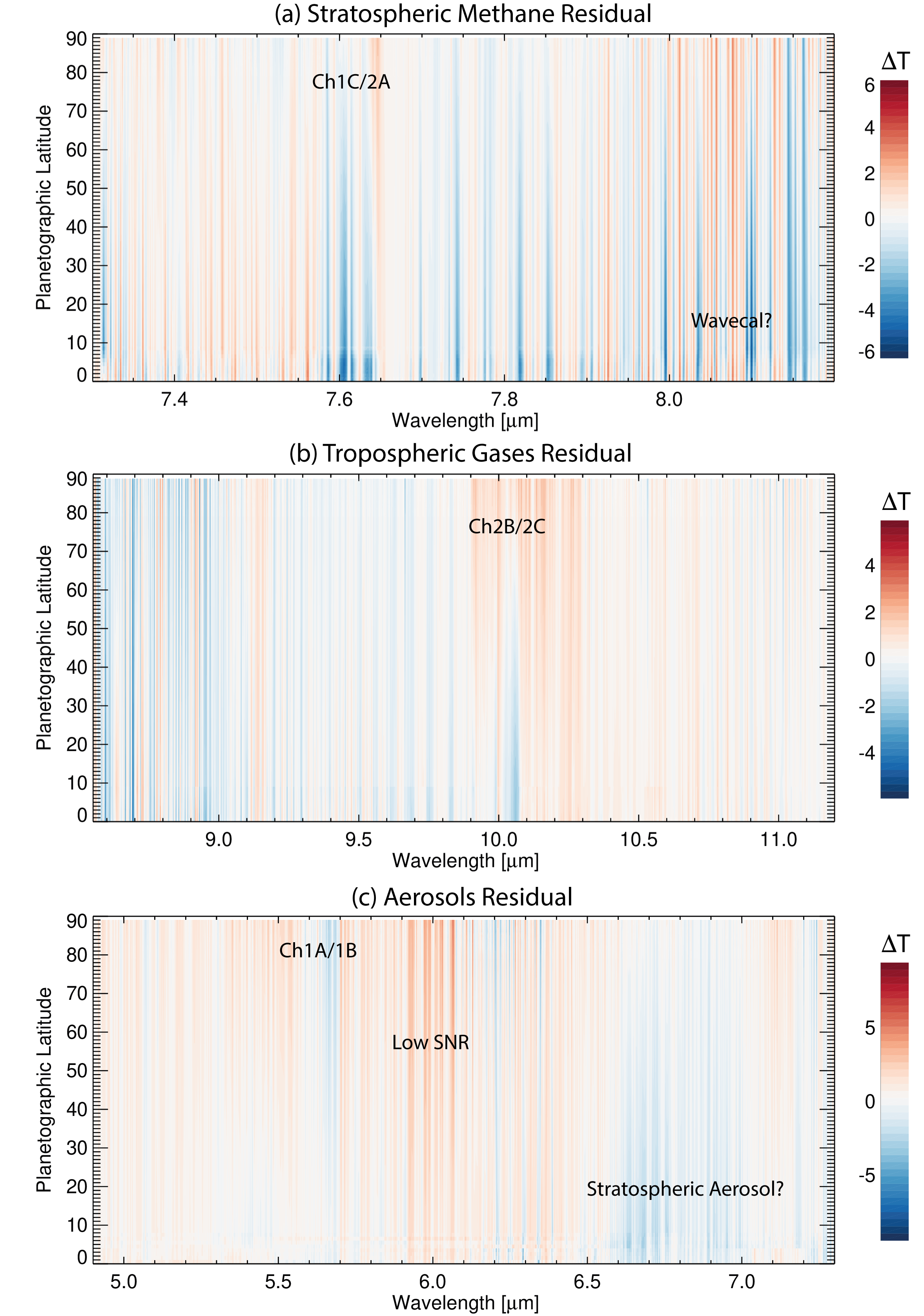}
\caption{Brightness temperature residuals in Kelvin between model and data as a function of latitude and wavelength after stage 2 and 3 of our multi-stage retrieval scheme.  Blue regions imply underfitting, red regions imply overfitting.  Possible causes of discrepancies are noted in the appropriate locations, including overlap regions between adjacent MIRI subbands, and persistent wavelength calibration issues.}
\label{fig:resid}
\end{figure}

\begin{figure}[h]
\centering
\includegraphics[width=\textwidth]{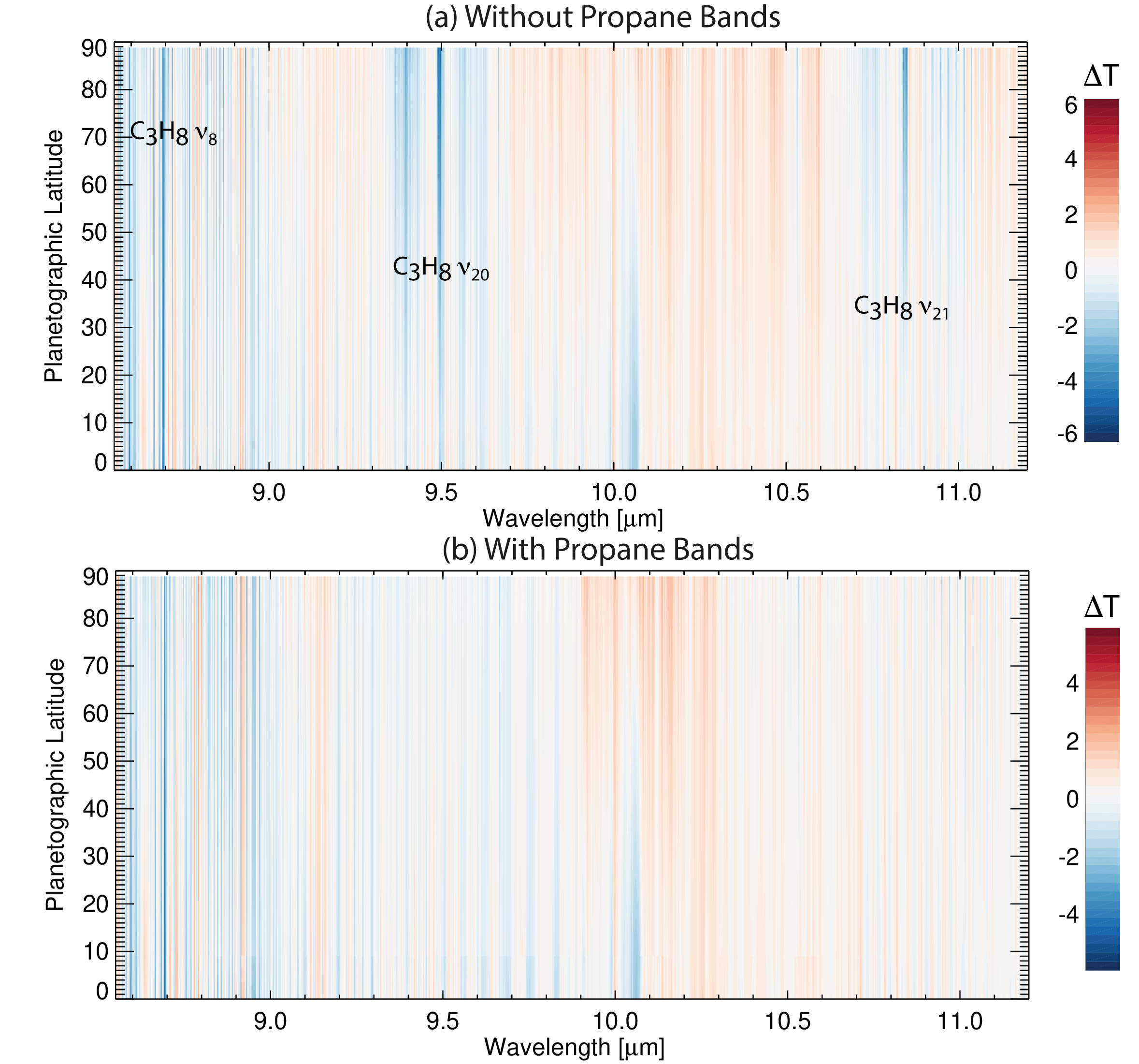}
\caption{Brightness temperature residuals in Kelvin before and after the addition of pseudolines of C$_3$H$_8$ from \citeA{13sung}. Blue regions imply underfitting, and two of the three propane bands in this range are clearly visible in panel (a), increasing in strength with latitude due to the warmer polar temperatures.  Residuals are much improved in panel (b), when propane is included.}
\label{fig:residc3h8}
\end{figure}

Figures \ref{fig:resid} and \ref{fig:residc3h8} show residuals between the NEMESIS models and the MRS measurements, where blue regions imply underfitting of the data, and red regions imply overfitting.  These figures were used to identify problems with wavelength calibration, the overlap between MIRI channels, or residuals implying missing spectral bands in our line database.

\begin{figure}[h]
\centering
\includegraphics[width=1.3\textwidth,center]{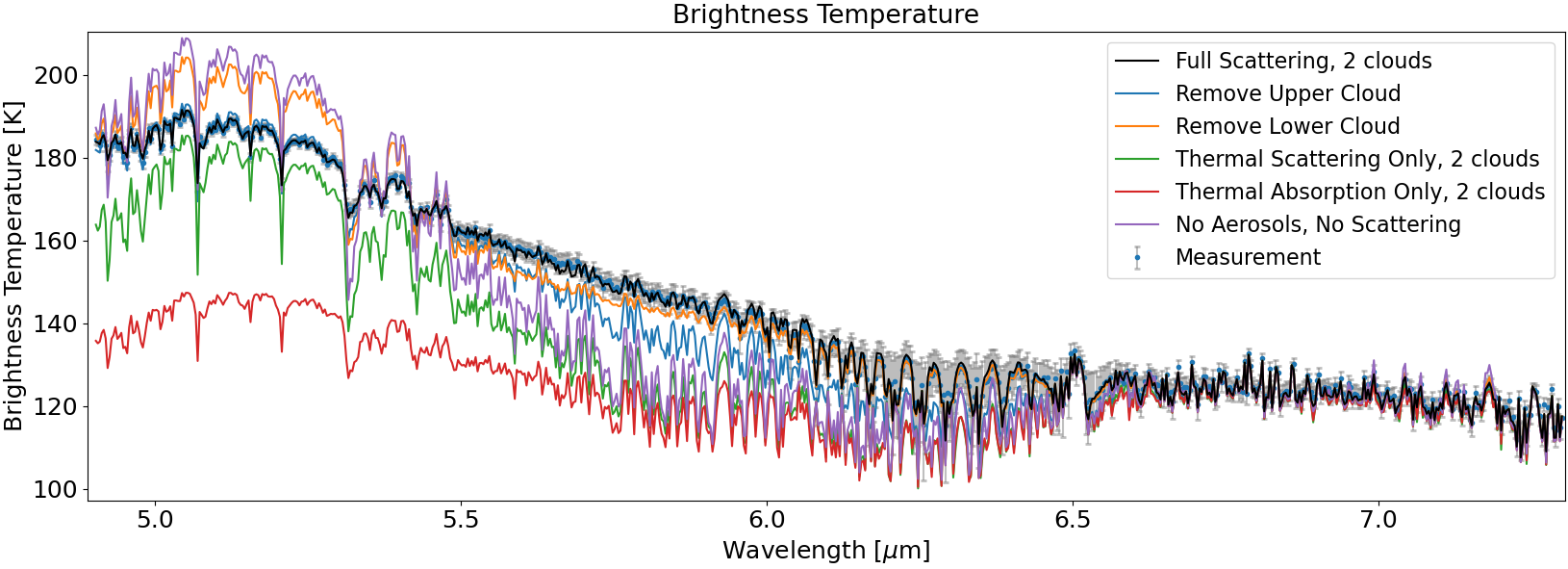}
\caption{Examples of forward models to the 4.9-7.3 $\mu$m spectral range (\ang{20}N) using different approaches to modelling aerosols.  The best fit with multiple scattering of scattered sunlight and thermal emission by our two-tier cloud model is shown by the black curve, the effects of removing either the upper (blue) or lower (orange) clouds is also shown.  If we turn off reflected sunlight but retain thermal scattering, we get the green line.  Turning off scattering entirely produces the red forward model (i.e., a purely absorbing cloud).  With no aerosols in the model, we get the high brightness temperatures of the purple line.  Beyond the CH$_4$ emission at 6.5 $\mu$m, the different models are largely indistinguishable from one another, supporting the exclusion of scattering and aerosols at longer wavelengths. }
\label{fig:cloud_tests}
\end{figure}

Figure \ref{fig:cloud_tests} shows a single spectrum at \ang{20}N, and the best-fitting NEMESIS model with multiple scattering of reflected sunlight and thermal emission.  We then remove parameters from the model to show their individual effects on the models - removing one of the two clouds; turning off reflected sunlight, and removing aerosols entirely.  This figure confirms that all models converge longward of $\sim6.5$ $\mu$m, such that longer wavelength MIRI channels can be modelled without the inclusion of aerosols or multiple scattering.

\begin{figure}[ht]
\centering
\includegraphics[width=0.5\textwidth]{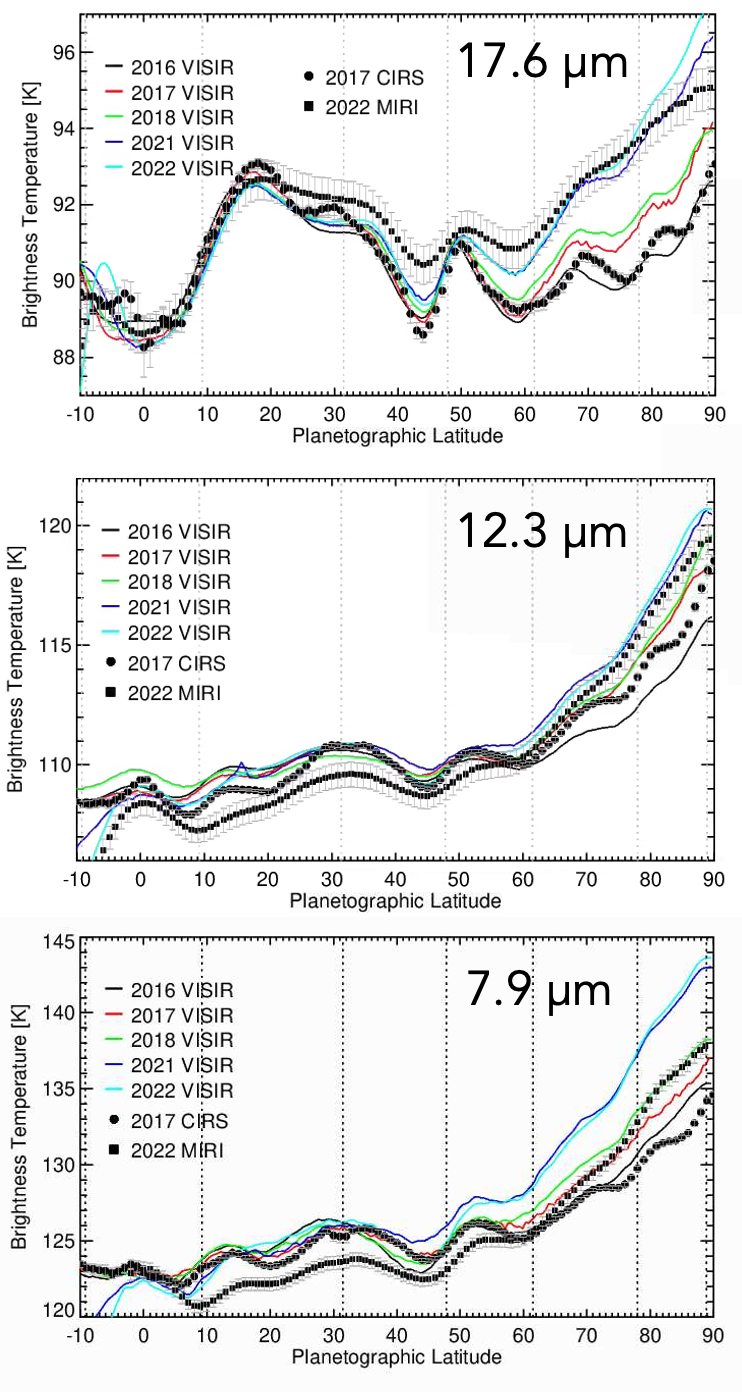}
\caption{\textbf{Comparison of VLT/VISIR observations from 2016-2022 to MIRI/MRS observations in 2022 and Cassini/CIRS in 2017.}  Ground-based observations were scaled to match a low-latitude CIRS average by \citeA{22blake}, which causes the mismatch at 7.9 $\mu$m where the low-latitudes are observed to have cooled with time. Nevertheless, the large-scale asymmetries and belt/zone contrasts are captured by the VISIR data, with the strongest brightness temperature gradients co-located with the peaks of eastward jets (vertical dotted lines in each panel).}
\label{fig:visir_comp}
\end{figure}

Finally, Fig. \ref{fig:visir_comp} compares Cassini/CIRS and MIRI/MRS data to central-meridian scans from ESO's VISIR instrument on the VLT between 2016 and 2022 \cite{22blake}.  CIRS and MRS data were convolved to match the VISIR filter functions with telluric contamination. As described in the main text, the VISIR observations reproduce the banded structure of Saturn revealed by Cassini and JWST, and highlight the challenges associated with calibrating ground-based datasets.

% \subsection*{VISIR Comparison} \label{app:visir_comp}

%Reference citation instructions and examples:
%
% Please use ONLY \cite and \citeA for reference citations.
% \cite for parenthetical references
% ...as shown in recent studies (Simpson et al., 2019)
% \citeA for in-text citations
% ...Simpson et al. (2019) have shown...
%
%
%...as shown by \citeA{jskilby}.
%...as shown by \citeA{lewin76}, \citeA{carson86}, \citeA{bartoldy02}, and \citeA{rinaldi03}.
%...has been shown \cite{jskilbye}.
%...has been shown \cite{lewin76,carson86,bartoldy02,rinaldi03}.
%... \cite <i.e.>[]{lewin76,carson86,bartoldy02,rinaldi03}.
%...has been shown by \cite <e.g.,>[and others]{lewin76}.
%
% apacite uses < > for prenotes and [ ] for postnotes
% DO NOT use other cite commands (e.g., \citeA, \cite, \citeyear, \nocite, \citealp, etc.).
%

\end{document}